# Insights into Xona Pulsar LEO PNT: Constellation, Signals, and Receiver Design

Jérôme Leclère, Thyagaraja Marathe, Tyler G. R. Reid, *Xona Space Systems*

**Biographies**

**Dr. Jérôme Leclère** is a GNSS FPGA Designer at Xona Space Systems in Montreal, Canada. He holds an engineering degree in Electronics and Signal Processing from ENSEEIHT (France, 2008) and a Ph.D. in GNSS from EPFL (Switzerland, 2014). Dr. Leclère has worked on hardware-based GNSS receivers and GNSS acquisition, as well as on signal processing algorithms for lidar and satellite phased antenna arrays. Currently, his work focuses on Pulsar signals and receiver development.

**Dr. Thyagaraja Marathe** is a Staff Research Engineer at Xona Space Systems, specializing in PNT using LEO satellite constellations. He previously worked at Rx Networks and Accord Software on GNSS corrections and receiver development. He holds a Ph.D. in Geomatics Engineering from the University of Calgary, where he also completed postdoctoral research. His interests include LEO PNT, high-accuracy positioning, and GNSS receiver design. He also holds degrees from BITS Pilani and VTU, India.

**Dr. Tyler G. R. Reid** is a co-founder and CTO of Xona Space Systems whose focus is commercial satellite navigation services from Low Earth Orbit. Tyler previously worked as a Research Engineer at Ford Motor Company in localization and mapping for self-driving cars. He was also a Software Engineer at Google and a lecturer at Stanford University where co-taught the course on GPS. Tyler received his Ph.D. ('17) and M.Sc. ('12) in Aeronautics and Astronautics from Stanford where he worked in the GPS Research Lab and his B.Eng in Mechanical Engineering from McGill ('10). He is a recipient of the RTCA Jackson Award.

**Abstract**

The landscape of global navigation satellite systems (GNSS) is expanding with the emergence of low Earth orbit (LEO) constellations such as Pulsar, which are expected to play a key role in the future of positioning, navigation, and timing (PNT). LEO-based systems provide advantages including stronger signals for greater robustness, faster dynamics that aid convergence and multipath mitigation, and shorter time to first fix (TTFF) enabled by high data rates. These benefits, however, come with changes in signal behavior and constellation geometry that require careful consideration in receiver design.

This paper investigates Pulsar properties using a GNSS simulator, analyzing parameters such as satellite pass duration, elevation, Doppler shift, Doppler rate, range, and number of satellites in view. Comparisons with GPS highlight the differences introduced by LEO operation. The analysis examines temporal evolution, statistical distributions, and maximum and minimum values. Beyond these statistical insights, the study explores interdependencies between parameters and differences across satellites, providing additional perspective. Evaluations are performed at multiple latitudes to ensure a worldwide perspective, and the impact of applying different elevation masks is discussed where relevant.

Building on these findings, the paper assesses Pulsar's impact on receiver design from two standpoints: design considerations, addressing expanded Doppler ranges, higher Doppler rates, and unique constellation structure; and design optimizations, exploiting parameter analyses and interdependencies (e.g., Doppler rate vs Doppler) to refine acquisition strategies and applying prediction and prioritization techniques to avoid unnecessary computations. Together, these optimizations can reduce acquisition time and lower receiver power consumption.

The results provide guidance for developing Pulsar-enabled and LEO-capable GNSS receivers, and contribute to understanding the evolving LEO-based navigation landscape.

## 1. INTRODUCTION

The global navigation satellite system (GNSS) landscape is undergoing a major transformation with the advent of low Earth orbit (LEO) constellations for positioning, navigation, and timing (PNT). Unlike legacy medium Earth orbit (MEO) GNSS systems such as GPS, Galileo, BeiDou or GLONASS, LEO-based systems offer several advantages that can significantly enhance navigation performance. Key benefits include stronger received signals, improving indoor positioning and resilience to jamming; higher orbital dynamics, which enhance multipath mitigation and accelerate convergence in precise point positioning (PPP) algorithms; and increased data rates, enabling faster cold-start times to first fix (TTFF) and supporting advanced correction services (Reid et al., 2020).

Alongside these advantages, LEO PNT constellations also introduce fundamental changes to signal characteristics and constellation properties compared to legacy GNSS, requiring adaptations in receiver design. This paper addresses this by analyzing the Pulsar PNT system developed by Xona Space Systems, both at the constellation and signal level.

The first objective of this paper is to provide insights into Pulsar's constellation and signal properties. A Pulsar-enabled GNSS simulator is used to evaluate parameters such as satellite pass duration, elevation, Doppler, Doppler rate, and the number of satellites in view. These characteristics are evaluated for the Pulsar Full Operational Capability (FOC) constellation (altitude near 1080 km), and compared to GPS to highlight the key differences introduced by LEO operation. When relevant, values for the Pulsar in-orbit-validation (IOV) satellite (Pulsar-0, launched in June 2025 at an altitude near 520 km) are also provided.

Beyond providing timing and statistical overviews, this analysis also investigates interdependencies between parameters. For example, while satellites at high elevation consistently exhibit low Doppler shifts, satellites at low elevation do not always correspond to high Doppler shifts. Similarly, the duration of a satellite pass is closely tied to the initial Doppler, influencing how long a receiver can track a satellite before it moves out of view. Identifying such relationships is crucial for receiver optimization.

The second objective is to assess the impact of LEO PNT signals on receiver design from two perspectives: (1) design considerations, focused on how key signal characteristics (higher Doppler and Doppler rate, stronger power, new features…) affect a receiver, especially the acquisition and tracking; and (2) design optimizations, which build on statistical insights and parameter interdependencies to improve acquisition efficiency and power consumption. For instance, the relationship between Doppler rate and Doppler can be exploited to refine the search space and reduce acquisition time.

The findings aim to support the GNSS community, particularly developers of Pulsar-enabled and LEO-capable receivers, and to contribute to understanding the emerging LEO-based navigation paradigm.

Section 2 presents an overview of the Pulsar constellation and signals. Section 3 analyzes Pulsar and GPS constellations, supported by plots in Appendix A. Section 4 analyzes Pulsar and GPS signals, with detailed plots in Appendix B. Section 5 discusses receiver adaptations required to process Pulsar signals. Section 6 outlines possible receiver optimizations. Section 7 concludes with a summary of findings.

## 2. XONA PULSAR OVERVIEW

### 2.1. Introducing Pulsar

Pulsar is a commercial 258-satellite LEO constellation planned by Xona Space Systems, specifically designed for PNT. The Pulsar IOV satellite was launched on June 23, 2025, and is currently in commissioning.

Pulsar is similar to legacy GNSS in many aspects:

- The overall principle is the same: satellites orbit the Earth, transmit data that allows computation of their position, and broadcast pseudorandom noise (PRN) codes. From these, receivers compute pseudoranges and then determine position, velocity, and time (PVT).
- Pulsar signals are transmitted in L1 and L5 bands, close to existing GNSS frequencies, enabling use by GNSS receivers without major changes to the RF chain.
- Pulsar signals include a pilot channel and a data channel.
- Pulsar signals use direct sequence spread spectrum (DSSS), with a primary PRN code. A secondary PRN code (overlay code) is added for the pilot channel, while the data channel overlays information bits.
- One of the two Pulsar signals uses binary phase shift keying (BPSK) modulation.

However, it also introduces several new or advanced features:

- Being in LEO, the distance at which the signals are transmitted from is significantly reduced, and consequently the received power on Earth is much higher (e.g. a factor of 20 in the distance means a difference of 26 dB in the free space path loss (FSPL)).

- For seamless operation with existing GNSS, a bandwidth efficient modulation is used (enhanced Feher's quadrature phase shift keying, or EFQPSK).
- One of the two Pulsar signals employs code shift keying (CSK) modulation, enabling a significantly higher data rate that contributes to faster TTFF.
- Flexible data structure, allowing future evolutions that are backward-compatible.
- Distribution of advanced GNSS corrections directly from satellites.
- Data encryption, data authentication and range authentication.

**2.2. Constellation**

To provide Earth coverage comparable to a MEO GNSS constellation, a LEO constellation must contain many more satellites (Reid, 2017). As currently planned, Pulsar will contain 258 satellites across 18 orbital planes. Two types of orbital planes will be used to provide a good worldwide coverage: inclined orbit with an inclination of 53°, and polar orbit with an inclination of 97°. The satellites are uniformly spaced in each plane, and the relative phasing between satellites in adjacent planes is fixed to distribute the satellites uniformly at the constellation level.

A nominal altitude of 1080 km was selected to provide a good balance between the number of satellites required and the dilution of precision (DOP) obtained (Reid et al., 2020). As for the Pulsar IOV satellite, it uses an inclination of 97° and an altitude around 520 km. To give an idea, the footprint radius is about 3864 km for a Pulsar FOC satellite, and 2626 km for the Pulsar IOV satellite (with a 0° elevation mask).

At this lower altitude, satellites move much faster, with an orbital period of less than 2 h compared to 12 h for GPS. This higher speed directly impacts Doppler observed on Earth, which is about 8 times greater for Pulsar than for GPS. The increase is even more pronounced for higher derivatives: the Doppler rate (rate of change of Doppler) is about 250 times higher for the Pulsar FOC constellation and about 500 times higher for Pulsar IOV. This has consequences on the design and performance of receivers, which will be discussed in Sections 5 and 6.

**TABLE 1** provides a summary of the main orbital parameters for the nominal constellations (Xona Space Systems, 2025; U.S. Department of Defense, 2020, Section 3.2), as well as the expected maximum Doppler and Doppler rate for a static receiver on Earth (the Doppler values consider the nominal constellations and no other effect than the rotation of the Earth), and **Figure 1** provides an illustration of Pulsar and GPS nominal constellations.

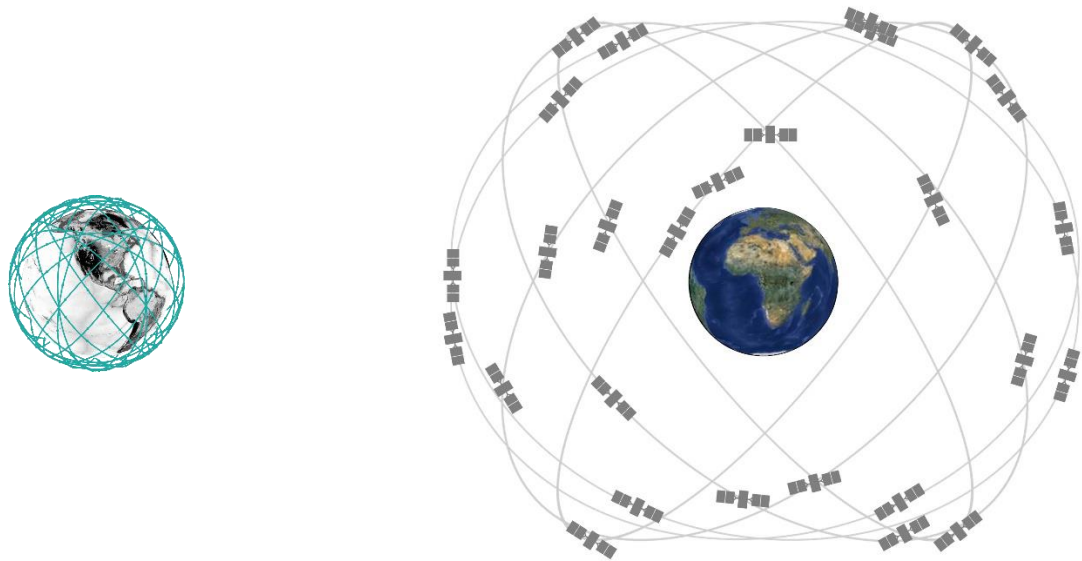

**Figure 1** Illustration of Pulsar (left) and GPS (right) nominal constellations at scales (U.S. Government, n.d.-a)

**TABLE 1** Nominal Pulsar IOV, Pulsar FOC and GPS Constellations Characteristics

| Parameters | Pulsar IOV | Pulsar FOC Polar Orbit | Pulsar FOC Inclined Orbit | GPS |
|---|---|---|---|---|
| Number of orbital planes | 1 | 6 | 12 | 6 |
| Orbital planes RAAN offset | - | 60° (360° / 6) | 30° (360° / 12) | 60° (360° / 6) |
| Number of satellites per plane | 1 | 11 | 16 | 4* |
| Satellites mean anomaly offset | - | ≈ 32.72 (360° / 11) | 22.5° (360° / 16) | Not unique** |
| Satellites phasing between planes | - | ≈ 5.45° (32.72 / 6) | 1.875° (22.5° / 12) | Not unique** |
| Number of satellites | 1 | 66 | 192 | 24* |
| Orbit inclination | 97° | 97° | 53° | 55° |
| Orbit eccentricity | 0 | 0 | 0 | 0 |
| Orbit altitude | 520 km | 1080 km | 1080 km | 20180 km |
| Orbit radius | 6891.0 km | 7451.0 km | 7451.0 km | 26551.0 km |
| Orbit perimeter | 43297.4 km | 46816.0 km | 46816.0 km | 166824.9 km |
| Orbital period | 94.9 min ≈ 1.6 h | 106.7 min ≈ 1.8 h | 106.7 min ≈ 1.8 h | 717.6 min ≈ 12.0 h |
| Satellite orbital speed | 7605.5 m/s | 7314.1 m/s | 7314.1 m/s | 3874.6 m/s |
| | 27379.8 km/h | 26330.8 km/h | 26330.8 km/h | 13948.6 km/h |
| Maximum ECEF satellite speed | 7682.9 m/s | 7400.0 m/s | 7000.6 m/s | 3186.8 m/s |
| | 27658.6 km/h | 26640.0 km/h | 25202.1 km/h | 11472.4 km/h |
| Maximum relative speed to a static user | 7103.2 m/s | 6327.4 m/s | 5985.9 m/s | 764.7 m/s |
| | 25571.5 km/h | 22778.6 km/h | 21549.1 km/h | 2752.8 km/h |
| Max. L1 band carrier Doppler | 37751.7 Hz | 33628.5 Hz | 31813.4 Hz | 4018.4 Hz |
| Max. L5 band carrier Doppler | 28207.7 Hz | 25126.9 Hz | 23770.7 Hz | 3000.8 Hz |

* The nominal 24-slot constellation considers 4 satellites per plane (Spilker & Parkinson, 1996; U.S. Department of Defense, 2020, Section 3.2), but the actual constellation has more satellites thanks to longer than expected satellites lifetime, and can have up to 6 satellites in an orbital plane (U.S. Coast Guard Navigation Center, n.d.).

** GPS satellites are not distributed evenly in each plane and across planes, because the GPS constellation has been optimized to minimize the effects of a single satellite failure on system degradation (Spilker, 1996).

### 2.3. Signals

Pulsar operates on the same principles as legacy GNSS, and its signals share many characteristics with existing systems. Pulsar satellites transmit two signals, one in L1 band and one in L5 band, called X1 and X5, respectively. Each signal contains a carrier, a primary PRN code, an overlay code, data bits via different modulations, and pilot and data channels. The main characteristics of Pulsar and GPS signals in these bands are provided in **TABLE 2** (Xona Space Systems, 2025; U.S. Department of Defense, 2022a, 2022b). Next subsections discuss the parameters in detail.

**TABLE 2** Pulsar and GPS Signals Characteristics

| Parameters | L1 band | | L5 band | |
| --- | --- | --- | --- | --- |
| | Pulsar X1 | GPS L1 C/A | Pulsar X5 | GPS L5 |
| Carrier frequency | 1593.3225 MHz ($155.75 \times 10.23$) | 1575.42 MHz ($154 \times 10.23$) | 1190.51625 MHz ($116.375 \times 10.23$) | 1176.45 MHz ($115 \times 10.23$) |
| Bandwidth[*] | 1.77 MHz | 2.046 MHz | 17.7 MHz | 20.46 MHz |
| Modulation | EFQPSK | BPSK | EFQPSK + CSK | QPSK |
| Minimum received power[**] | −148.2 dBW | −158.5 dBW | −144.9 dBW | −154.0 dBW |
| Maximum received power | −139.1 dBW | −153.0 dBW | −136.2 dBW | −150.0 dBW |
| Data/pilot channel combination | Quadrature (I/Q) | - | Quadrature (I/Q) | |
| Data/pilot channel power split | 50 % / 50 % | - | 50 % / 50 % | |
| Primary PRN code family | Kasami (small set) | Gold | Extended Gold | |
| Primary PRN code chip rate | 1.023 Mchip/s | | 10.23 Mchip/s | |
| Primary PRN code length | 1023 chip = 1 ms | | 10230 chip = 1 ms | |
| Overlay code chip rate | 1000 chip/s | - | 1000 chip/s | |
| Overlay code length (pilot) | 100 chip = 100 ms | - | 100 chip = 100 ms | 10 chip = 10 ms |
| Overlay code length (data) | - | - | - | 20 chip = 20 ms |
| Symbol duration (data) | 1 ms | 20 ms | 2 ms | 10 ms |
| Symbol rate (data) | 1000 symbol/s | 50 symbol/s | 500 symbol/s | 100 symbol/s |
| Bit rate (data) | 1000 bit/s | 50 bit/s | 4000 bit/s | 100 bit/s |

[*] Indicated bandwidths contain 99.5 % of the total power for Pulsar and corresponds to the main lobe for GPS.
[**] Minimum received powers are specified for an elevation angle of 10° or more for Pulsar and 5° or more for GPS.

### 2.3.1. Power

Pulsar satellites range is between 1080 km and 3900 km, while GPS satellites range is between 20187 km and 25788 km (see **TABLE 4** in Section 4.1). Therefore, the range is reduced by a factor of between 6.6 and 18.7, corresponding to a FSPL reduction of 16.4 dB to 25.4 dB. This, combined with the satellite antenna gain pattern, leads to the expected received power of Pulsar signals provided in **TABLE 2**.

### 2.3.2. How to Coexist with Legacy GNSS

As explained previously and shown in **TABLE 2**, Pulsar signals are received with a much higher power than legacy GNSS signals. To not interfere with legacy GNSS, Pulsar relies on two mechanisms.

First, the carrier frequencies are slightly offset (17.9025 MHz on the L1 band, and 14.06625 MHz on the L5 band), which make Pulsar signals not on top of the main lobe of legacy GNSS signals.

Second, an advanced pulse shaping is used to reduce the bandwidth of the main lobe and the level of the side lobes, namely the EFQPSK modulation (Simon & Yan, 1999). The time domain I/Q components and the spectrum of the X1 and X5 signals can be found in (Xona Space Systems, 2025; Reid et al., 2025). This pulse shaping is almost transparent for a receiver, since the local code replicas can still be rectangular waveforms; the only particularity is that the pilot channel has an offset of 0.5 chip compared to the data channel (Xona Space Systems, 2025). A detailed compatibility analysis of Pulsar with legacy GNSS can be found in (Reid et al., 2025).

*2.3.3. PRN Codes - Scalability for Large Constellations*

The Pulsar X1 and X5 primary PRN codes use the same chip rate and length as the GPS L1 C/A and L5 codes. The X5 primary PRN codes generation is identical to the GPS L5 primary PRN codes generation, whereas X1 primary PRN codes generation is very similar to the GPS L1 C/A PRN codes generation (see details in Section 5.2.1).

In Pulsar, a notable difference compared to legacy GNSS is the PRN assignation. In legacy GNSS, each satellite has its own primary PRN (except for frequency division multiplexing access (FDMA) GLONASS signals which use a unique PRN). This would not be practical for a constellation with hundreds of satellites, because the acquisition would need to search too many PRNs. In Pulsar, all satellites in the same orbital plane share the same primary PRN. This is feasible because their signals always exhibit large Doppler differences (see Sections B.12 and B.13), preventing interference after correlation. This sharing of PRN will require adaptations on receivers, which are discussed in Section 5.2.

Since there are 18 orbital planes, 18 different pairs of primary PRN codes will be transmitted (one pair consists of 1 PRN for the data channel and 1 PRN for the pilot channel). Among these 36 codes, 32 are Kasami codes (small set) and 4 are Gold codes. Planes 1 to 16 (236 satellites) will have Kasami codes, and planes 17 and 18 (22 satellites) will have Gold codes. Therefore, only 8.5 % of the satellites will transmit Gold codes (Xona Space Systems, 2025).

The small set of Kasami codes has been chosen because it offers near optimal auto and cross-correlation performance, with a correlation maximum of 33 instead of 65 for Gold codes, i.e. −29.83 dB instead of −23.94 dB, a roughly 6 dB improvement (Wallner & Ávila-Rodríguez, 2011). This extra cross-correlation margin is useful because the dynamic range of the received power level is higher than legacy GNSS. A drawback of the Kasami small set is its limited family size (32), requiring the use of Gold codes to complete the required code set for Pulsar. From a receiver's perspective, this is not an issue, since Gold codes can also be generated using a Kasami code generator (see Section 5.2.1).

With respect to the overlay codes, they are longer than those used in GPS, but have the same length as Galileo E5 codes. Each satellite within an orbital plane uses a distinct overlay code, although the same set is reused across orbital planes. Since there are at most 16 satellites per plane, the constellation transmits 16 unique overlay codes. These are memory codes, and the same set is used for both X1 and X5 signals (Xona Space Systems, 2025). The maximum cross-correlation between any two overlay codes ranges from −13.98 dB to −9.37 dB.

Therefore, each satellite has a unique combination of primary code and overlay code, and the assignment between SVID (space vehicle Identifier), primary PRN ID and overlay PRN ID can be found in (Xona Space Systems, 2025).

*2.3.4. Data*

To offer significantly higher data rate than legacy GNSS, Pulsar leverages two mechanisms: higher power, and advanced modulation for X5.

The detection, tracking, and decoding capability depends on the signal power and on the integration time used by the receiver. For the data channel, the integration time is limited by the data symbol duration, therefore, a higher symbol rate implies a lower integration time. However, Pulsar leverages its significantly higher power to compensate for the shorter integration time, resulting in similar or improved decoding performance. This is used for both X1 and X5.

The X1 signal does not include any additional modulation components. It transmits one bit per symbol, like traditional BPSK signals, with one symbol transmitted per primary PRN code period.

On the other hand, the X5 signal was designed to provide additional data, such as GNSS corrections, requiring a higher data rate. To achieve this, it transmits 8 bit/symbol via CSK modulation, and one symbol is transmitted for two primary code periods. This configuration is denoted CSK(256,2).

The CSK modulation scheme involves transmitting the PRN code with a specific circular shift applied (Garcia Peña et al., 2015). For example, to transmit the bit pattern 00000000, the PRN code is transmitted without any shift; to transmit 00000001, it is shifted by one chip; and so on, up to 11111111, which corresponds to a shift of 255 chips. This technique enables the transmission of 8 bits per symbol without reducing the integration time, thereby preserving signal sensitivity.

The primary drawback is the increased receiver-side complexity, as the receiver must search across 256 possible delays simultaneously. However, this is expected to be manageable with future-generation GNSS receivers. Notably, CSK modulation is also employed in the L6 signal of Japan's Quasi-Zenith Satellite System (QZSS) (Cabinet Office, Government of Japan, 2025).

## 3. XONA PULSAR CONSTELLATION ANALYSIS

### 3.1. Constellations Configuration for Simulation

Sections 3 and 4 analyze the characteristics of Pulsar and GPS constellations and signals. A commercial Pulsar-enabled GNSS simulator was used to generate CSV log files containing parameters such as time, elevation, azimuth, range, and Doppler for each satellite. Several GNSS simulators are currently supporting Pulsar (Safran, 2024; Spirent, 2024) or plan to support it (Syntony, 2023). For this study, Safran's Skydel GNSS Simulator Software was used.

Since the Pulsar constellation is not yet fully deployed, the simulator initially used default parameters based on the nominal constellation, with some randomness added. Without correction, however, long simulations (several days) can show noticeable deviations in certain characteristics. To avoid this, the nominal Pulsar and GPS constellations defined in Section 2.2 were used.

Using the nominal constellations gives characteristics very similar to those that would be obtained with the actual constellation in space (some plots would just look slightly less regular with the actual constellation). Only the number of satellites for GPS will be different, since the actual constellation has more satellites in orbit than the initial configuration. There are currently 31 operational satellites, but five of them are in orbit for more than 20 years (U.S. Coast Guard Navigation Center, n.d.; U.S. Government, n.d.-b), and only two launches are planned before 2027 and two others in 2027 (Wikipedia., n.d.), therefore, the number of operational GPS satellites may slightly change in the future.

Here are the common parameters of the simulations:

- The duration of the simulation is 3 days, because it is long enough to capture unbiased statistics and metrics, and it is a multiple of the revisit period of the GPS constellation (1 day).
- Because latitude strongly affects observations, the analysis was performed at several latitudes to ensure worldwide coverage: 0°, 15°, 30°, 45°, 60°, 75°, 90°, –45° (the latter to illustrate differences at negative latitudes).
- The longitude is 0° and the altitude 0 m. The choice of longitude does not matter over the full constellation period.
- An elevation mask of 0° is considered, to be as generic as possible when showing time domain and distribution plots. There are also plots showing parameters maximum, average and minimum as a function of the elevation mask, and when useful some analyses will provide additional values as a function of the elevation mask.

Note that GNSS simulators comply with the interface control documents (ICDs), but do not necessarily have an exact representation of the satellite's antenna gain pattern. For this reason, the received power has not been analyzed.

To not overload the core of the article, the detailed plots are provided in Appendices A and B, and the subsections of Sections 3 and 4 contain only the analysis of these plots and a few additional plots to highlight specific points. For Pulsar, certain plots are shown separately for the inclined and polar orbital planes when the results significantly differ.

### 3.2. Sky Plot

A sky plot shows satellite positions in the sky over time. It is useful for revealing potential weaknesses of a constellation and how these may affect users in specific environments. Section A.1 shows the sky plots for Pulsar inclined, Pulsar polar and GPS satellites over 3 days.

Looking at **Figure A.3** (GPS), there is always one or two gaps in the north/south direction of the sky plot, where no GPS satellites are present. These gaps can be relatively large, e.g., at a latitude of 45°, it represents more than one fifth of the sky view. This can be quite problematic in urban canyons, with many buildings around already blocking a significant portion of the sky. For example, consider a user on a north–south street in an urban canyon, where surrounding buildings block the sky up to 30° except along the street direction. In this case, more than one third of the visible sky contains no satellites. This is a significant drawback because the number of visible satellites will be quite limited, and all the visible satellites will be in a small portion of the sky, which will lead to a poor DOP. The other MEO GNSS constellations have similar holes in similar directions, only GLONASS has slightly smaller gaps due to its higher inclination (Kou et al., 2021; Li et al., 2015). Therefore, a multi-constellation MEO GNSS receiver can improve the overall satellite availability, however, does not improve DOP.

In the full Pulsar constellation, such gaps appear only at very high latitudes and are smaller than those of GPS (**Figure A.2**). Therefore, with Pulsar, the issue mentioned above will not appear in most of the dense cities. When looking separately at the inclined and polar orbits, the behavior differs significantly. Inclined orbits start to have a gap at a latitude near ±30°, and the gap gets larger as the latitude gets higher (**Figure A.1**), until a point where the inclined orbit becomes not visible (near ±84.2° without any elevation mask). Instead, polar orbits start to have a gap at a higher latitude (near ±60°), and the gap is smaller compared to MEO GNSS (**Figure A.2**). Therefore, it is thanks to the polar orbit that the sky plot coverage is excellent, and this shows the complementary nature of these orbits.

This also shows how Pulsar can complement MEO GNSS in urban areas. Even a single additional satellite in view can significantly improve geometry, DOP, and positioning accuracy. This was demonstrated with simulations in (Takayama et al., 2025), where tight integration of LEO with MEO GNSS only or with MEO GNSS/INS significantly improved the overall positioning error.

The plots in Section A.1 show the behavior over a long duration, but it is also interesting to observe what happens for short term durations. **Figure 2** shows Pulsar and GPS sky plots for a 1-hour observation. During this period, GPS satellites move very little in the sky, which can be inconvenient in certain situations. For instance, consider a user on a long street in a urban canyon, where a satellite is blocked by surrounding buildings. If the user does not change their direction, this satellite will not be visible for a long time. Instead, Pulsar satellites move very fast, therefore, a satellite cannot stay blocked for a long time. Even if an orbital plane is blocked, since it moves due to the rotation of the Earth, it should not stay blocked as long as an MEO GNSS satellite.

As mentioned in the introduction, faster satellites have other advantages. For example, in a multipath environment, where signals may bounce off surfaces and cause errors, faster-moving satellites help reduce the duration of multipath effects, as the signal path changes more quickly, minimizing their impact. Faster satellites also improve PPP algorithms by enhancing satellite geometry and increasing the frequency of observations, leading to faster convergence and more accurate positioning.

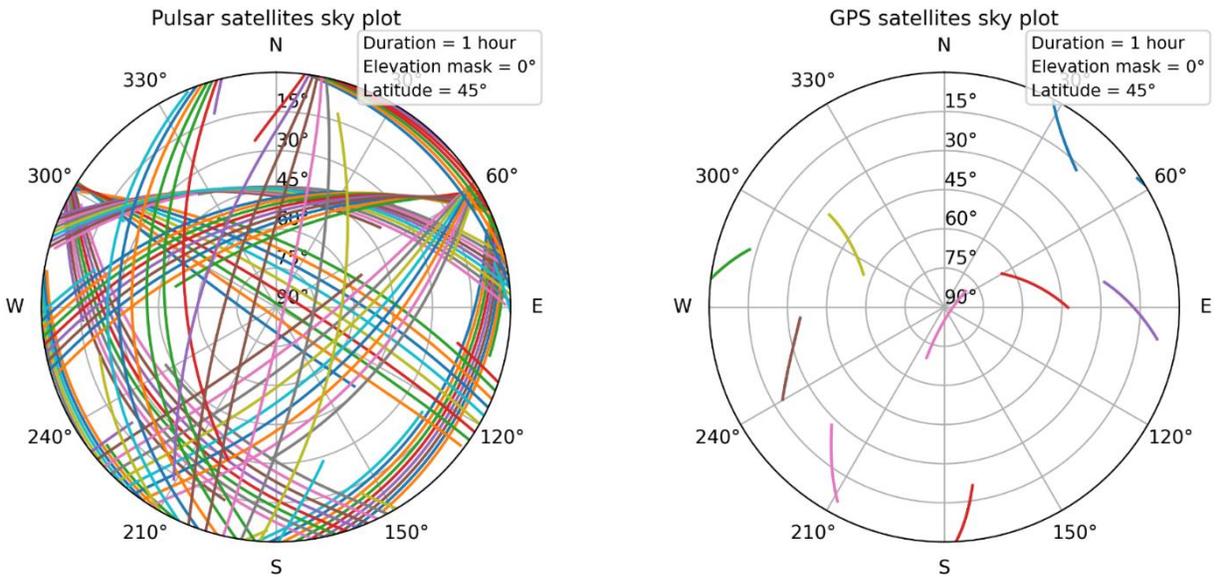

**Figure 2** Sky plot of the Pulsar and GPS constellations over 1 hour

### 3.3. Elevation

Understanding satellite elevation trends is important because environments pose different challenges depending on elevation angle. Low-elevation satellites may be obstructed in urban canyons, while high-elevation satellites may be less useful for applications such as indoor positioning. As such, we will analyze both the evolution of the elevation of the satellites over time and their distribution.

Section A.2 shows the satellites elevation over time for Pulsar and GPS constellations (on different time scales, because Pulsar satellites move faster). For GPS, many satellites have an elevation that increases to a peak and then decreases, but there are different shapes, and sometimes the elevation can increase again after a decrease when being not too far from the equator (**Figure A.6**). Therefore, there is not a global pattern observable, except at the poles. Whereas for Pulsar, there is a clear pattern related to the orbital planes in view (**Figure A.4** and **Figure A.5**). When a new orbital plane comes in view, it will be near the horizon, thus, the satellites will have a relatively low elevation. Then, over time, the orbital plane appears higher and higher in the sky, thus, the satellites will have a higher elevation. And then the elevation of the orbital plane and of its satellites will decrease until they disappear.

Section A.3 shows the distribution of the elevation of the satellites. For both constellations, satellites are more present at low elevations than high elevations (except at the poles), but the decrease is kind of linear for GPS, whereas it is faster for Pulsar. This leads to a lower average elevation for Pulsar compared to GPS (note that the average value will increase if an elevation mask is considered, as shown in **TABLE 3**. This table shows results for all latitudes and excluding those above 60° where the inclined satellites have very low elevation, affecting the average). This is obviously an inconvenience for dense urban environments, but it is not specific to Pulsar, it would be inherent to any LEO constellation, and the lower the constellation altitude, the lower the average elevation. **Figure 3** illustrates this by showing that the ratio between the time above a certain elevation (green lines) and the time a satellite is below that elevation (red lines) is smaller at lower orbit than at higher orbit, hence a smaller average elevation.

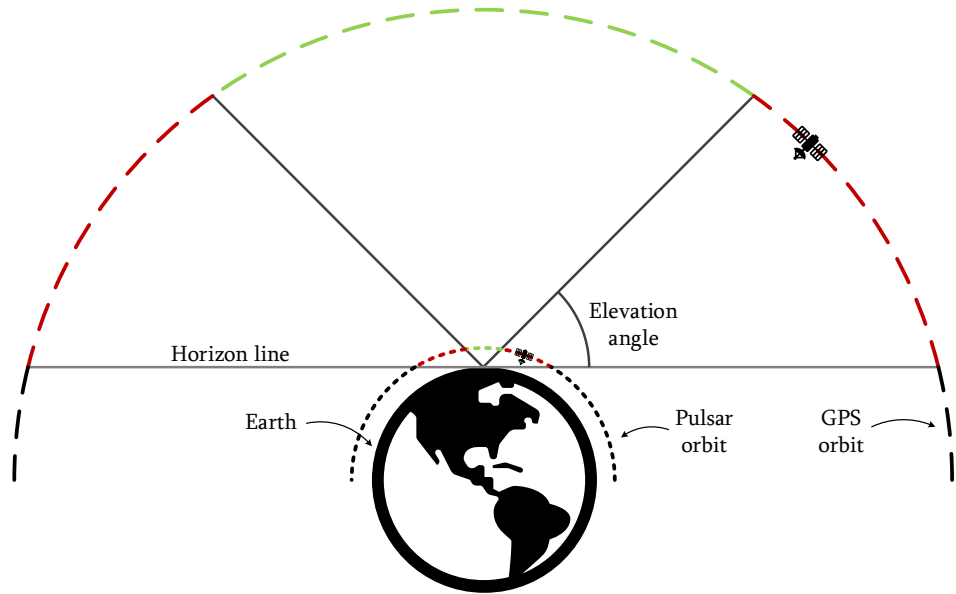

**Figure 3** Illustration of the time a satellite is above a certain elevation (green lines) vs below a certain elevation (red lines)

**TABLE 3** Pulsar and GPS Average Elevation

| Context | Elevation mask | 0° | 5° | 10° | 15° | 20° |
|---|---|---|---|---|---|---|
| Average elevation for all latitudes | Pulsar | 16.5° | 21.2° | 26.3° | 31.4° | 35.8° |
| | Pulsar inclined | 12.3° | 16.1° | 19.9° | 21.9° | 25.2° |
| | Pulsar polar | 17.0° | 21.6° | 26.4° | 31.3° | 36.0° |
| | GPS | 29.1° | 32.1° | 35.3° | 38.5° | 41.6° |
| Average elevation for latitudes up to 60° | Pulsar | 16.0° | 20.8° | 25.8° | 30.7° | 35.5° |
| | Pulsar inclined | 16.2° | 20.9° | 25.8° | 30.6° | 35.2° |
| | Pulsar polar | 15.4° | 20.4° | 25.5° | 30.7° | 35.7° |
| | GPS | 29.8° | 33.1° | 36.7° | 40.2° | 43.7° |

### 3.4. SVID

Section A.4 provides the SVID in view over time for Pulsar and GPS constellations on the same time scale. Remember that SVID 1 to 192 are in inclined orbits and SVID 193 to 258 are in polar orbits, hence the difference in behavior observed for these sets of SVID in **Figure A.10**. Notably, the SVIDs for each orbital plane are visible in descending order (e.g. for plane 1, after SVID 8, the next one is SVID 7, then 6, etc.), which can be used by a receiver to predict the next satellites to come as discussed in Section **Error! Reference source not found.**. **Figure A.11** provides the same plot for GPS for comparison.

### 3.5. PRN ID

Section A.5 provides the primary PRN ID in view over time for Pulsar constellation. Remember that PRN ID 1 to 12 are for inclined orbits and PRN ID 13 to 18 are for polar orbits, hence the difference in behavior observed for these sets of PRN ID in **Figure A.12**. Notably, the PRN IDs are visible in ascending order (e.g. after PRN ID 8, the next one is PRN ID 9, then 10, etc.), which can be used by a receiver to predict the next PRN to come as discussed in Section **Error! Reference source not found.**.

**Figure A.12** also show that the duration for which an orbital plane is in view varies significantly with the latitude.

### 3.6. Satellite Pass Duration

As shown earlier, LEO satellites move much faster than MEO satellites. Section A.4 provides the maximum, average and minimum pass duration as a function of the elevation mask. While GPS satellites are in view for several hours (**Figure A.16**), Pulsar satellites are in view for about ten minutes typically. (**Figure A.13**). This presents a noticeable change for GNSS receivers, because the acquisition and tracking should be fast enough to use the satellite for positioning as long as possible, and the positioning algorithm will see new satellites and satellites dropping much more frequently.

### 3.7. Number of Satellites in View

Section A.7 provides the maximum, average and minimum number of satellites in view as a function of the elevation mask. The average and maximum number of satellites is higher for Pulsar (**Figure A.17**) than GPS (**Figure A.20**). Note, however, that the GPS plots are based on a 24-slot constellation, whereas the constellation has generally more satellites than that (U.S. Coast Guard Navigation Center, n.d.). **Figure A.18** and **Figure A.19** show that there are more satellites in view from inclined orbit than polar orbit except when getting close to the poles (as expected, because there are 192 satellites in inclined orbital planes against 66 satellites in polar orbital planes).

### 3.8. Number of Orbital Planes in View

Section A.8 provides the maximum, average and minimum number of orbital planes in view as a function of the elevation mask. As mentioned previously, each orbital plane is associated with a primary PRN ID, therefore, the plots in **Figure A.21** show the number of different primary PRN codes that can be received simultaneously by a receiver. Again, **Figure A.22** and **Figure A.23** show that there are more inclined orbital planes in view than polar orbital planes in view except when getting close to the poles (as expected, because there are 12 inclined orbital planes against 6 polar orbital planes).

### 3.9. Number of Satellites in View from a Same Orbital Plane

Section A.9 provides the maximum, average and minimum number of satellites in view from the same orbital plane as a function of the elevation mask. As mentioned previously, all of the satellites in the same orbital plane share the same primary PRN code, therefore, these plots show the number of possible true detections when searching a primary PRN ID. **Figure A.24** and **Figure A.25** show that there is slightly more satellites in view from the same orbital plane for inclined orbital planes than polar orbital planes (as expected, because there are 16 satellites per inclined orbital plane against 11 satellites per polar orbital plane).

## 4. XONA PULSAR SIGNALS ANALYSIS

### 4.1. Range

For both LEO and MEO constellations, the user-satellite range closely follows satellite elevation (**Figure 4**). Since detailed elevation plots are already shown, detailed range plots are not provided. In summary, over a pass the range typically decreases from a maximum to a minimum and returns to the maximum, following an approximately quadratic profile.

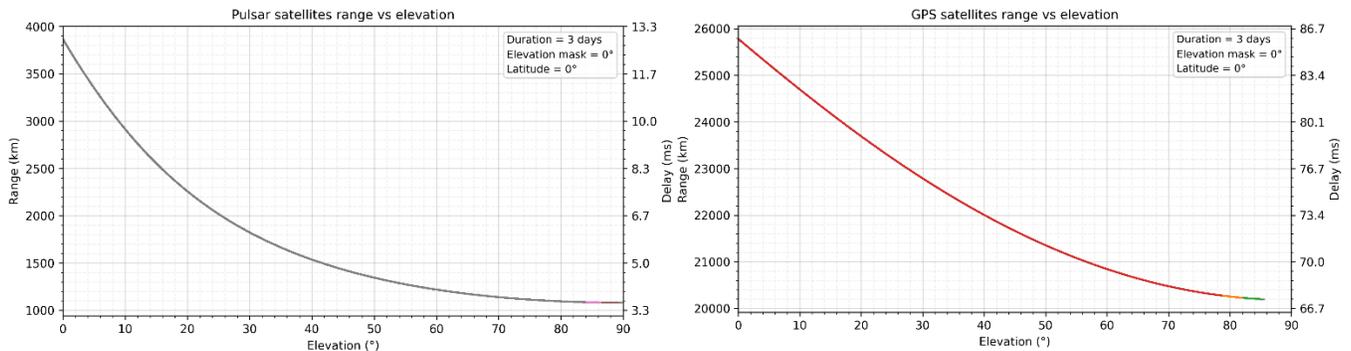

**Figure 4** Range as a function of the elevation for Pulsar and GPS constellations

TABLE 4 lists minimum ranges by latitude (because it does not change with the elevation mask, as long as the satellite is still visible), and TABLE 6 lists maximum ranges by elevation mask (because it varies very little with the latitude, as long as the satellite is still visible; the values given are the maximum among all the latitudes). TABLE 5 and TABLE 7 give the corresponding propagation delays. As expected, Pulsar LEO ranges are far shorter than GPS MEO ranges: Pulsar ranges span roughly 1080–3901 km (≈ 3.6–13.0 ms), while GPS ranges are ≈ 20187–25788 km (≈ 67.3–86.0 ms). Section 6.3 will show that this information can be used for faster acquisition in some cases. Later, Section 4.5 completes this analysis with the difference of range between satellites.

**TABLE 4** Minimum range (km) for Pulsar and GPS Constellations, for different latitudes.

| Latitude (°) | 0 | 15 | 30 | 45 | 60 | 75 | 90 |
|---|---|---|---|---|---|---|---|
| Pulsar inclined | 1080.0 | 1081.4 | 1085.3 | 1090.8 | 1369.4 | 2838.3 | - |
| Pulsar polar | 1080.0 | 1081.5 | 1085.3 | 1090.7 | 1096.0 | 1099.9 | 1385.6 |
| GPS | 20196 | 20187 | 20188 | 20196 | 20227 | 20695 | 21662 |

**TABLE 5** Minimum propagation delay (ms) for Pulsar and GPS Constellations, for different latitudes.

| Latitude (°) | 0 | 15 | 30 | 45 | 60 | 75 | 90 |
|---|---|---|---|---|---|---|---|
| Pulsar inclined | 3.6 | 3.6 | 3.6 | 3.6 | 4.6 | 9.5 | - |
| Pulsar polar | 3.6 | 3.6 | 3.6 | 3.6 | 3.7 | 3.7 | 4.6 |
| GPS | 67.4 | 67.3 | 67.3 | 67.4 | 67.5 | 69.0 | 72.3 |

**TABLE 6** Maximum range (km) for Pulsar and GPS Constellations, for different elevation masks.

| Elevation mask (°) | 0 | 5 | 10 | 15 | 20 |
|---|---|---|---|---|---|
| Pulsar inclined | 3897.0 | 3379.0 | 2941.4 | 2578.5 | 2281.4 |
| Pulsar polar | 3900.7 | 3393.0 | 2955.7 | 2592.2 | 2294.2 |
| GPS | 25788.0 | 25257.0 | 24723.0 | 24208.0 | 23716.0 |

**TABLE 7** Maximum propagation delay (ms) for Pulsar and GPS Constellations, for different elevation masks.

| Elevation mask (°) | 0 | 5 | 10 | 15 | 20 |
|---|---|---|---|---|---|
| Pulsar inclined | 13.0 | 11.3 | 9.8 | 8.6 | 7.6 |
| Pulsar polar | 13.0 | 11.3 | 9.9 | 8.6 | 7.7 |
| GPS | 86.0 | 84.2 | 82.5 | 80.7 | 79.1 |

### 4.2. Range Rate / Relative Speed / Doppler

Range rate (the time derivative of range) equals the relative speed between user and satellite and directly produces Doppler shifts on carrier and code. Since the local replicas must match the received signals, knowing the Doppler characteristics is of major importance to design receivers. The figures concentrate on carrier Doppler; readers can convert carrier Doppler to code Doppler or range rate using the carrier frequency and code chip rates given in **TABLE 2**.

Sections B.1 shows the carrier Doppler over time. For Pulsar, carrier Doppler during a pass follows a smooth, quasi-symmetric trajectory: it begins at a positive extreme, decreases through zero and reaches an opposite negative extreme (**Figure B.1** and **Figure B.2**). GPS displays a similar but less regular pattern near the equator (occasional direction changes at pass edges) (**Figure B.3**).

For Pulsar, the starting and ending value depends on where the satellite is within the orbital plane pass (a similar pattern for the elevation was observed in Section 3.3). Maximum X1 carrier Doppler magnitudes for Pulsar are near 32–34 kHz, whereas GPS L1 maximum is near 4 kHz. This large Doppler expansion materially increases the acquisition frequency search space for Pulsar signals.

Section B.2 shows the distribution of the carrier Doppler. We can see that statistically there is a higher chance to get a large Doppler (either positive or negative) than a small Doppler for Pulsar (**Figure B.4** and **Figure B.5**), while GPS is more evenly distributed, except when getting close to the poles (**Figure B.6**). The use of such information will be discussed in Section 6.1. The plots do not show it, but a higher elevation mask reduces the peaks at large Doppler values, even if they do not vanish entirely.

Finally, Section B.3 shows the Doppler as a function of the elevation. For Pulsar there are clear patterns observable, which are relatively stable with the latitudes (except when getting close to the poles). The higher the elevation, the smaller the maximum Doppler (**Figure B.7** and **Figure B.8**). But a low elevation does not necessarily mean a large Doppler (except at the poles, where there is a 1-to-1 relation between the Doppler and the elevation), since we saw previously that it depends where we are within an orbital plane pass.

It is similar for GPS, except it is less regular when getting closer to the equator (**Figure B.9**).

### 4.3. Range Rate Derivative / Relative Acceleration / Doppler Rate

The Doppler rate (the rate of change of the Doppler frequency due to relative acceleration between a user and a satellite) matters for any long-integration processing because receiver replicas must follow carrier and code drift. Otherwise, the correlation peak may shift across acquisition bins, degrading detection performance.

For Pulsar, Doppler rates can be up to 250 times those typical of MEO GNSS, making accurate Doppler-rate modelling essential for long coherent integrations (see Section 6.2 for receiver mitigations).

Sections B.4 to B.7 provide the Doppler rate over time, the Doppler rate distribution, and the Doppler rate as a function of the elevation, and the Doppler rate as a function of the Doppler.

For Pulsar, similarly to the Doppler, a clear trend can be observed in the Doppler rate, the starting/ending values and the highest value depending on where the satellite is within an orbital plane pass (**Figure B.10** and **Figure B.11**). The value is always negative, except very briefly at the start and end of some passes (only for latitudes far from the poles and without elevation mask). GPS has a similar pattern, except it is less regular and can have positive values when getting closer to the equator, as already mentioned in the previous section (**Figure B.12**).

Section B.5 shows that statistically there is a higher chance to get a moderate Doppler rate for Pulsar (**Figure B.13** and **Figure B.14**), while very high Doppler rates are quite rare, since it happens at high elevation which does not happens often and does not last long. Also, when getting close to the poles, the largest negative value decreases. As for GPS, the Doppler rate is rather evenly distributed (except when getting close to the poles), being more negative than positive (**Figure B.15**).

Section B.6 shows the Doppler rate as a function of the elevation. For Pulsar, there are clear patterns observable, which are relatively stable with the latitudes (except when getting close to the poles). Low elevations imply a low to moderate Doppler rate, and high elevations imply a high Doppler rate (**Figure B.16** and **Figure B.17**). Near the pole, there is a 1-to-1 relation between the Doppler rate and the elevation. It is similar for GPS, except it is much less regular when getting farther from the pole (**Figure B.18**).

Finally, Section B.7 presents the relation between the Doppler rate and the Doppler. Again, there is a clear pattern for Pulsar, where overall the Doppler rate amplitude is inversely proportional to the Doppler amplitude (**Figure B.19** and **Figure B.20**). A large Doppler implies a low Doppler rate, but a small Doppler can mean a moderate to high Doppler rate. Near the pole, there is a 1-to-1 relation between the Doppler rate and the Doppler. As before, it is similar for GPS, except it is much less regular when getting farther from the pole (**Figure B.21**).

### 4.4. Range Rate Second Derivative / Relative Jerk / Doppler Rate Derivative

The Doppler rate derivative (the rate of change of the Doppler rate due to relative jerk between a user and a satellite) is typically negligible for short integrations but can influence very long coherent processing. Simulations show a maximum carrier Doppler rate derivative of ±1.26 Hz/s² for Pulsar X1 and ±0.0006 Hz/s² for GPS L1, i.e., Pulsar can exhibit jerk roughly 2500 times larger than GPS in extreme cases. Receivers that attempt unusually long coherent integrations should account for this.

### 4.5. Range Difference

The difference in range between satellites is information that can be used by receivers to optimize some processing (discussed in Section 6.3). Hence the interest in looking in its properties. Section B.8 and B.9 show the range difference between any visible satellites: Pulsar satellites, Pulsar inclined satellites, Pulsar polar satellites, and GPS satellites. Whereas Sections B.10 and B.11 show the range difference between satellites within the same orbital plane (for Pulsar only).

Sections B.8 and B.10 show the distribution of the range difference between satellites in the same orbital plane, which is useful to understand the behavior of this parameter and what is more common. Section B.9 and B.11 show the maximum, average and minimum range difference as a function of the elevation mask, to know the possible interval according to the context.

Sections B.8 and B.10 show that it is more likely to observe small differences than large differences (**Figure B.22** to **Figure B.25**, and **Figure B.30** and **Figure B.31**). Section B.9 and B.11 show that the range difference decreases with the elevation mask, as it could be expected (**Figure B.26** to **Figure B.29**, and **Figure B.32** and **Figure B.33**). The curves are relatively similar for latitudes up to 60°. **TABLE 8** provides the maximum range difference observed among all the latitudes for different elevation masks, and **TABLE 9** the corresponding delay difference. Section 6.3 will show that this information can be used for faster acquisition in some cases.

**TABLE 8** Maximum range difference (km) for Pulsar and GPS Constellations, for different elevation masks.

| Elevation mask (°) | 0 | 5 | 10 | 15 | 20 |
|---|---|---|---|---|---|
| Pulsar | 2795.7 | 2294.0 | 1855.0 | 1489.6 | 1193.9 |
| Pulsar inclined | 2785.3 | 2286.8 | 1850.6 | 1471.2 | 1182.9 |
| Pulsar polar | 2792.6 | 2286.8 | 1853.3 | 1458.9 | 1193.9 |
| Pulsar inclined (same plane) | 2360.0 | 2169.9 | 1848.6 | 1428.6 | 985.9 |
| Pulsar polar (same plane) | 2788.4 | 2094.3 | 1349.7 | 673.3 | 92.5 |
| GPS | 5586.4 | 5054.8 | 4523.3 | 4000.1 | 3500.9 |

**TABLE 9** Maximum delay difference (ms) for Pulsar and GPS Constellations, for different elevation masks.

| Elevation mask (°) | 0 | 5 | 10 | 15 | 20 |
|---|---|---|---|---|---|
| Pulsar | 9.3 | 7.7 | 6.2 | 5.0 | 4.0 |
| Pulsar inclined | 9.3 | 7.6 | 6.2 | 4.9 | 3.9 |
| Pulsar polar | 9.3 | 7.6 | 6.2 | 4.9 | 4.0 |
| Pulsar inclined (same plane) | 7.9 | 7.2 | 6.2 | 4.8 | 3.3 |
| Pulsar polar (same plane) | 9.3 | 7.0 | 4.5 | 2.2 | 0.3 |
| GPS | 18.6 | 16.9 | 15.1 | 13.3 | 11.7 |

### 4.6. Doppler Difference

Doppler separation between satellites in the same plane is a key reason the primary PRN can be reused per plane. This parameter can also be used by receivers to optimize some processing (discussed in Section 6.1). Hence the interest in investigating these properties.

Section B.12 shows the carrier Doppler difference between satellites in the same orbital plane over time. Near the equator, the lowest and highest values clearly depends on where we are within the orbital plane pass, but as the latitude increases it becomes less and less obvious (**B.38** and **B.39**), as it was observed with other patterns.

Section B.13 shows the maximum, average and minimum carrier Doppler difference between satellites in the same orbital plane as a function of the elevation mask. The minimum difference is 7875 Hz for X1 and 5884 Hz for X5 for inclined orbits (**Figure B.32**) and 32601 Hz for X1 and 24359 Hz for X5 for polar orbits (**Figure B.33**), and these differences increase with the elevation mask.

The smallest difference is higher for polar satellites than inclined satellites, because there are less satellites per plane (11 instead of 16), therefore, they are more far from each other in the sky.

### 4.7. Doppler Rate Difference

Finally, to complete the analysis, we look at the difference in Doppler rate between satellites within the same orbital plane, which can also be used by receivers to optimize some processing (discussed in Section 6.2).

Section B.14 shows the carrier Doppler rate difference between satellites in the same orbital plane over time. Near the equator, the highest values clearly depend on where we are within the orbital plane pass, but as the latitude increases it becomes less and less obvious (**Figure B.38** and **Figure B.39**), as observed before for other patterns.

Section B.15 shows the maximum, average and minimum carrier Doppler rate difference between satellites in the same orbital plane as a function of the elevation mask. The minimum difference is 0 Hz/s, and the maximum is about 200 Hz/s for X1 and 149 Hz/s for X5 for an elevation mask below 10° (the maximum decreases when the latitude increases) and then it decreases with the elevation mask (**Figure B.40** and **Figure B.41**).

## 5. PULSAR RECEIVER DESIGN CONSIDERATIONS

This section examines the fundamental differences between legacy GNSS receivers and Pulsar receivers, focusing on the signal processing across the different stages inside a receiver. The comparison is structured across three domains: the analog domain, where the initial signal conditioning occurs; the high-rate digital domain, which extends up to the correlator outputs; and the low-rate digital domain, where post-correlation processing takes place. Each of these domains presents distinct challenges and design considerations.

### 5.1. Analog Domain

#### 5.1.1. Carrier Frequency

As described in Section 2.3, the carrier frequencies of the X1 and X5 signals are slightly offset from those of legacy GNSS signals, by 17.9025 MHz and 14.06625 MHz, respectively, relative to the L1 and L5 signals. Consequently, both the antenna and the RF front-end must be designed or configured to accommodate these shifts in center frequency, and potentially any changes in the required bandwidth.

#### 5.1.2. RF Power

As shown in Section 2.3, Pulsar signals are received at significantly higher power levels than legacy GNSS signals. As a result, the RF front-end (which performs filtering, amplification, downconversion, and digitization) must be designed to handle these elevated input powers without distortion or saturation.

### 5.2. Digital Domain – High-Rate Processing

#### 5.2.1. PRN Code Generation

A receiver should of course be able to generate the PRN codes for acquisition and tracking, which should not be an obstacle thanks to the similarity with GPS.

The X1 primary PRN codes generator is a slight variation of the L1 C/A generator. The L1 C/A PRN generator is composed of two 10-bit LFSR registers (called G1 and G2 in (U.S. Department of Defense, 2022a)), whose outputs are XOR'd. The X1 PRN generator is composed of the same two registers, but also of a third 5-bit register, and their outputs are XOR'd. These three registers are called XA, XB and XC in (Xona Space Systems, 2025). The initial state of each register will define the PRN generated. The first 16 pairs of PRN codes are Kasami codes, where the initial state of XA is full of 1s and the initial state of XB is full of 0s. The last pairs of PRN codes are C/A codes, where the initial state of XC is full of 0s, and the initial state of X1 is either full of 1s or full of 0s.

The X5 primary PRN codes generator is the same as the L5 primary PRN codes generator (U.S. Department of Defense, 2022b), composed of two 13-bit generators, called XA and XB in both ICDs. Only the initial state of the XB register is different (Xona Space Systems, 2025).

As for the overlay codes, they are memory codes like most GNSS overlay codes, and they are the same length as Galileo E5 overlay codes (100 chips), but only 16 are used in the nominal constellation. Their definition can be found in (Xona Space Systems, 2025).

### 5.2.2. Data Rate

The symbol rate for X1 and X5 is much higher than legacy GNSS, with one symbol every millisecond or two milliseconds. Therefore, the receiver should be able to deal with this, but it does not represent a particular difficulty.

On the contrary, the higher data rate is overall positive for a receiver because it can get the data needed to compute the PVT faster, which can save battery for some applications.

### 5.2.3. Number of Satellites

The number of satellites in the Pulsar is significantly higher than in legacy GNSS, with up to 258 satellites. As a result, receivers must be capable of handling a much larger set of satellite identifiers. In particular, receivers that store the SVID using only 8 bits (allowing for a maximum of 256 unique values) may not be able to support the full Pulsar constellation without modification.

### 5.2.4. Doppler

As discussed in Sections 2.2 and 4.2, the Doppler shift affecting Pulsar signals is approximately 8 times greater than the one observed with MEO GNSS signals. This has a significant impact on both the acquisition process and the transition from acquisition to tracking.

In the absence of assistance or prior information, the higher Doppler shift requires searching over a much wider frequency range. This results in a greater number of frequency bins to evaluate compared to MEO GNSS systems, which directly increases acquisition time. Moreover, the expanded search space also affects the probability of false alarms (Borio, 2008). Fortunately, there are still ways to reduce the search space using basic knowledge of the constellation, which are discussed in Section 6.1.

A further consequence of high dynamics is the code Doppler, which introduces a shift in code phase over time. Denoting $\dot{\tau}$ as the code Doppler, the code shift introduced after a duration $T$ is $\dot{\tau}\, T$. Considering the worst-case code Doppler on X1 of 21.6 chip/s, there is a shift of one chip after 46.3 ms, half a chip after 23.1 ms, and a quarter of a chip after 11.6 ms. This impacts both the acquisition and the transition from acquisition to tracking.

During acquisition, for moderate and long integration times (coherent or non-coherent), the code Doppler cannot be neglected, otherwise the correlation peak will spread on two or more code bins, reducing the detection performance.

Since acquisition typically operates on buffered data (van Diggelen, 2021), there may be a significant delay (often hundreds of milliseconds or more) between the time the data was recorded and the start of tracking the live signal after a successful acquisition. Therefore, again, the code Doppler cannot be neglected, and it is essential to account for code Doppler when determining the initial tracking parameters.

For MEO GNSS, the code Doppler is about eight times lower than Pulsar; therefore, there is much more margin, but only the context can tell if the code Doppler can be neglected or not.

Note that for receivers performing acquisition directly on L5 band signals, the code Doppler is 10 times higher, making this consideration even more critical. With X5, the worst-case code Doppler implies a shift of a quarter of a chip after only 1.2 ms, so it is not negligible.

### 5.2.5. Doppler Rate

As seen in Sections 2.2 and 5.2.5, the Doppler effect creates a drift about 250 times higher on Pulsar signals than MEO GNSS signals (even 500 with the Pulsar IOV satellite). Such high Doppler rate impacts both acquisition and tracking.

Denoting $\dot{f}_\mathrm{D}$ the carrier Doppler rate, the frequency shift introduced by the Doppler rate after a duration $T$ is $\dot{f}_\mathrm{D}\, T$. The Doppler rate on the carrier has a similar impact as the Doppler on the code.

The impact during acquisition depends on the coherent integration time (CIT), since the frequency step is inversely proportional to it. Let us look at two cases, minimum CIT and extended CIT, assuming a frequency step of $1/(2\, T_\mathrm{C})$ where $T_\mathrm{C}$ is the coherent integration time, and a worst-case carrier Doppler rate on X1 of 230 Hz/s.

Considering a CIT of 1 ms, the frequency shift due to the Doppler rate would be one frequency bin after 2.2 s, half a frequency bin after 1.1 s, and a quarter of frequency bin after 0.54 s. Therefore, in this case, a very long noncoherent integration time can be used without taking into account the carrier Doppler rate.

Considering an extended CIT, the frequency shift due to the carrier Doppler rate would be one frequency bin after $1/(2\,T_C) = 230\,T_C \Rightarrow T_C = \sqrt{1/460} \approx 46.6$ ms (which gives a frequency step of about 10.7 Hz), half a frequency bin after 33.0 ms (frequency step of 15.2 Hz), a quarter of frequency bin after 23.3 ms (frequency step of 21.4 Hz). In this case, the carrier Doppler rate can be neglected only for small to moderate CIT.

For the transition between acquisition and tracking, as mentioned in the Section 5.2.4, there may be several hundreds of milliseconds, thus, with a carrier Doppler rate up to 230 Hz/s it may be expected to observe a shift by a few dozens of hertz. So, it is the CIT that will tell if this can be neglected or not.

There is another application where the code Doppler and the carrier Doppler rate must be carefully taken into account, for the exact same reason as the transition between acquisition and tracking: receivers performing duty cycling. The gap in time between switch off and switch on is large enough to observe a drift in the input signal, which must be compensated by the receiver.

As for the code Doppler rate, with a maximum of 0.15 chip/s² on X1 and of 1.48 chip/s² on X5, it is likely negligible for X1 (except with extremely long CIT), and it depends on the context for X5.

Note that the Doppler rate Pulsar IOV is twice that of Pulsar, therefore, the number above should be adapted accordingly to check the application with Pulsar IOV.

For the tracking, in order for the phase lock loop (PLL) to be stable, the phase error should stay under a certain threshold. This phase error is composed of thermal noise error, vibration induced error, Allan deviation induced error, and dynamic stress error. The dynamic stress error is obtained from the steady state error, which is inversely proportional to the *n*th power of the tracking loop bandwidth and directly proportional to the *n*th derivative of range, where *n* is the loop filter order (Ward, 2017). Since Pulsar has much higher dynamics than MEO GNSS (8 times more for Doppler, 250 times more for Doppler rate, and more than 2500 more for Doppler rate derivative), the dynamic stress error will be much bigger. Therefore, compared to MEO GNSS, Pulsar tracking loops will likely need different parameters.

*5.2.6. PRN Codes Assignment*

As seen in Section 2.3.3, several satellites share the same PRN code. This impacts mostly the acquisition and the high-level management.

For acquisition, it means there are less PRNs to search, which is a good thing. But it also means that several true peaks will be observed when acquiring a PRN (the number of peaks depends on the number of satellites in view from the same orbital plane, as shown in Sections 3.9 and A.9). This does not represent a particular difficulty, it is just a change in the operation of the acquisition, such that when a true peak is detected, the acquisition should not stop as usual, but should continue to browse the search space.

The difference is also that when starting a tracking channel after a detection in acquisition, we may not know yet which satellite is tracked if no overlay code synchronization was performed. The SVID will be known only after overlay code synchronization or data decoding. This also means that several tracking channels will track the same PRN, with different Doppler and delay which is not a major difficulty, but receivers must handle it accordingly.

Except at the first acquisition where no tracking channel is working, future acquisitions will search for new satellites with the same PRN of satellites that were already tracked. Since we do not want to reacquire tracked satellites, it means information from the tracking is required to inform acquisition. The most efficient way is to simply not search frequency regions where we are already tracking a satellite. However, if a receiver is not capable of excluding frequency regions in acquisition, then the entire search space will be browsed, and since some of the detections will correspond to satellites being tracked, the controller should identify them and to not start a new tracking.

*5.2.7. RF Power*

We previously saw that the higher RF power had an impact in the analog domain, but it also has an impact in the digital domain, a positive one this time.

The higher RF power means a higher $C/N_0$. Therefore, for a same integration time, the correlation peak will be much higher; or seen differently, to get a similar correlation peak, a smaller integration time can be used. This can speed-up the acquisition process, or require less processing in general, leading to a significant power saving, which is of major importance for many applications.

It also positively impacts the tracking. As mentioned in Section 5.2.5, the phase error should stay under a certain threshold to keep the PLL stable, and this phase error depends, amongst other parameters, on the thermal noise error. The thermal noise error which is proportional to the square root of the noise bandwidth and inversely proportional to the square root of the $C/N_0$ (Ward, 2017). For Pulsar, since the $C/N_0$ is higher compared to MEO GNSS, the same noise bandwidth will lead to a smaller thermal noise error, or by targeting a similar error, the noise bandwidth could be increased, which is good for the dynamic stress error (see Section 5.2.5).

### 5.2.8. CSK Modulation

The use of the CSK modulation on the X5 signal is among the most significant changes for a receiver. Indeed, to decode the X5 data, it is required to perform 256 correlations for each symbol (i.e. each 2 ms), continuously during tracking. Other methods can be used, such as FFT-based correlation, but it stays a relatively heavy operation, compared to current GNSS signals.

### 5.3. Digital domain – Low-Rate Processing

### 5.3.1. TPC

All modern GNSS signals have advanced Forward Error Correction (FEC). Pulsar has also an advanced one, based on Turbo Product Code (TPC). TPC presents multiple benefits as a Forward Error Correction (FEC) technique, including strong bit error rate (BER) performance, ease of implementing error correction, absence of error floors, and excellent reliability in environments with multipath interference and signal fading. Additionally, it typically requires no more than four Soft-In-Soft-Out (SISO) decoding iterations to reach convergence. Various decoding methods are available for TPC frames, involving combinations of soft-decision and hard-decision algorithms. More details about the TPC can be found in (Xona Space Systems, 2025; Gunning, 2024).

### 5.3.2. Flexible Data Structure

In legacy GNSS, the data bit structure is typically organized in messages, subframes, frames, pages, and everything is fixed. Such structure does not allow evolution without being backward compatible. Instead, Pulsar defines messages, and packets for these messages, but does not strictly define the order of the messages. This completely opens the path to new messages in the future, to evolve with new needs that may be identified later. Such new messages would simply be ignored by older Pulsar receivers, as they would not be able to understand them. More details about the data content and structure can be found in (Xona Space Systems, 2025).

### 5.3.3. Data Encryption

Pulsar encrypts some data bits within certain message types, requiring an access key to interpret those bits. Those without the access keys will still be able to determine their authenticity. Pulsar's encryption system is flexible, resilient, and convenient. Pulsar will support over the air rekeying while also supporting wrong key detection to ensure information is never decrypted by the receiver with the wrong key (up to the applicable authentication security level). More details can be found in (Xona Space Systems, 2025).

### 5.3.4. Data Authentication

Pulsar provides cryptographic data authentication utilizing Time Efficient Stream Loss Tolerant Authentication (TESLA). The protocol is similar to Galileo's OSNMA (Fernández-Hernández et al., 2021; European Union, 2023), but improves the time to authentication by utilizing Pulsar's higher bandwidth and techniques from (Anderson, 2024). Each X1 message and each X5 dataframe and each transport layer frame receives a message authentication code, providing at least 32-bit security on each message's authenticity and integrity. More details can be found in (Xona Space Systems, 2025).

### 5.3.5. Ranging Authentication

Pulsar's ranging signal includes a cryptographic watermark that inverts several of the primary PRN code chips. The selection of the watermarked chips is based on the same TESLA protocol supporting data authentication. Because these inverted chips are not known in advance of broadcast, the receiver must record and store the baseband IQ samples (and accompanying phase/Doppler information). Initially the tracking correlation will be degraded by a small amount. After distribution of the watermark chip inversion seed on the data channel, the receiver will need to process the stored IQ samples again to determine the ranging authenticity.

Pulsar's watermark requires aggregating 1 second of baseband samples to make an authenticity determination. The watermark provides 32-bit security while also minimizing the needed degradation to achieve that security. More details can be found in (Anderson, 2024, 2025; Xona Space Systems, 2025).

*5.3.6. LEO Positioning*

Being in a much lower altitude than MEO GNSS, positioning algorithms may need to be adapted, because of issues related to the initial point estimate for Iterative Descent algorithms (Van Uytsel et al., 2025), or related to the convergence (Chen et al., 2025). Also, satellites will appear and disappear much more frequently, and the positioning engine may need to be adapted for that.

# 6. PULSAR RECEIVER DESIGN OPTIMIZATIONS

The Doppler and Doppler rate being much larger with Pulsar compared to MEO GNSS implies there are much more bins to search during acquisition. The long overlay code may also bring challenges when using long coherent integration time. If assistance is provided, or almanac and coarse time are available, the Doppler, Doppler rate, and overlay code delay search spaces can be reduced in a similar way to MEO GNSS, solving the issue. Therefore, next, we are investigating how the knowledge developed in Sections 3 and 4 can be used to optimize the acquisition of Pulsar signals without relying on external recent information.

## 6.1. Optimizing Acquisition Regarding the Doppler

This section discusses how to optimize the Doppler search during acquisition, considering two different cases:

- At start-up, when the receiver is in a cold start. In this case, the goal of the receiver is to acquire all of the visible satellites, which may appear anywhere in the sky (unless blocked).
- During operation, when the receiver is already running and tracking satellites. In this case, the goal of the receiver is to acquire only new satellites coming in view, which cannot emerge from anywhere in the sky, but normally from the horizon (except if blocked).

We will also look at the difference between an open sky environment and a urban environment where elevations below 30° are typically blocked.

*6.1.1. At Start-Up*

**Open Sky.** As mentioned before, when starting a receiver in cold start, satellites can be anywhere in the sky in an open sky environment. Therefore, the only tool usable at this stage is the distribution of the carrier Doppler to determine if some Doppler frequencies are more likely to happen than others. Section B.2 provides this information. **Figure B.4** and **Figure B.5** show that there is a higher chance to observe a large Doppler (positive or negative) than a small one for Pulsar satellites. As mentioned in Section 4.2, applying an elevation mask reduces the peaks at the extremes of the distribution, though they remain present.

Therefore, in this case, if the Doppler frequencies are searched sequentially, it is more efficient to start searching satellites from large Doppler to small Doppler, rather than starting at 0 Hz and going to large Doppler for example. Note that the largest Doppler is slightly lower for inclined satellites than polar ones, therefore, to avoid wasting energy searching for something that does not exist in the signal, the frequency search space should be tailored for each type of orbit, which is easy to do since the type of orbit is directly related to the primary PRN ID (PRN 1 to 12 = inclined orbit, PRN 13 to 18 = polar orbit).

Nevertheless, the plots in Section B.1 show that the largest positive Doppler are at the beginning of the satellite pass, and the largest negative value are at the end of the pass. Finding a satellite close to the end of its pass is not interesting, because it will soon disappear and provide limited utility. Therefore, we can conclude that the most efficient way to acquire satellite is to start from the largest positive value, and then search in decreasing order.

There is an additional reason to start searching from the largest positive Doppler. Looking simultaneously at Sections A.2 and B.1, it can be seen that for a satellite pass, the larger the initial Doppler, the higher the maximum elevation, and thus the longer the pass. A long pass is effective because the effort spent in acquiring the satellite is made worthwhile by the extended tracking and positioning, and a higher elevation means best performance in urban areas and stronger power. Therefore, searching first for large positive Doppler means searching for satellites that will be available longer and that will be higher in the sky later.

**Figure B.6** shows that a similar trend can be observed for GPS, but only for relatively high latitudes (and the peak is not at the extreme value but slightly before, except at the pole where it is at the extreme value), so such strategy would be less effective.

**Urban Area.** For an urban environment, there are two possible strategies: Keep trying to acquire satellites with the largest positive Doppler for the same reasons as before (longer pass duration and higher maximum elevation later), because Sections A.2 and B.2 show this is where there is a higher chance to find satellites. In this case, Section B.3 shows that the largest Doppler observable will be lower compared to open sky, e.g. $\approx 29$ kHz for a mask of $30°$ instead of $\approx 34$ kHz, but the reduction is minor. Or the receiver can try first to acquire satellites very high in the sky, because they would still be visible if the receiver changes direction and they may provide higher power. In this case, acquiring satellites with a negative Doppler is correct, and thus the search could start form 0 Hz (this is what a $90°$ elevation satellite will have, but satellites at other elevations may have this Doppler as well, see Section B.3) up to a limit given by the minimum elevation searched (e.g. $\approx 12$ kHz for a mask of $70°$). This last strategy is probably the most interesting, because even if there is less chance to find a LEO satellite very high in the sky, even a single satellite can significantly improve the position error when used with MEO GNSS as discussed in Section 3.2.

*6.1.2. During Operation*

As mentioned before, when the receiver is already running and tracking satellites, the new satellites to acquire cannot emerge from anywhere in the sky.

**Open Sky.** In the case of clear open skies, new satellites will always emerge from the horizon. Section B.1 shows that the initial Doppler depends on where we are in the orbital plane pass. Therefore, the strategy depends on if we try to acquire a new PRN or not.

If the receiver tries to acquire a new PRN that is not yet being tracked, then it means the orbital plane itself will emerge from the horizon, and Section B.1 shows that in this case the Doppler will be relatively small (except when getting close to the poles), therefore it makes more sense to start searching from 0 Hz and then increasing the frequency.

If the receiver tries to acquire a new PRN that is already being tracked, then knowledge from the tracked satellites can be used. E.g. the Doppler rate amplitude would tell if the orbital plane is high or low in the sky (Section B.4 and B.6); or the evolution of the elevation can be analyzed to determine if the orbital plane is going up or down in the sky (Section A.2) to deduce if the next initial Doppler would be higher or lower; or at least the Doppler of the tracked satellites could be used to reduce the search space using the possible range for the Doppler difference (Section B.13).

**Urban Area.** In the case of urban areas, there are more unknown since a new satellite can emerge at different elevations depending on the buildings around, but the strategy still depends on if we try to acquire a new PRN or not.

If the receiver tries to acquire a new PRN that is not yet being tracked, if we assume that the orbital plane itself will be near the elevation mask, then Sections A.2 and B.1 show that the initial Doppler will be already quite large, therefore it makes sense to start searching from a large positive Doppler and then decreasing the frequency.

If the receiver tries to acquire a new PRN that is already being tracked, the same ideas as for open sky can be used.

**6.2. Optimizing Acquisition Regarding the Doppler Rate**

As shown in Section 2.2, the Doppler rate is much higher for Pulsar than for MEO GNSS (approximately 250 times more). The largest carrier Doppler rate is up to $-230$ Hz/s on X1 and $-172$ Hz/s on X5, and the code Doppler rate is up to $-21.6$ chip/s on X1 and $-216$ chip/s on X5.

When it cannot be neglected (see Section 5.2.5), it becomes a new dimension of the acquisition search space, which significantly complicates the acquisition (basically the number of cells to test is multiplied by the number of Doppler rate bins to test). Therefore, limiting the Doppler rate search space as much as possible, and browsing it in the most efficient way is essential. However, we will see that the Doppler rate may not ultimately be a major issue for receivers.

Regarding the search space, Section B.7 gives the relation between the Doppler rate and the Doppler. Basically, the Doppler rate minimum and maximum amplitude are inversely proportional to the Doppler amplitude with a quadratic relation, which is very easy to model. Therefore, when the receiver is searching a specific Doppler bin, it can adapt the Doppler rate bins searched.

In the previous section, it was shown that the most efficient strategy was sometimes to start with a large positive Doppler and to start with a small Doppler when targeting high elevation. Fortunately, when the Doppler is large, the Doppler rate is always small (see Section B.7), and then the elevation is high, the Doppler rate is always high (see Section B.6). Therefore, the number Doppler rate bins can be drastically reduced, or even neglected depending on the context. Then, if a further analysis is required, the same elements discussed for the Doppler in Section 6.1 can be adapted for the Doppler rate.

### 6.3. Optimizing Acquisition Regarding the Overlay Code

To reach high sensitivity, a receiver needs to extend the CIT (van Diggelen, 2021), and in this situation the overlay code becomes an obstacle and must be managed. Some techniques exist for short coherent integration time (Borio at al., 2010) and moderate coherent integration time (Leclère & Landry, 2019, 2018), but for long integration coherent time, it is necessary to correlate with the secondary code as well (Leclère et al., 2017; Di Grazia et al., 2023). This significantly increases the number of operations required because it adds a new dimension to the search space (in addition, the Doppler search space becomes denser because the step gets reduced as the CIT extends).

However, LEO constellations are much closer to the Earth compared to MEO constellations, therefore, the user-satellite range is smaller, and the difference of range between several satellites are much smaller. This is shown in detail in Section B.10 to B.13.

**TABLE 8** and **TABLE 9** show that the maximum range difference between Pulsar satellites is between 2795.7 km ↔ 9.3 ms for an elevation mask of 0° and 1193.9 km ↔ 4.0 ms for an elevation mask of 20°. Therefore, when a receiver is tracking a satellite, it knows that other satellites can be shifted by at most 4 to 9 milliseconds. Consequently, only 4 to 9 delays need to be searched since the others are not physically possible. With intelligence in the receiver, it may be possible to predict the sign of the difference, and halve this. Then, Sections B.8 and B.10 show that there is a higher chance to have a small difference rather than a large one, therefore, it is best to start the search from zero delay and go to the larger delays.

This could be applied for GPS as well, but the range difference is larger (5586.4 km ↔ 18.6 ms for an elevation mask of 0° and 3500.9 km ↔ 11.7 ms for an elevation mask of 20°) and the overlay codes are shorter on L5 (10 chips and 20 chips). Therefore, it is not really as applicable for L5, but it would be useful for the L1C signal that has a 1800 chip secondary code.

If the receiver is starting and does not yet have any information regarding any satellite, but knows the time very precisely (millisecond level), it is still possible to reduce the overlay code delay search space. Since the satellites transmit the first chip of the PRN at the start of each millisecond, and that the range is quite constrained, then the unknown arrival time is also quite constrained. For an elevation mask of 0°, the possible range is between 1080 km ↔ 3.6 ms and 3900.7 km ↔ 13.0 ms(see **TABLE 4** to **TABLE 7**), thus, the unknown arrival time is at most 2820.7 km ↔ 9.4 ms. Therefore, 9 delays (plus the uncertainty on the local time) need to be searched. This goes down to 6.3 ms with an elevation mask of 10°, and 4.0 ms with an elevation mask of 20°.

This could be applied for GPS as well, but the interval will be larger since the range interval is bigger (between 11.8 ms for an elevation mask of 20° and 18.7 ms for an elevation mask of 0°), therefore, the applicability is limited except for L1C signal, as discussed above with the range difference.

### 7. CONCLUSION

This paper presented a comprehensive description of the Pulsar LEO PNT system, covering both the constellation (orbital parameters, number of satellites, maximum relative speed, etc.) and the signals (carrier frequencies, PRN codes, data rates, etc.), and investigated their implications for receiver design while proposing optimizations.

Using a GNSS simulator, key parameters were evaluated at both the constellation level (sky plot, elevation, pass duration, number of satellites or planes in view) and the signal level (Doppler, Doppler rate, range, and their differences between satellites), considering temporal trends, statistical distributions, elevation dependence, and latitude variations.

All the figures presented in this work provide a consolidated view of both the Pulsar and GPS constellations for comparison purposes. By bringing together multiple perspectives in a single collection, these figures offer a practical reference for understanding the operational characteristics and unique dynamics of both Pulsar and MEO GNSS satellites.

The analyses revealed important interdependencies between parameters, such as between Doppler and Doppler rate, and highlighted how parameters change with the elevation or with the orbital plane view. Differences in parameter values across satellites were also observed. Doppler values demonstrate that satellites do not interfere with each other despite the reuse of primary PRN codes, while range values open new opportunities. These findings have clear implications for the efficient design of Pulsar-enabled and LEO-capable GNSS receivers.

The differences between a Pulsar receiver and a GNSS receiver were discussed, focusing on the impact of PRN reuse, higher power, increased Doppler, and higher Doppler rates on acquisition, tracking, and the transition between the two. From this comparison, it was seen that a Pulsar receiver is quite similar to a GNSS receiver, the main differences being the higher dynamics and the additional features (CSK, TPC, encryption & authentication).

Finally, design optimizations that exploit parameter relationships and employ prediction techniques can reduce acquisition time and lower receiver power consumption. For example, the acquisition search space can be reduced along Doppler, Doppler rate, and overlay code delay dimensions, leveraging knowledge obtained from all the plots and the analyses. These optimizations were considered separately for receivers at startup and receivers in operation, and in open sky and urban area environments.

Before the full constellation arrives, there is room for the GNSS community to explore many directions: from building Pulsar-capable receivers, to measuring power savings, testing resilience against jamming and spoofing, validating with realistic simulations or the IOV signal. The present study offers a foundation for these efforts, supporting the development of efficient LEO-capable GNSS receivers and contributing to a deeper understanding of the emerging LEO-based navigation landscape.

## ACKNOWLEDGMENTS

We acknowledge the support of the Canadian Space Agency (CSA) 24STDPU43.

## REFERENCES

Anderson, J. (2024). *Designing cryptography systems for GNSS data and ranging authentication* [Doctoral dissertation, Stanford University]. https://stacks.stanford.edu/file/pj787wh6240/ja_dissertation_a15c57d-augmented.pdf

Anderson, J. (2025). *World's first authenticated satellite pseudorange from orbit* [Conference presentation]. ION GNSS+ 2025, Baltimore, MD, United States.

Borio, D. (2008). *A statistical theory for GNSS signal acquisition* [Doctoral dissertation, Politecnico di Torino]. https://theses.eurasip.org/theses/569/a-statistical-theory-for-gnss-signal-acquisition

Borio, D., O'Driscoll, C., & Lachapelle, G. (2010). Composite GNSS signal acquisition over multiple code periods. *IEEE Transactions on Aerospace and Electronic Systems*, *46*(1), 193–206. https://doi.org/10.1109/TAES.2010.5417156

Cabinet Office, Government of Japan. (2025). *Quasi-Zenith Satellite System Interface Specification: Centimeter Level Augmentation Service (IS-QZSS-L6-007)*. https://qzss.go.jp/en/technical/ps-is-qzss/ps-is-qzss.html

Chen, Y., Zhang, Y., Cai, X., Tang, R., & Jiang, W. (2025). An improved LEO positioning solution algorithm based on the LM algorithm. In *2025 6th International Conference on Electrical, Electronic Information and Communication Engineering (EEICE)* (pp. 634–641). IEEE. https://doi.org/10.1109/EEICE65049.2025.11033727

Di Grazia, D., Pisoni, F., Crasta, S., Napolitano, A., Darsena, D., & Ardiero, S. (2023). Long GNSS secondary codes acquisition by characteristic length method. In *2023 IEEE International Workshop on Metrology for Automotive (MetroAutomotive)* (pp. 105–110). IEEE. https://doi.org/10.1109/MetroAutomotive57488.2023.10219112

European Union. (2023). *Galileo open service navigation message authentication (OSNMA) signal-in-space interface control document (SIS ICD)*, Issue 1.1. https://www.gsc-europa.eu/electronic-library/programme-reference-documents

Fernández-Hernández, I., Rijmen, V., Seco-Granados, G., Simón, J., Rodríguez, I., & David Calle, J. (2021). A Navigation Message Authentication Proposal for the Galileo Open Service. *Navigation*, 63(1), 85–102.

Garcia Peña, A. J., Aubault-Roudier, M., Ries, L., Boucheret, M.-L., Poulliat, C., & Julien, O. (2015). Code shift keying: Prospects for improving GNSS signal designs. *Inside GNSS, 10*(6), 52–62. https://insidegnss.com/auto/novdec15-WP.pdf

Gunning, K. (2024). Xona Partner Paper: Decoding Turbo Product Codes [Technical Report]. Xona Space Systems.

Kou, R., Yang, B., Dong, Z., Liang, F., & Yang, S. (2021). Mapping the spatio-temporal visibility of global navigation satellites in the urban road areas based on panoramic imagery. *International Journal of Digital Earth*, *14*(7), 807–820. https://doi.org/10.1080/17538947.2021.1886357

Leclère, J., Botteron, C., & Farine, P.-A. (2017). High sensitivity acquisition of GNSS signals with secondary code on FPGAs. *IEEE Aerospace and Electronic Systems Magazine*, *32*(8), 46–63. https://doi.org/10.1109/MAES.2017.160176

Leclère, J., & Landry, R., Jr. (2018). Combining secondary code correlations for fast GNSS signal acquisition. In *2018 IEEE/ION Position, Location and Navigation Symposium (PLANS)* (pp. 46–55). https://doi.org/10.1109/PLANS.2018.8373364

Leclère, J., & Landry, R, Jr. (2019). Galileo E5 Signal Acquisition using Intermediate Coherent Integration Time. *Journal of Navigation*, *72*(3), 555–574. https://doi:10.1017/S0373463318001054

Li, X., Zhang, X., Ren, X., Fritsche, M., Wickert, J. & Schuh, H. (2015) Precise positioning with current multi-constellation Global Navigation Satellite Systems: GPS, GLONASS, Galileo and BeiDou. *Scientific Reports, 5*, Article 8328. https://doi.org/10.1038/srep08328

Reid, T. G. R. (2017). *Orbital diversity for global navigation satellite systems* [Doctoral dissertation, Stanford University]. https://stacks.stanford.edu/file/dc409wn9227/ORBITAL_DIVERSITY_FOR_GNSS_VFinal_Submit-augmented.pdf

Reid, T. G. R., Chan, B., Goel, A., Gunning, K., Manning, B., Martin, J., Neish, A., Perkins, A., & Tarantino, P. (2020). Satellite navigation for the age of autonomy. In *2020 IEEE/ION Position, Location and Navigation Symposium (PLANS)* (pp. 342–352). https://doi.org/10.1109/PLANS46316.2020.9109938

Reid, T. G. R., Gala, M., Kriezis, A., O'Meara, M., Pant, A., Tarantino, P., & Youn, C. (2025). *Xona Pulsar compatibility with GNSS* [Conference presentation]. ION GNSS+ 2025, Baltimore, MD, United States.

Safran. (2024). *Safran Electronics & Defense Adds Xona PULSAR™ to GNSS Simulation Capability*. https://safran-navigation-timing.com/xona-pulsar-gnss-simulation-capability

Simon, M. K. & Yan, T.-Y. (1999). Performance Evaluation and Interpretation of Unfiltered Feher-Patented Quadrature-Phase-Shift Keying (FQPSK), *Telecommunications and Mission Operations Progress Report*, 137, 1–29. https://tda.jpl.nasa.gov/1990-1999/progress_report/42-137/137C.pdf

Spilker, J. J., Jr. (1996). Satellite constellation and geometric dilution of precision. In B. W. Parkinson & J. J. Spilker Jr. (Eds.), *Global positioning system: Theory and applications* (Vol. I, pp. 177–208). American Institute of Aeronautics and Astronautics.

Spilker, J. J., Jr., & Parkinson, B. W. (1996). Overview of GPS operation and design. In B. W. Parkinson & J. J. Spilker Jr. (Eds.), *Global positioning system: Theory and applications* (Vol. I, pp. 29–54). American Institute of Aeronautics and Astronautics.

Spirent. (2024). *Simulate and test the Xona LEO PNT constellation now, before operational deployment*. https://spirentfederal.com/simxona

Syntony. (2023). *Syntony GNSS, a leading provider of SDR Positioning, Navigation, and Timing (PNT) solutions, has announced the integration of Xona Space Systems' LEO PNT constellation to its GNSS simulators and receiver solutions.* https://syntony-gnss.com/news/our-press-release/syntony-gnss-partners-with-xona-space-systems

Takayama, Y., Osugi, A., Marathe, T., Leclère, J., & Chan, B. (2025). *Tight integration of GNSS/LEO/INS in dense urban environments* [Conference presentation]. ION GNSS+ 2025, Baltimore, MD, United States.

U.S. Coast Guard Navigation Center. (n.d.). *GPS constellation*. Retrieved September 19, 2025. https://www.navcen.uscg.gov/gps-constellation

U.S. Department of Defense. (2020). *Global positioning system standard positioning service performance standard (5th ed.)*. https://www.gps.gov/performance-standards-specifications

U.S. Department of Defense. (2022a). *Navstar GPS Space Segment/Navigation User Segment interfaces (IS-GPS-200 Rev. N)*. https://www.gps.gov/interface-control-documents-icds-interface-specifications-iss

U.S. Department of Defense. (2022b). *Navstar GPS Space Segment/User Segment L5 interfaces (IS-GPS-705 Rev. J)*. https://www.gps.gov/interface-control-documents-icds-interface-specifications-iss

U.S. Government. (n.d.-a). GPS Constellation (Expandable 24-slot configuration, as defined in SPS Performance Standard). Retrieved September 19, 2025. https://www.gps.gov/images

U.S. Government. (n.d.-b). GPS space segment. Retrieved September 19, 2025. https://www.gps.gov/space-segment

van Diggelen, F. (2021). High-sensitivity GNSS. In Y. T. J. Morton, F. van Diggelen, J. J. Spilker, Jr., & B. W. Parkinson (Eds.), *Position, navigation, and timing technologies in the 21st century* (Vol. 1, pp. 445–480). Wiley-IEEE Press. https://doi.org/10.1002/9781119458449.ch18

Van Uytsel, W., Majorana, A. M., Janssen, T., & Weyn, M. (2025). Convergence of iterative descent algorithms for LEO-PNT. *IEEE Access*. https://doi.org/10.1109/ACCESS.2025.3602499

Wallner, S., & Ávila-Rodríguez, J.-Á. (2011). Codes: The PRN family grows again. *Inside GNSS*, 6(5), 83–92. https://insidegnss.com/auto/sepoct11-wp.pdf

Ward, P. W. (2017). GNSS receivers. In E. D. Kaplan & C. J. Hegarty (Eds.), *Understanding GPS/GNSS: Principles and Applications* (3rd ed., pp. 339–548). Artech House.

Wikipedia. (n.d.). List of GPS satellites — Planned launches. Retrieved September 19, 2025. [https://en.wikipedia.org/wiki/List_of_GPS_satellites#Planned_launches](https://en.wikipedia.org/wiki/List_of_GPS_satellites#Planned_launches)

Xona Space Systems. (2025). *Xona Pulsar Navigation Signal Interface Control Document (v1.0.2)*.

**A. DETAILED PLOTS RELATED TO THE CONSTELLATION**

## A.1. Sky Plots

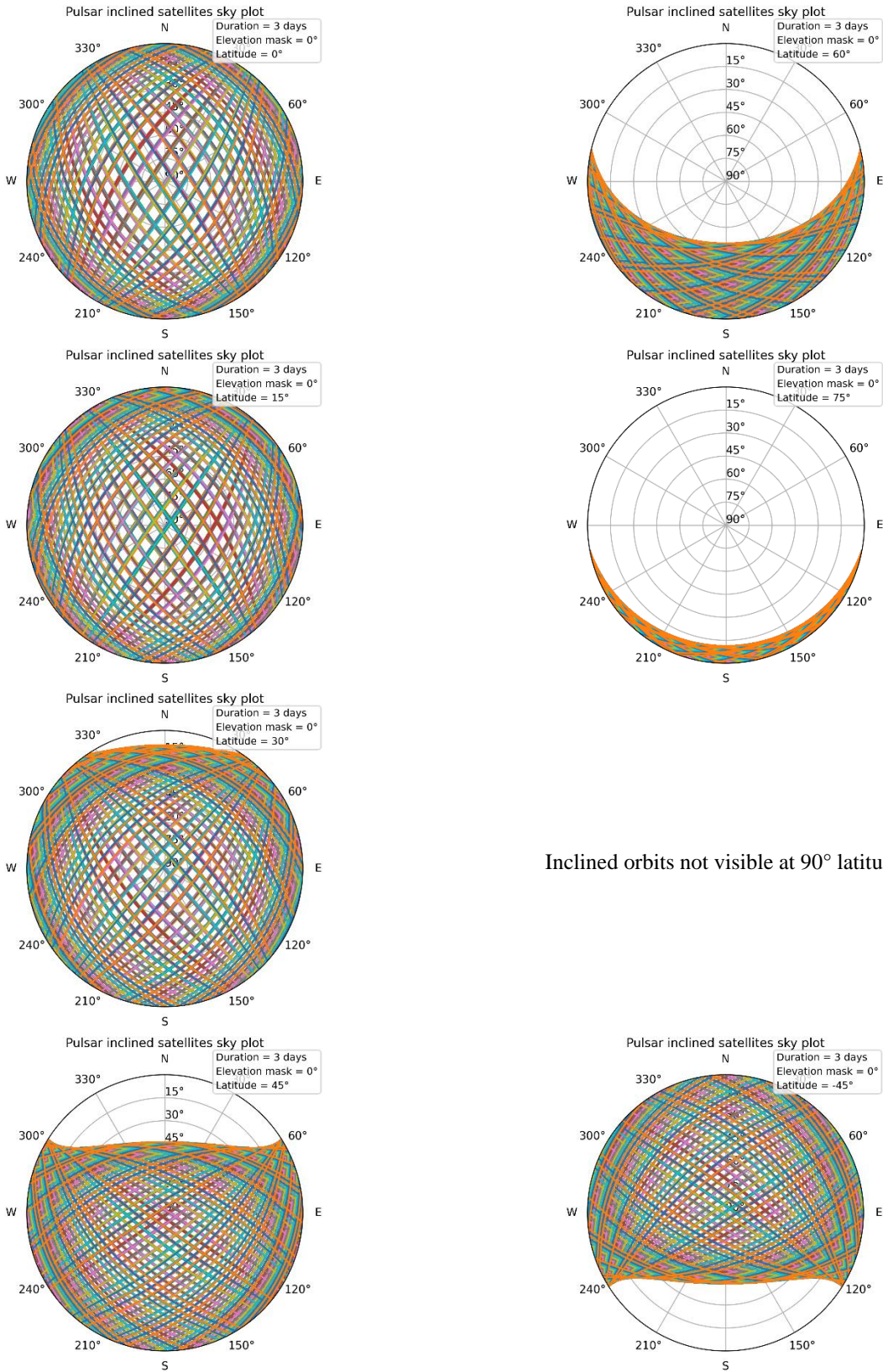

**Figure A.1** Pulsar inclined satellites sky plot for different latitudes

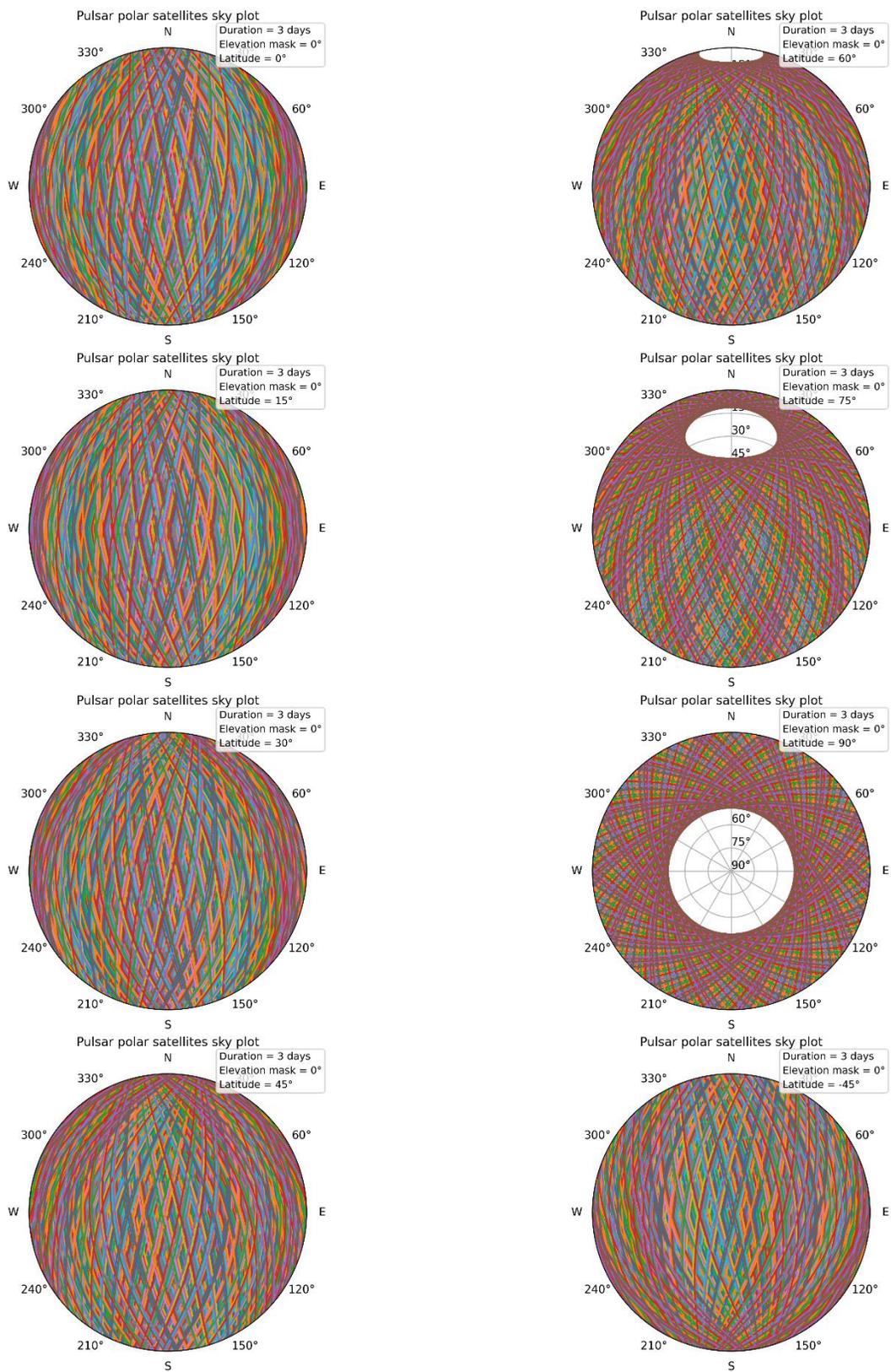

**Figure A.2** Pulsar polar satellites sky plot for different latitudes

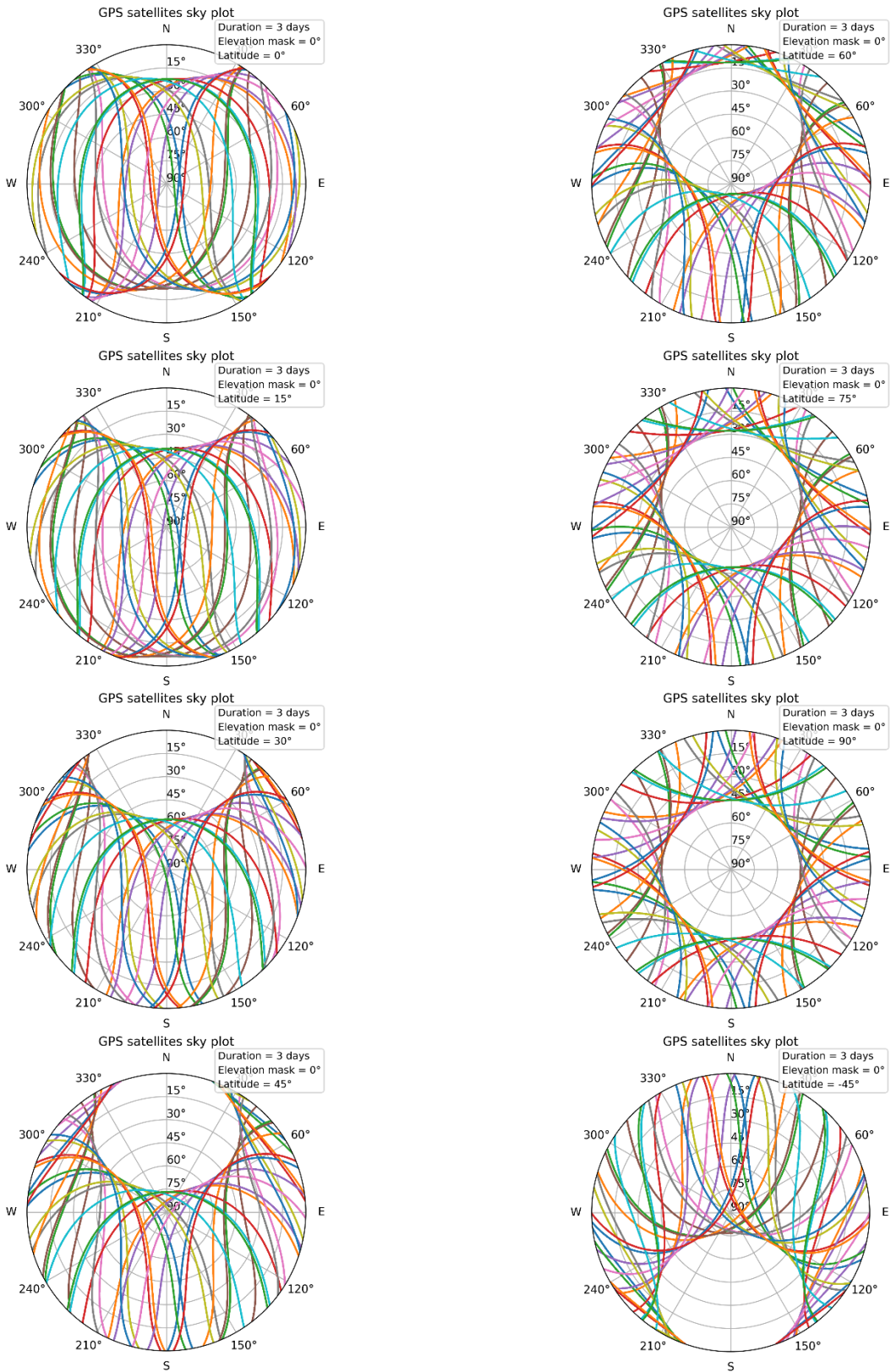

**Figure A.3** GPS satellites sky plot for different latitudes

## A.2. Elevation Over Time

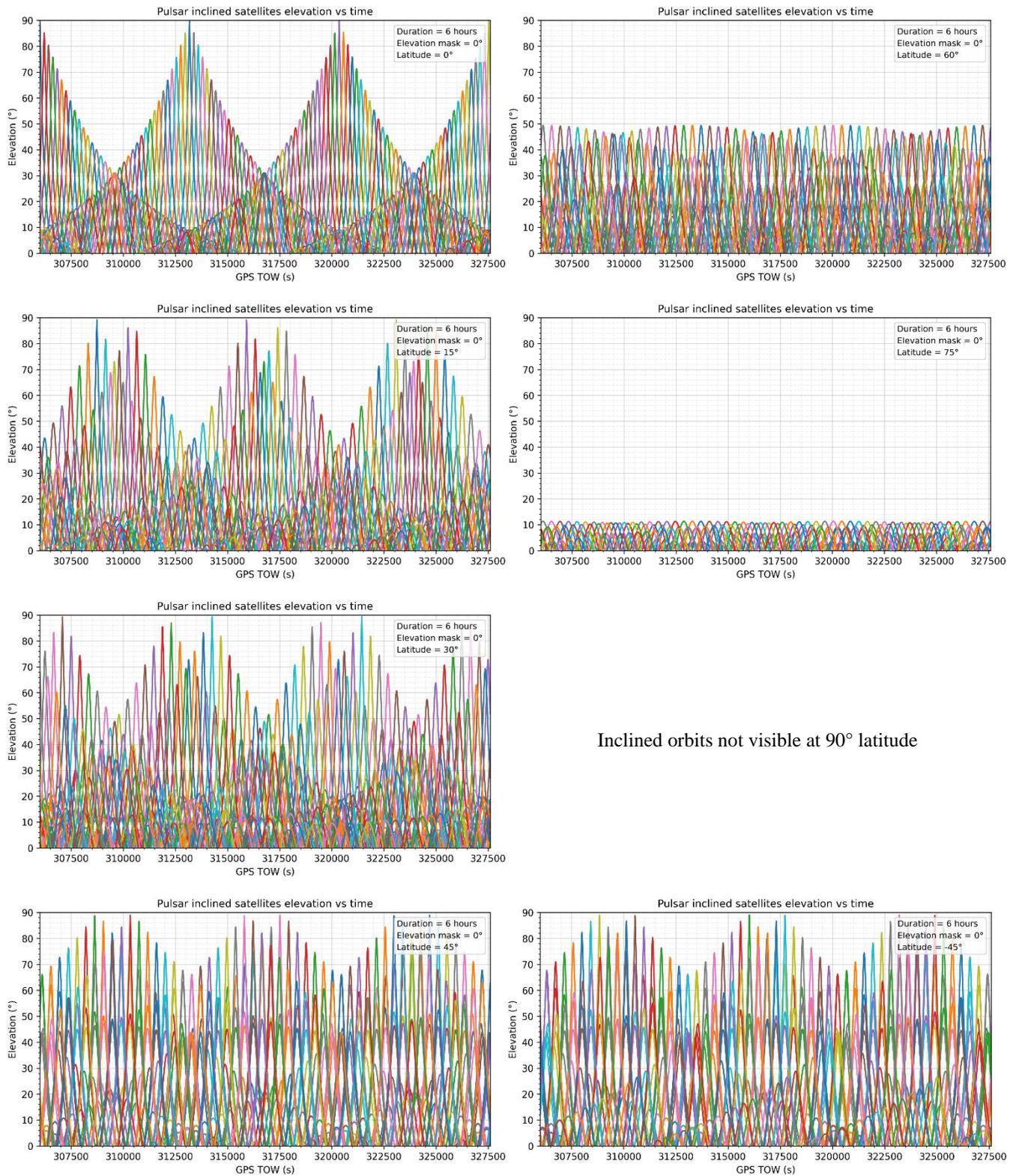

**Figure A.4** Pulsar inclined satellites elevation over time for different latitudes

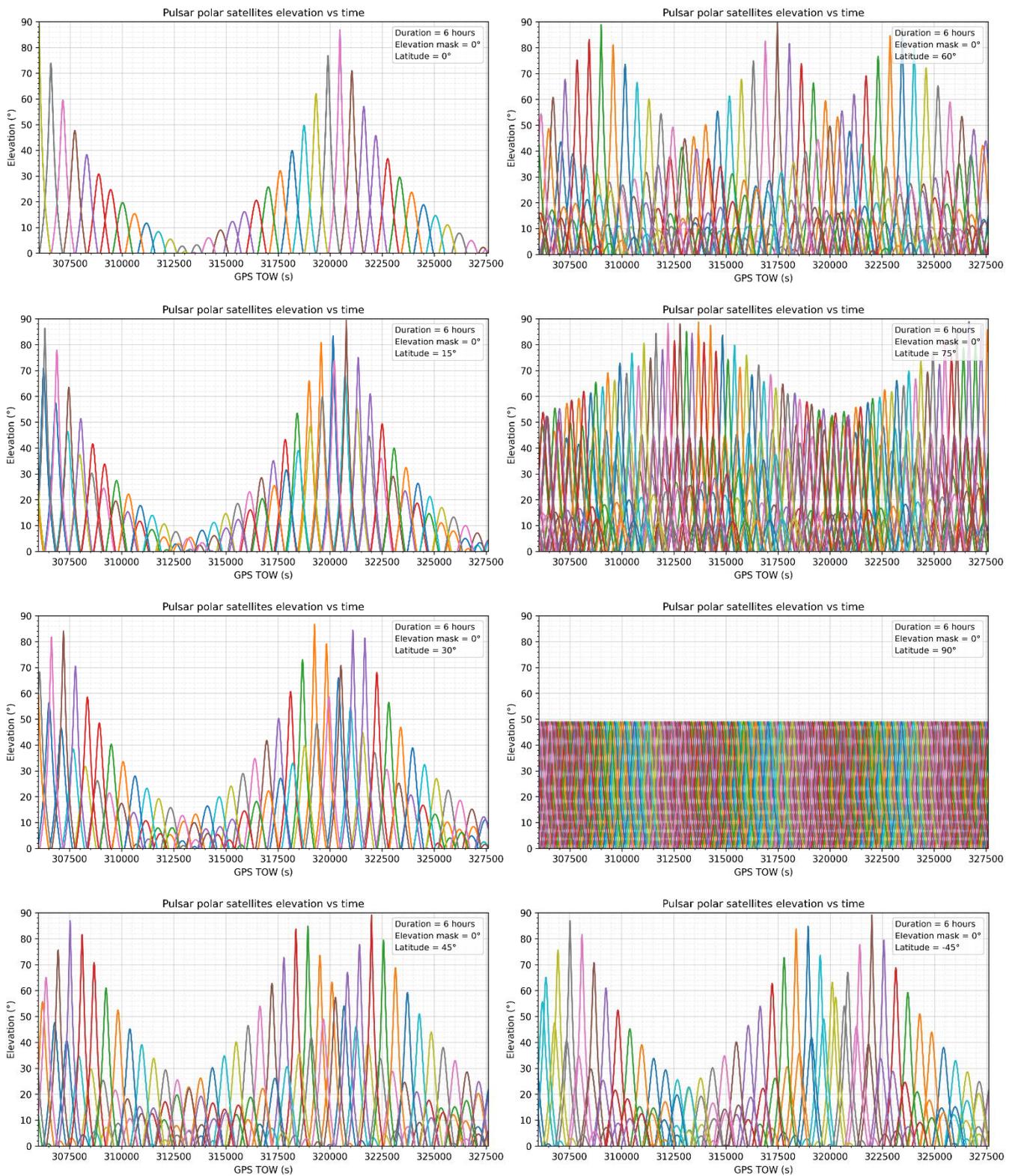

**Figure A.5** Pulsar polar satellites elevation over time for different latitudes

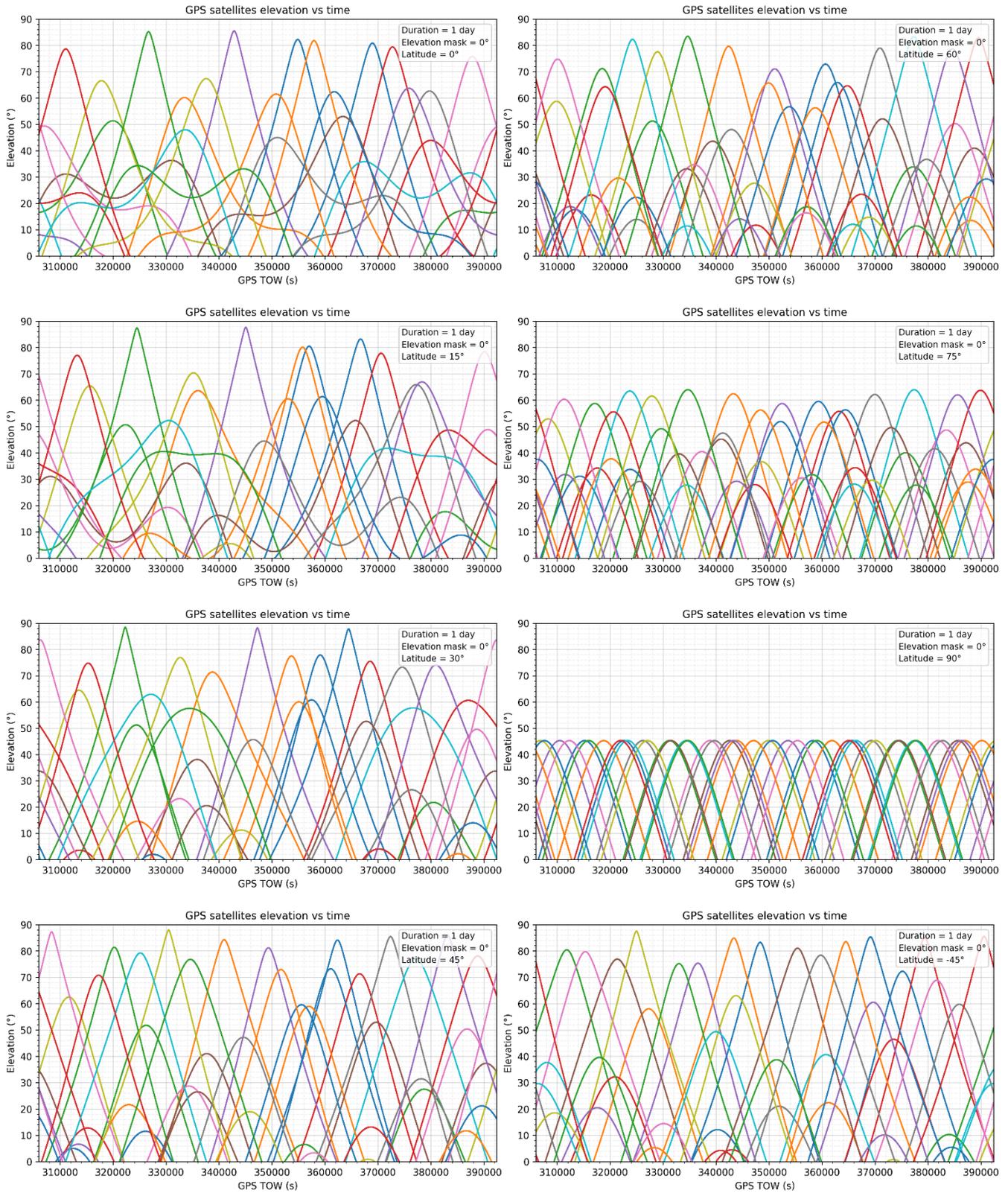

**Figure A.6** GPS satellites elevation over time for different latitudes

### A.3. Elevation Distribution

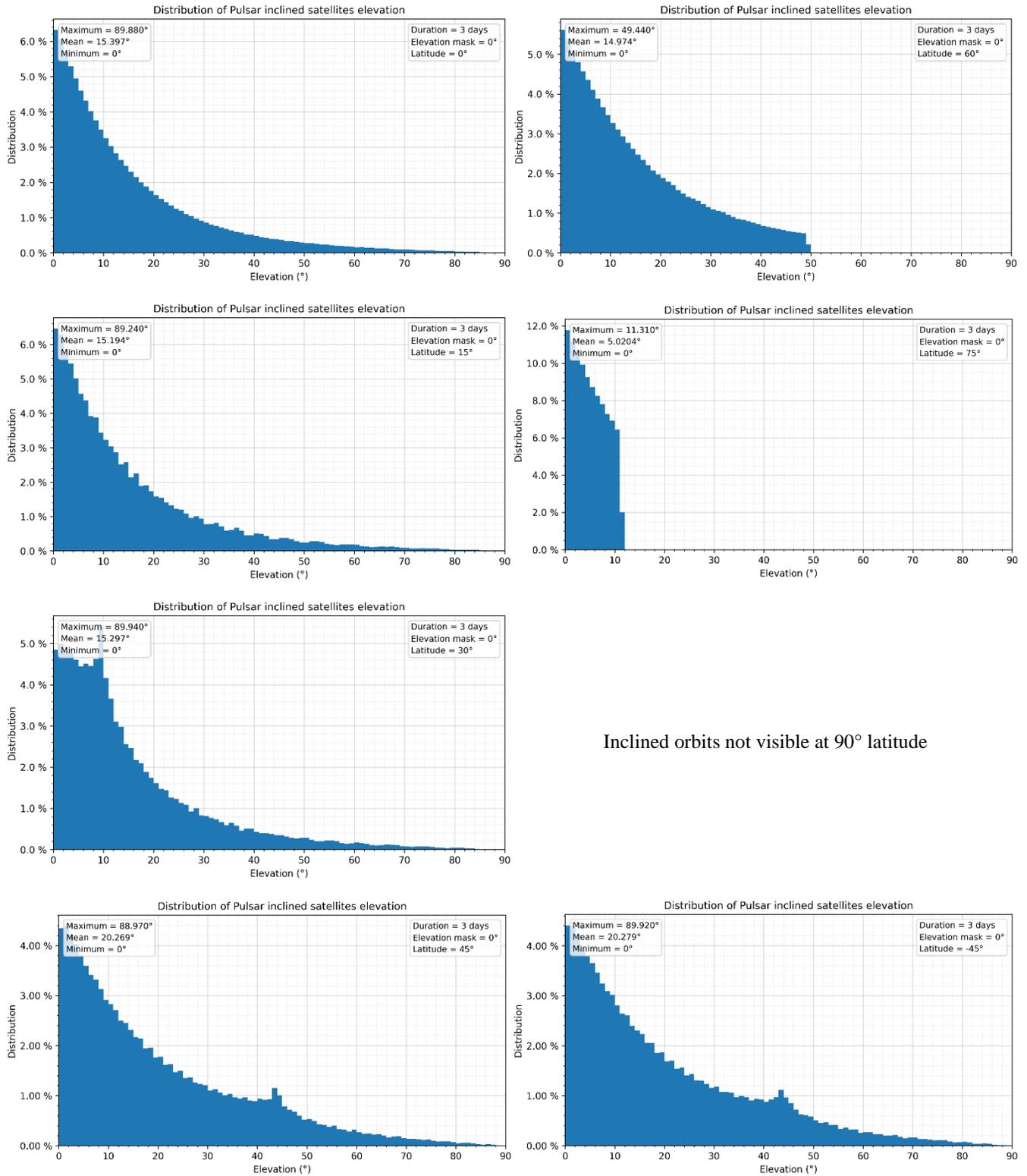

**Figure A.7** Distribution of Pulsar inclined satellites elevation for different latitudes

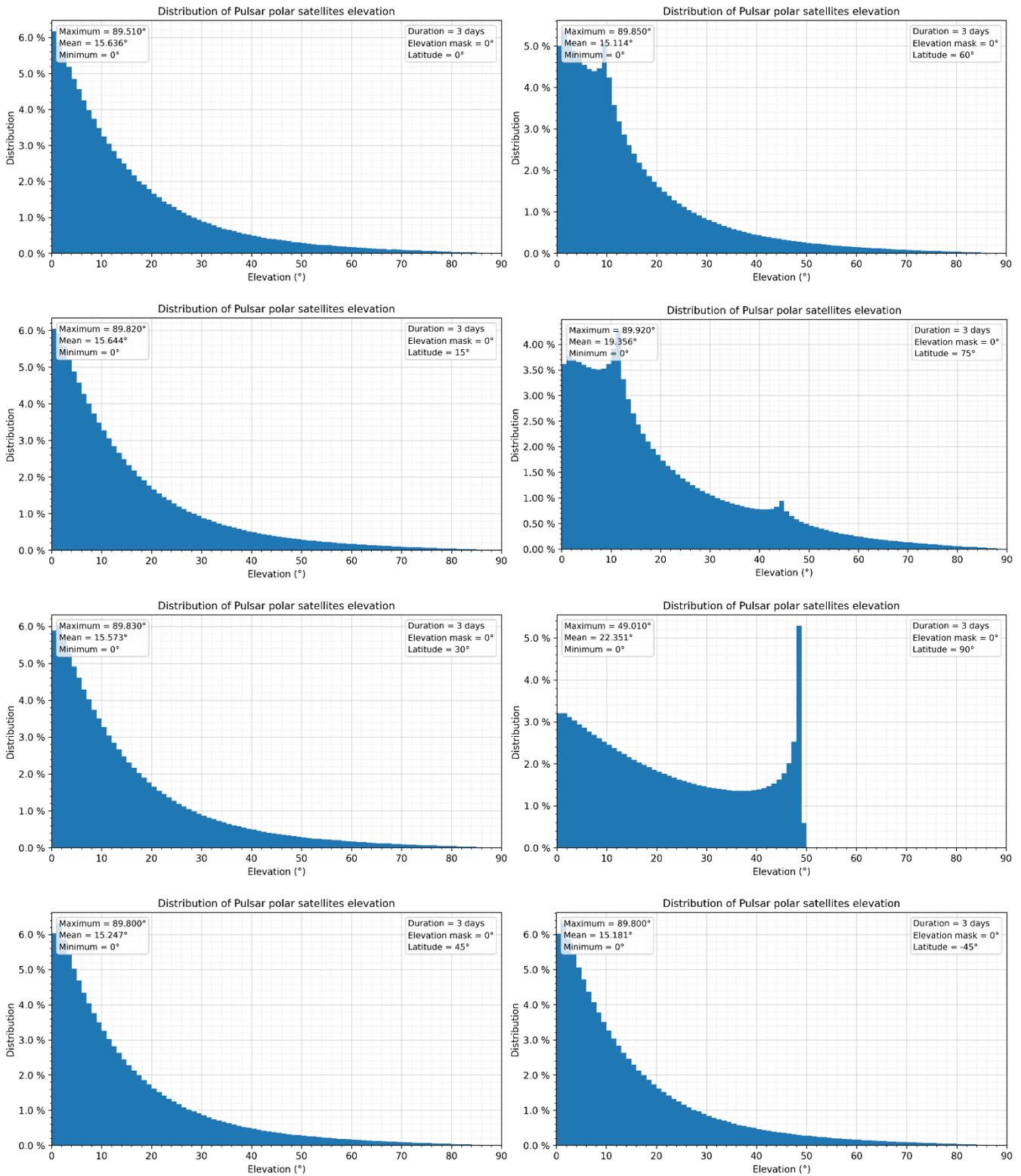

**Figure A.8** Distribution of Pulsar polar satellites elevation for different latitudes

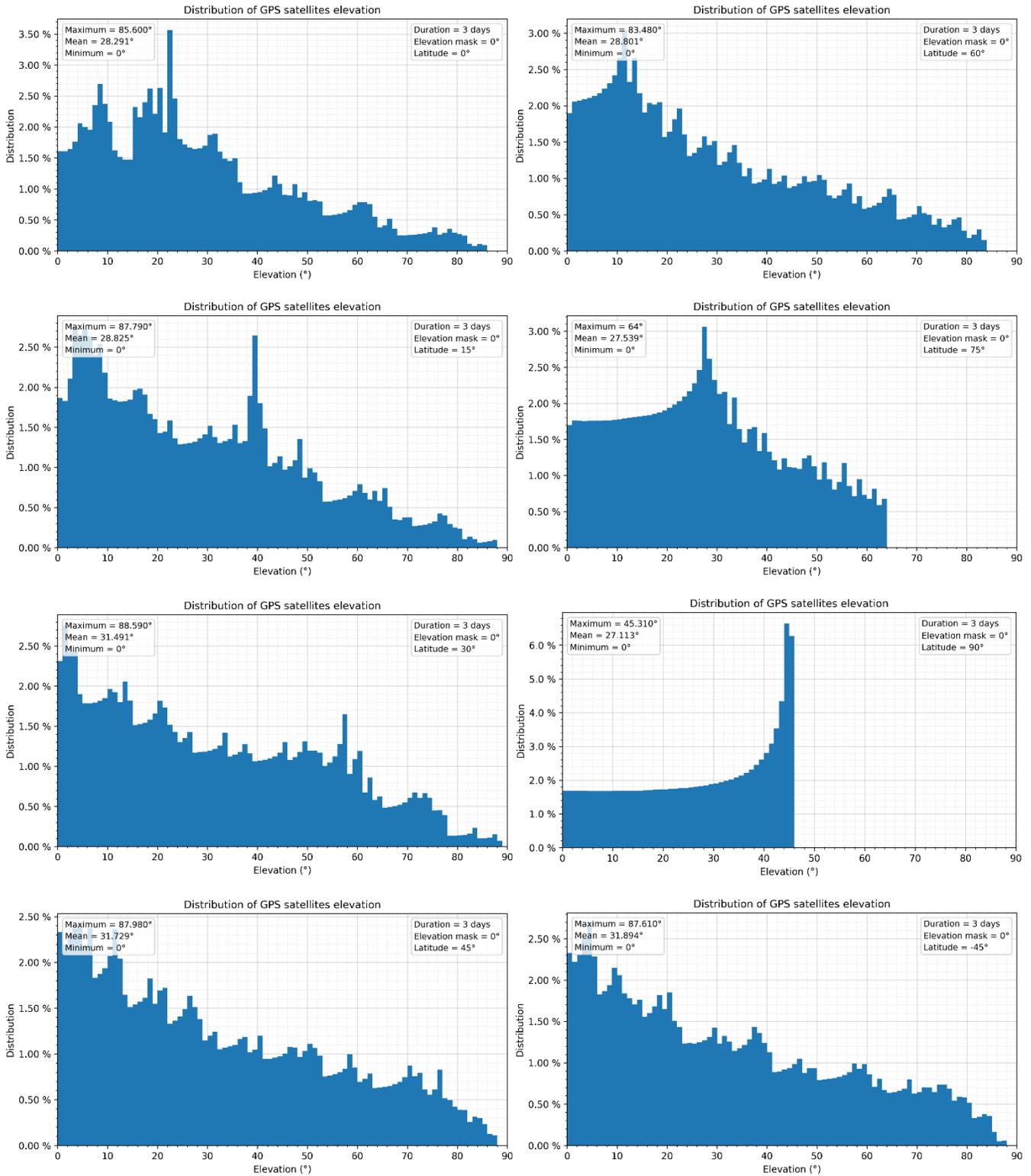

**Figure A.9** Distribution of GPS satellites elevation for different latitudes

## A.4. SVID

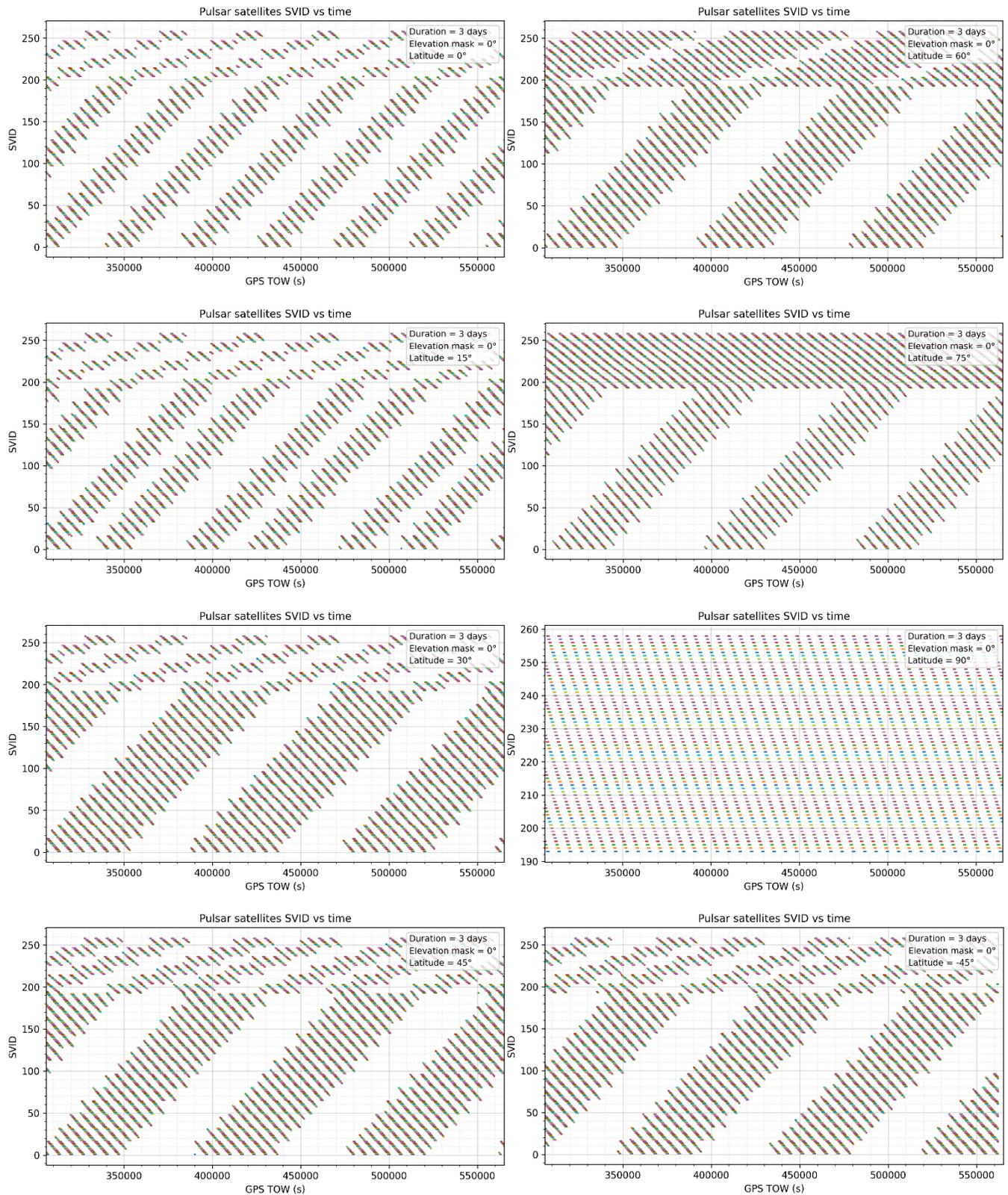

**Figure A.10** Pulsar satellites SVID over time for different latitudes

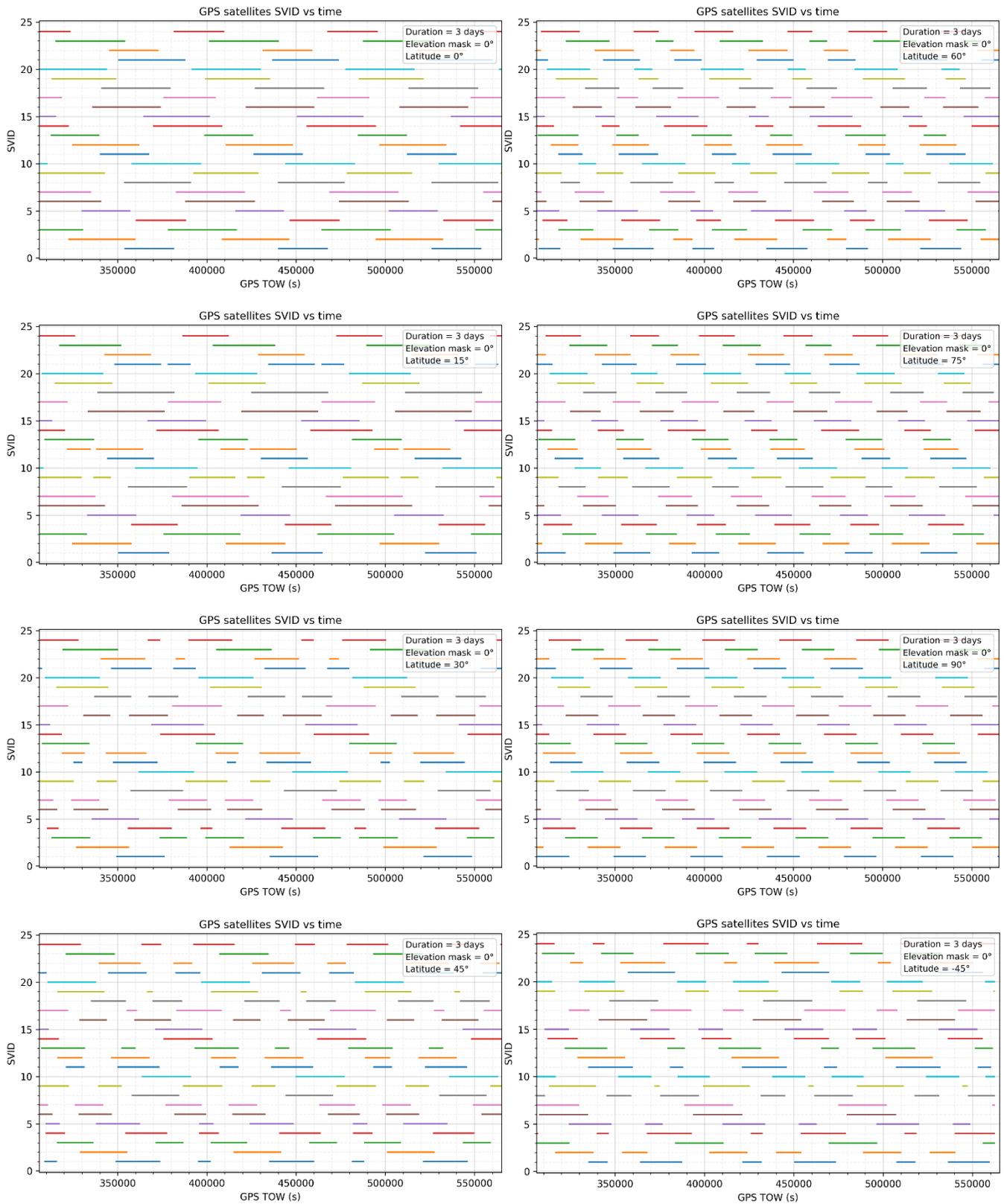

**Figure A.11** GPS satellites SVID over time for different latitudes

## A.5. PRN ID

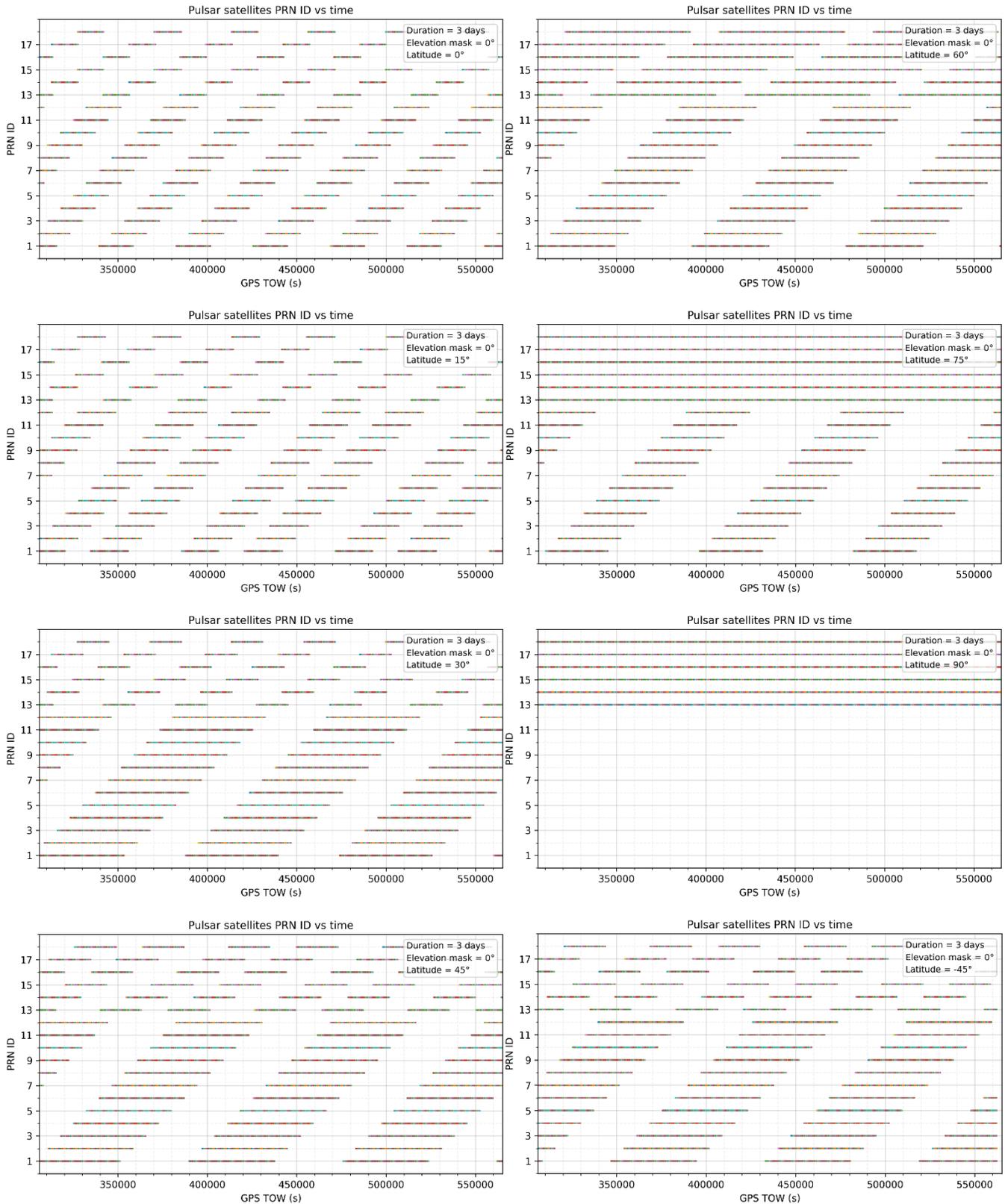

**Figure A.12** Pulsar satellites PRN ID over time for different latitudes

## A.6. Satellites Pass Duration

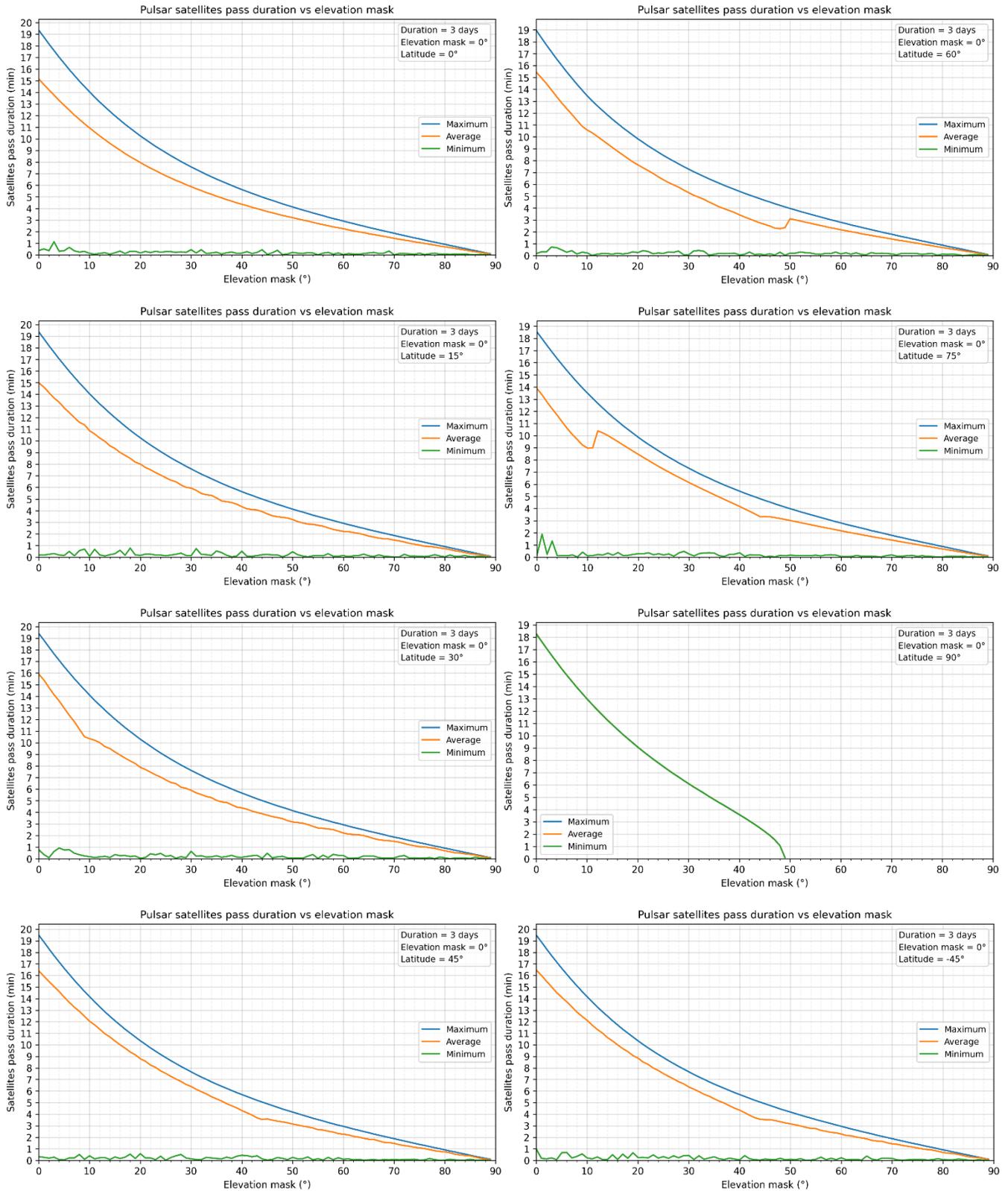

**Figure A.13** Pulsar satellites pass duration vs elevation mask for different latitudes

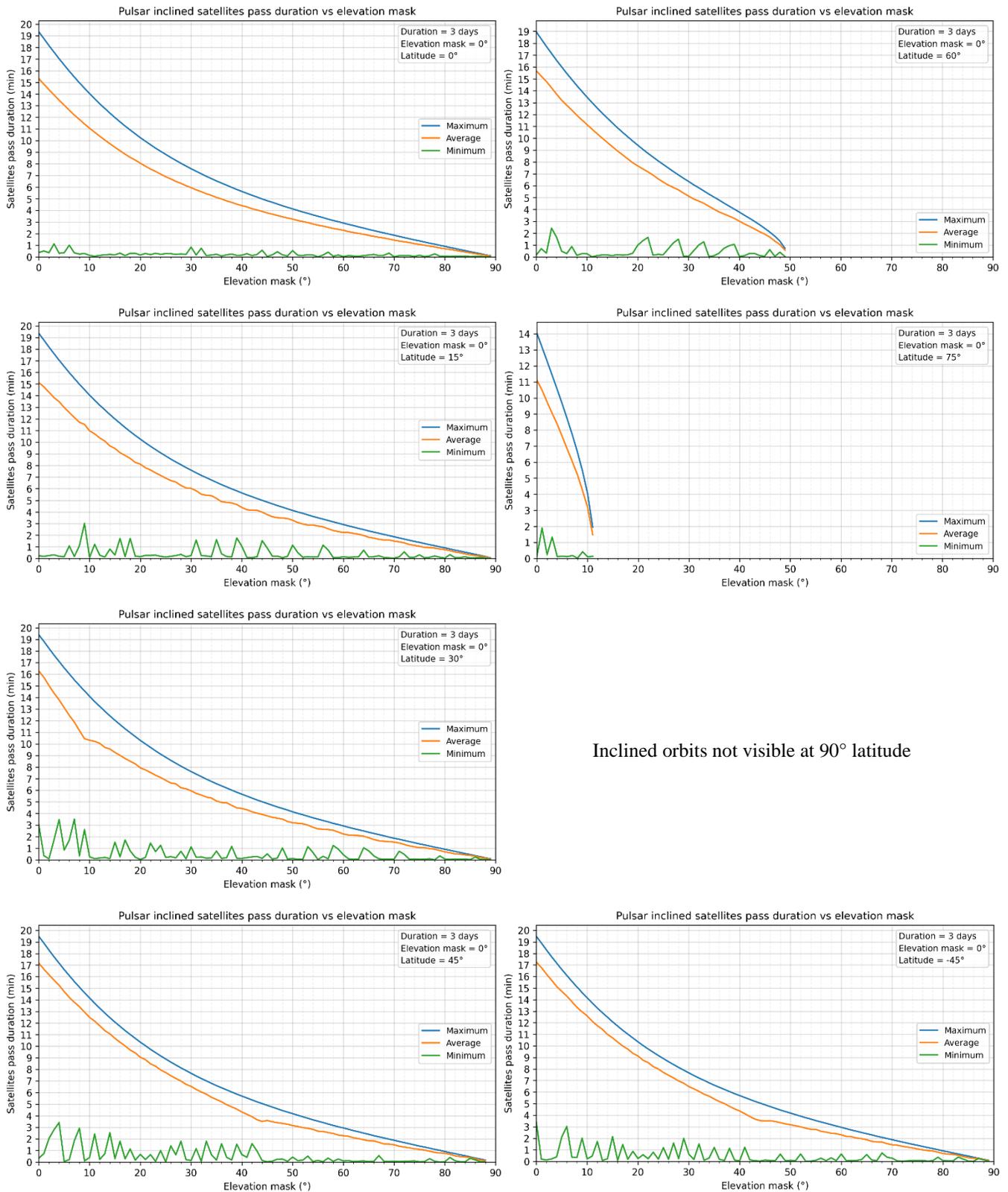

**Figure A.14** Pulsar inclined satellites pass duration vs elevation mask for different latitudes

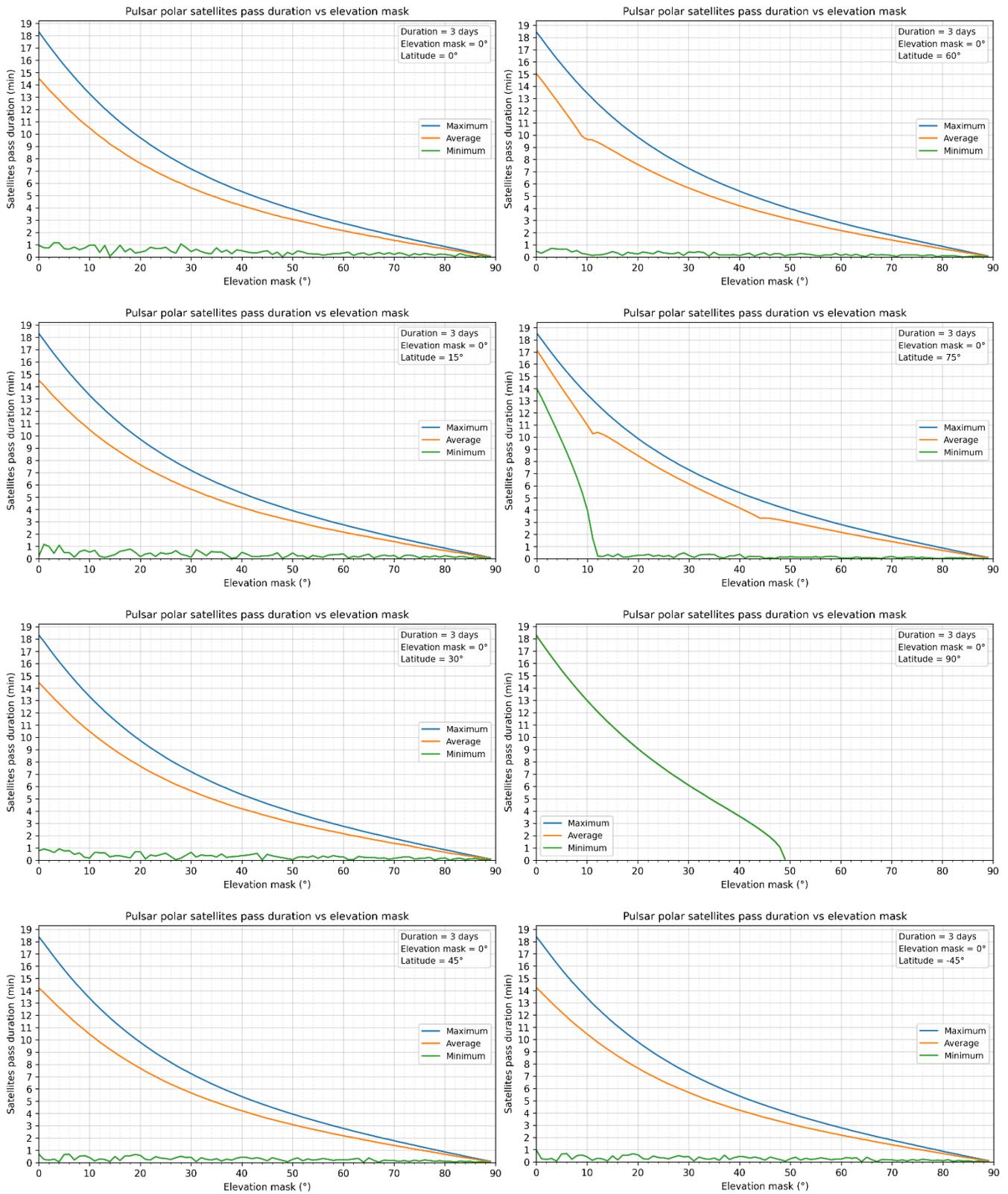

**Figure A.15** Pulsar polar satellites pass duration vs elevation mask for different latitudes

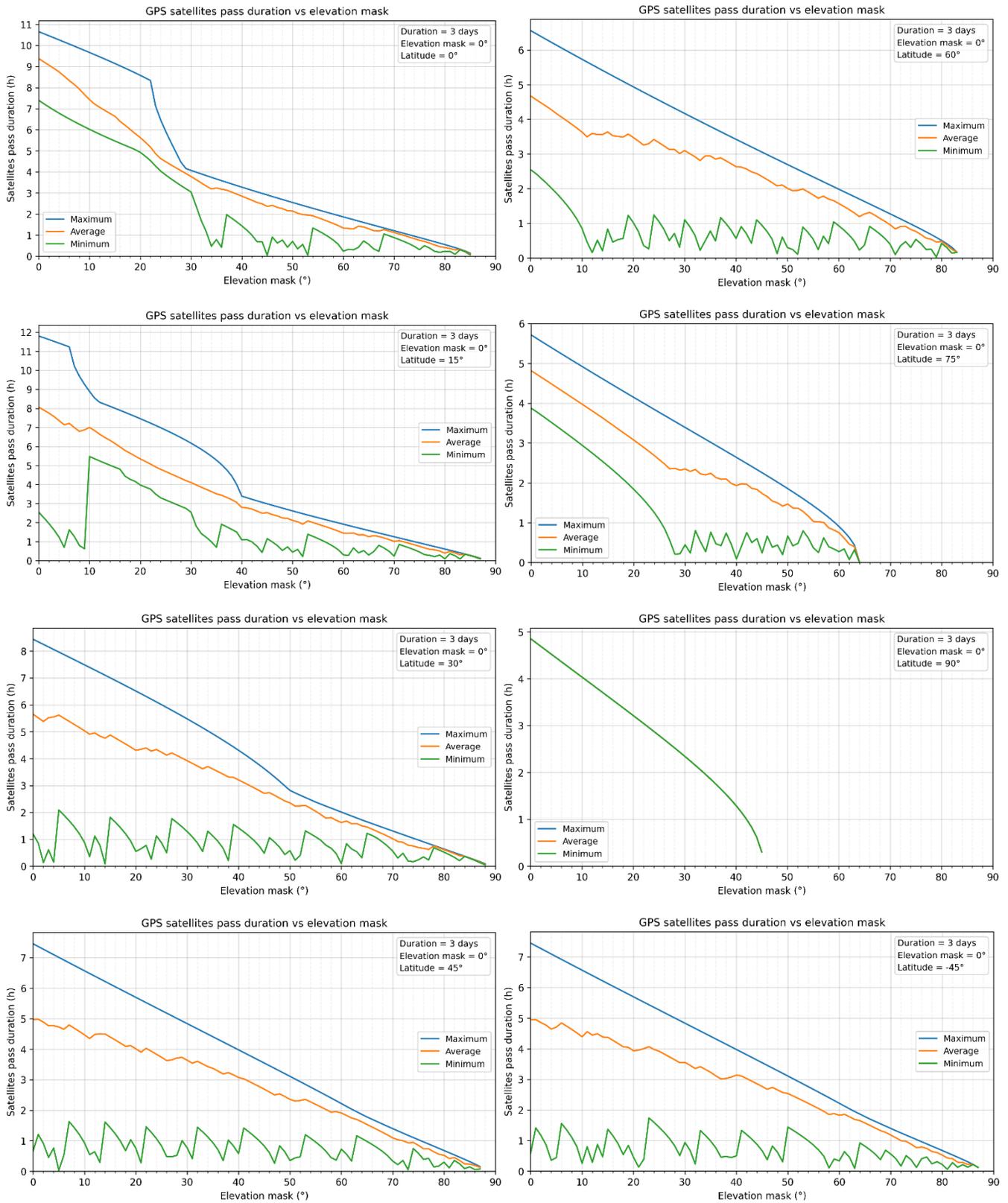

**Figure A.16** GPS satellites pass duration vs elevation mask for different latitudes

## A.7. Number of Satellites in View

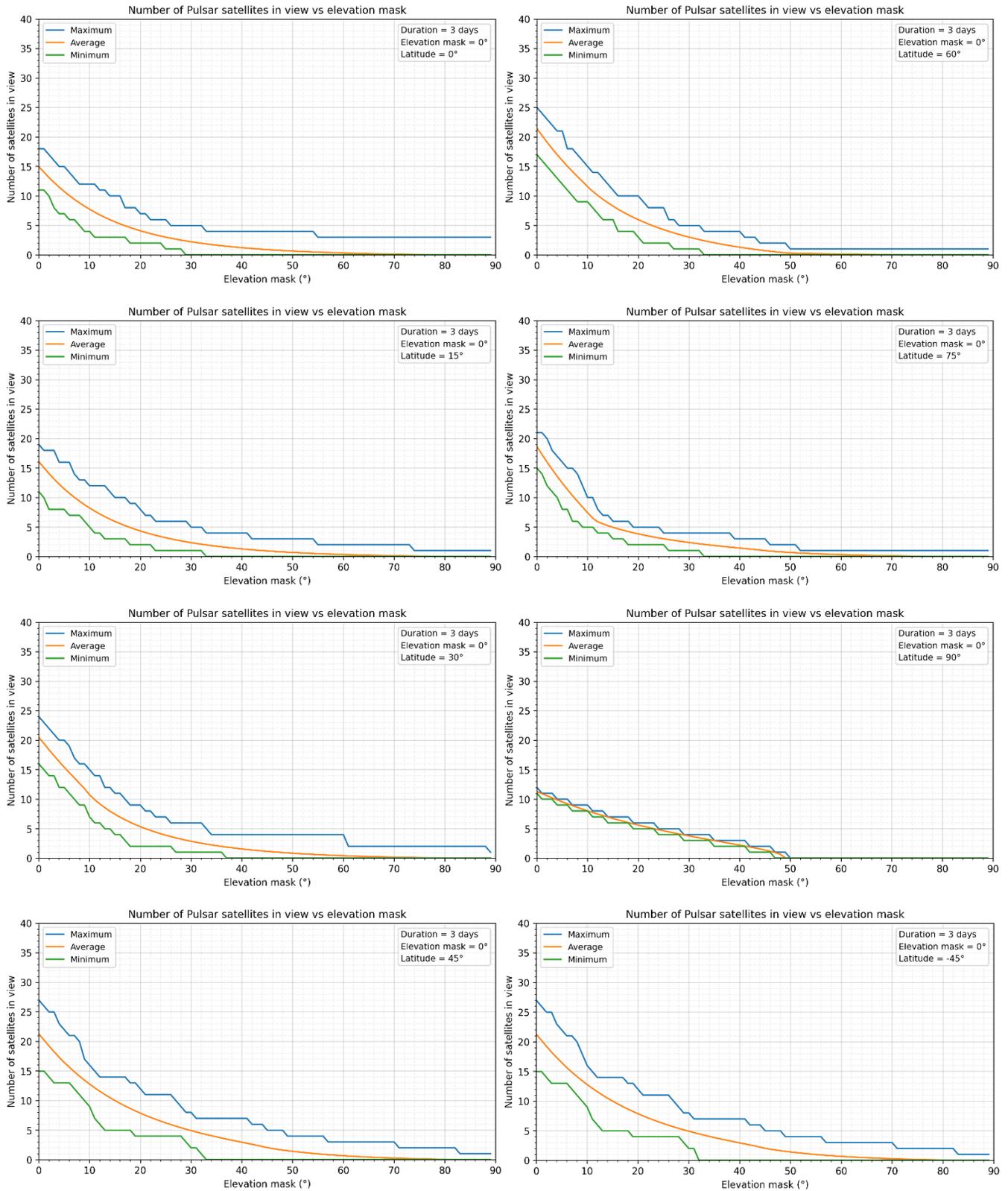

**Figure A.17** Number of Pulsar satellites in view vs elevation mask for different latitudes

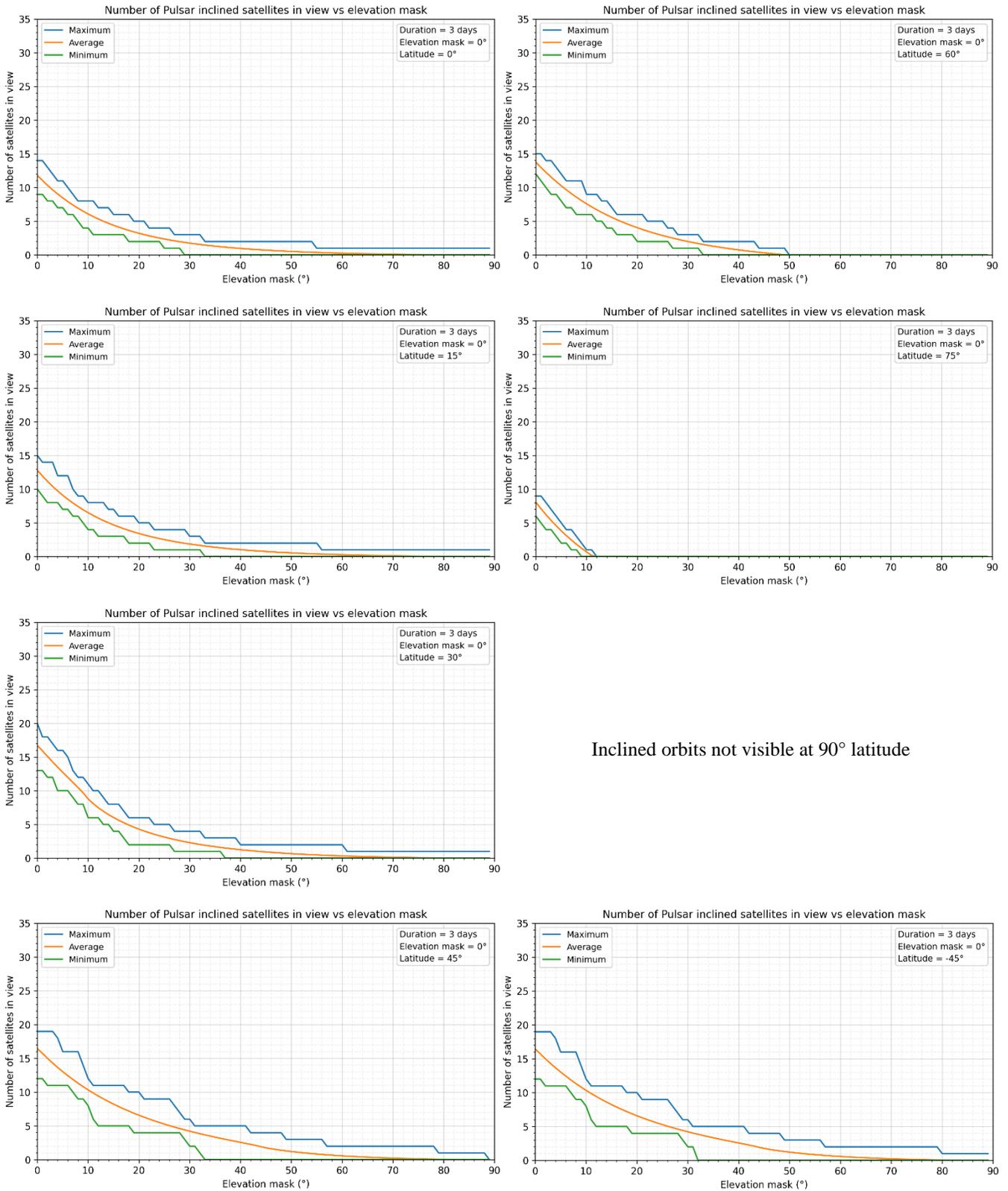

**Figure A.18** Number of Pulsar inclined satellites in view vs elevation mask for different latitudes

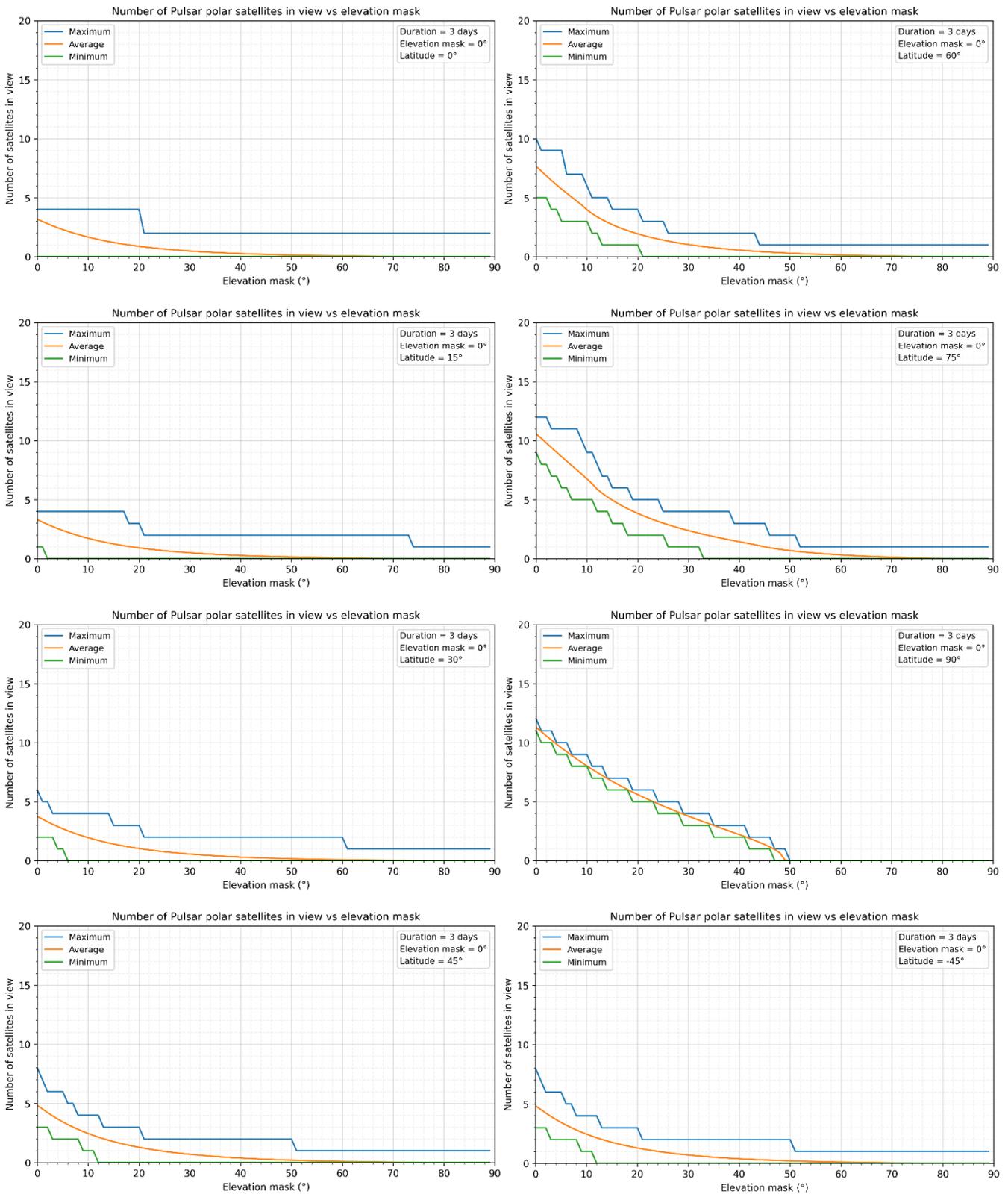

**Figure A.19** Number of Pulsar polar satellites in view vs elevation mask for different latitudes

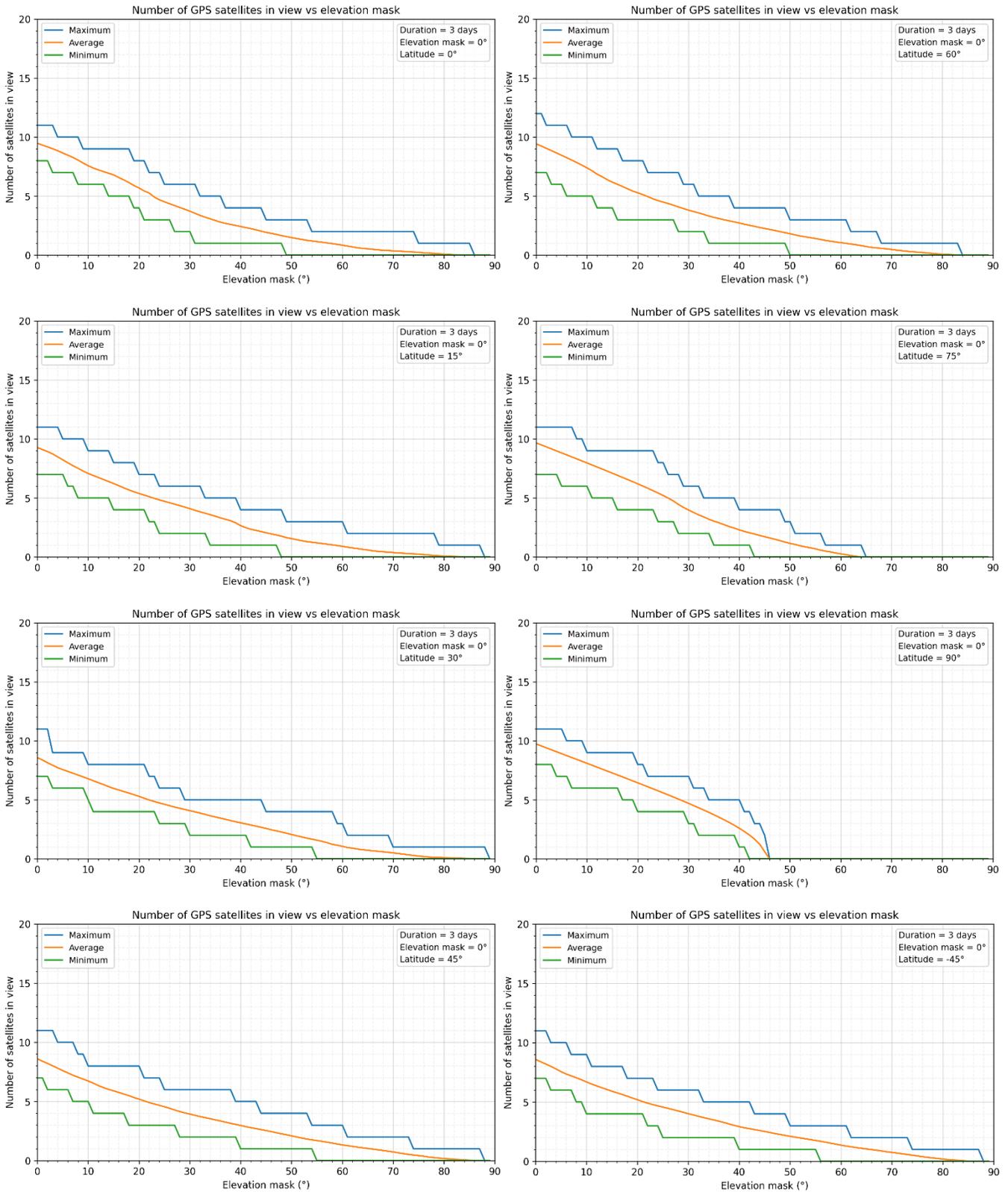

**Figure A.20** Number of GPS satellites in view vs elevation mask for different latitudes

## A.8. Number of Orbital Planes in View

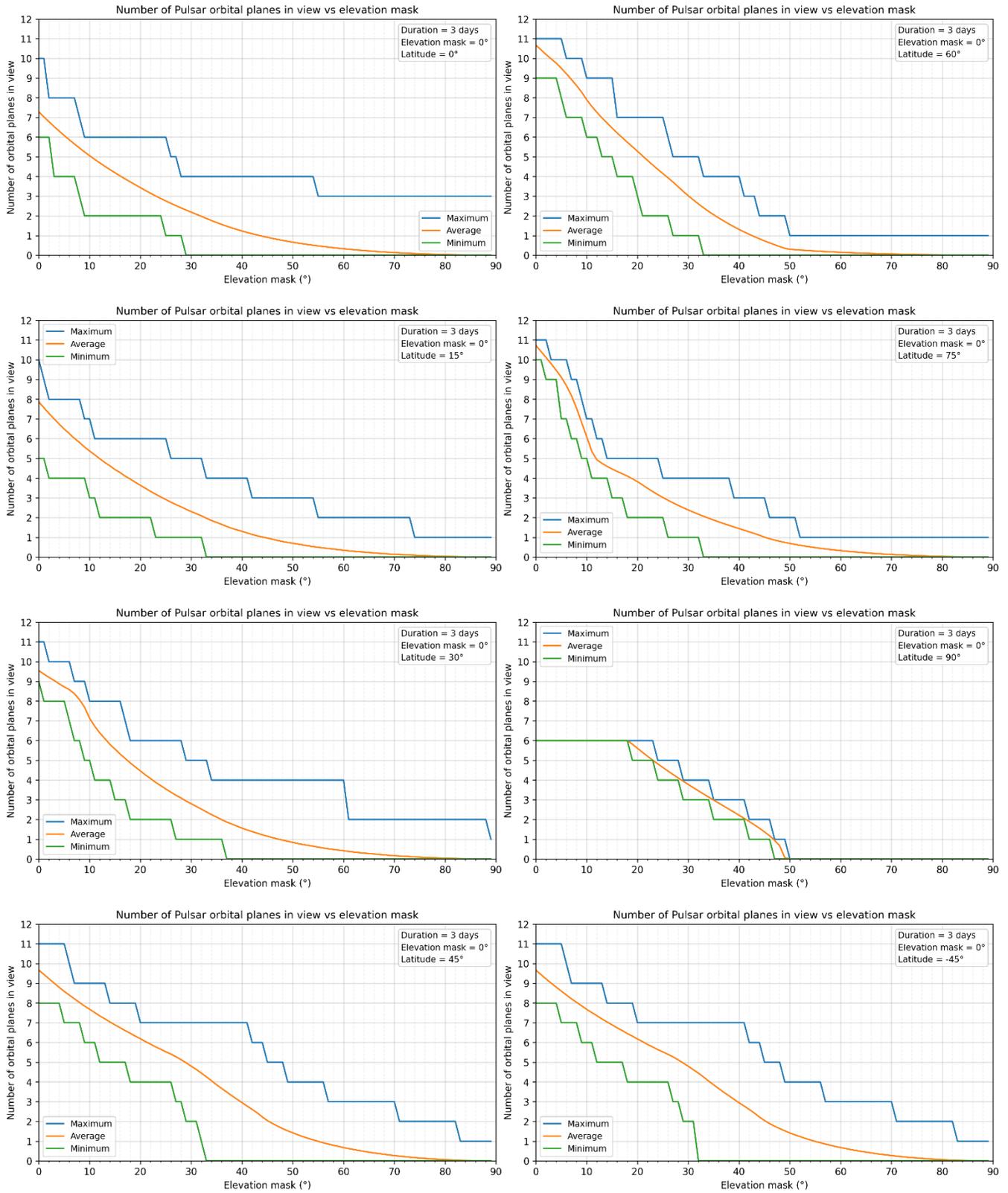

**Figure A.21** Number of Pulsar orbital planes in view vs elevation mask for different latitudes

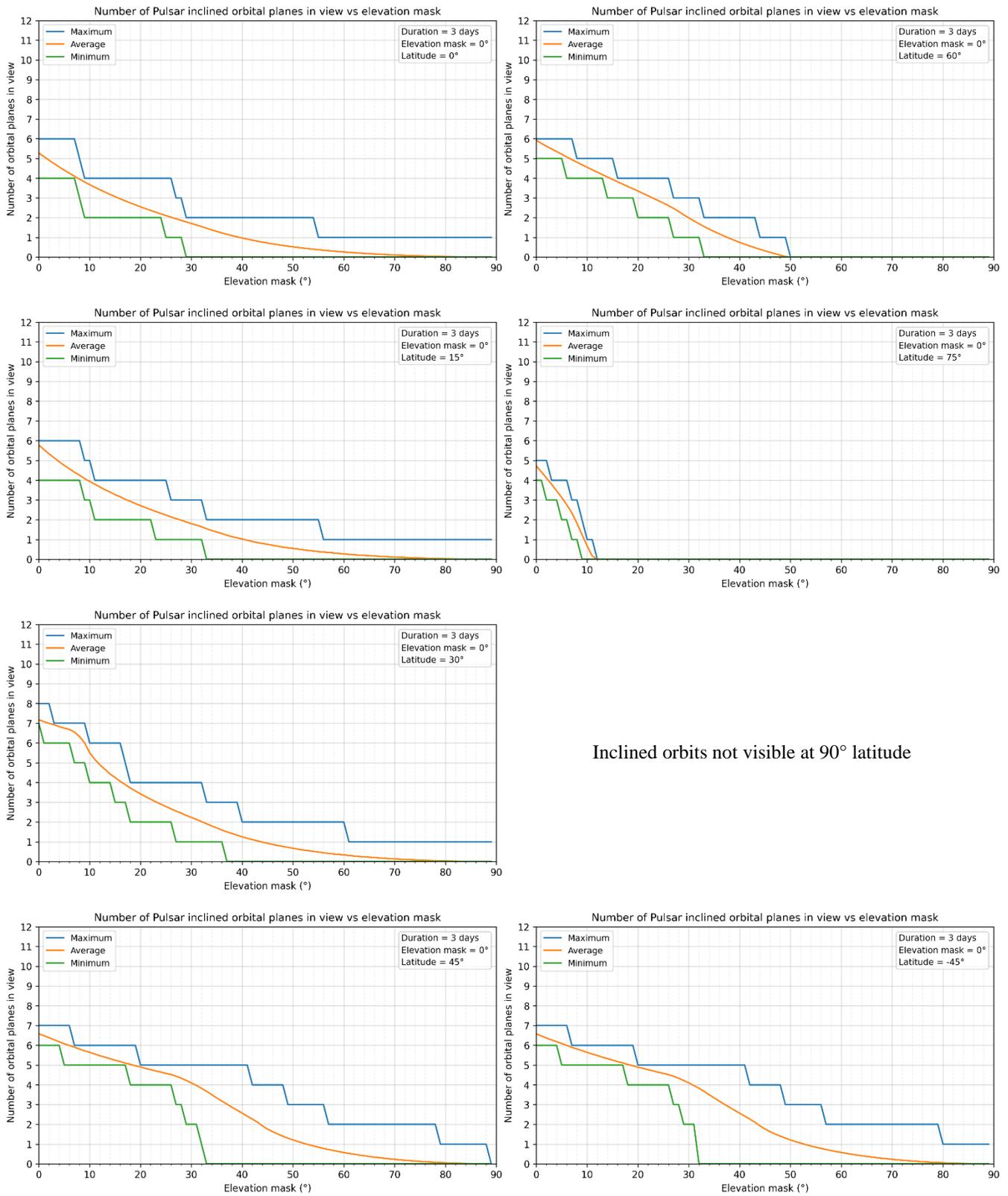

**Figure A.22** Number of Pulsar inclined orbital planes in view vs elevation mask for different latitudes

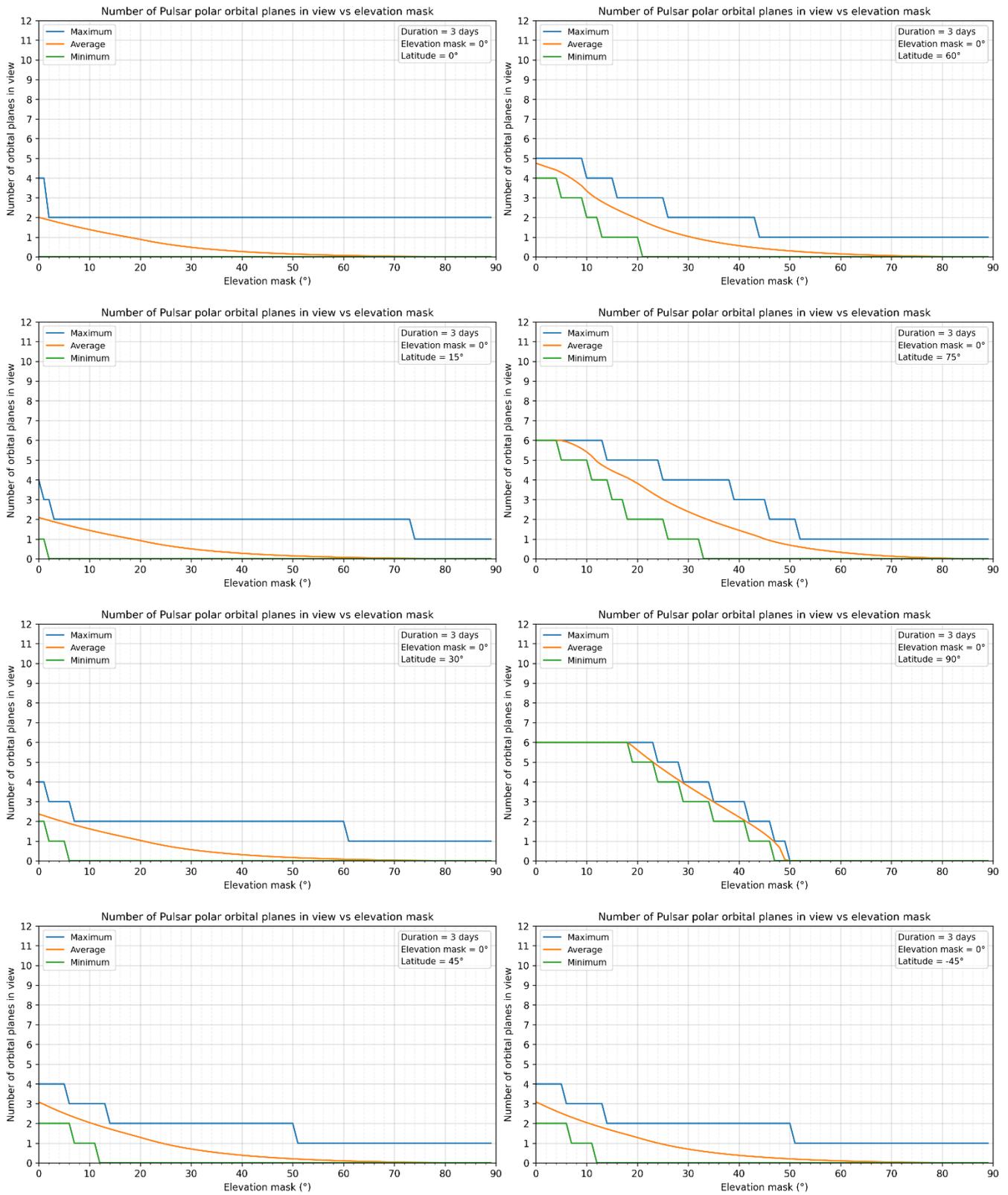

**Figure A.23** Number of Pulsar polar orbital planes in view vs elevation mask for different latitudes

## A.9. Number of Satellites in View from a Same Orbital Plane

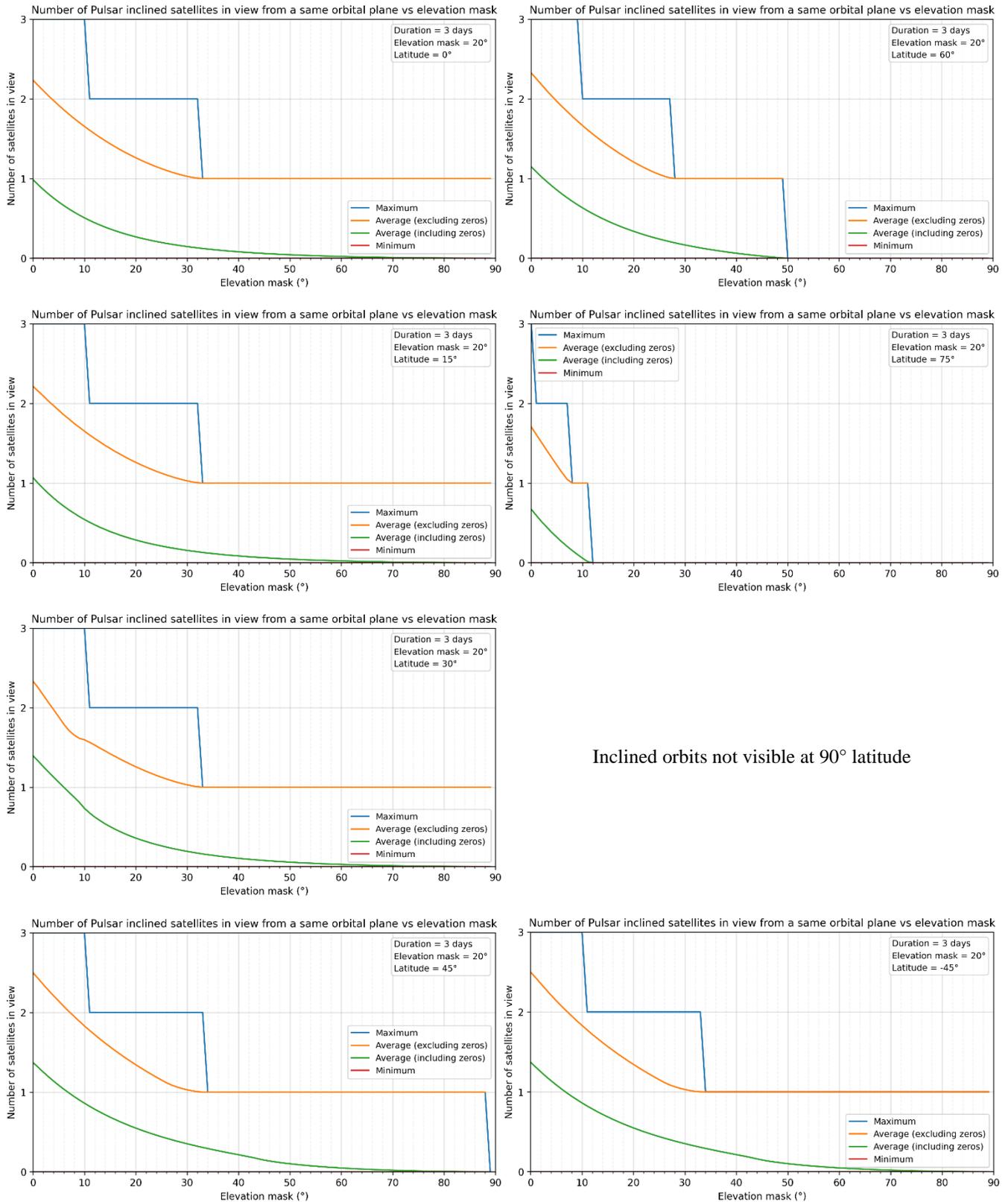

**Figure A.24** Number of Pulsar inclined satellites in view from a same orbital plane vs elevation mask for different latitudes

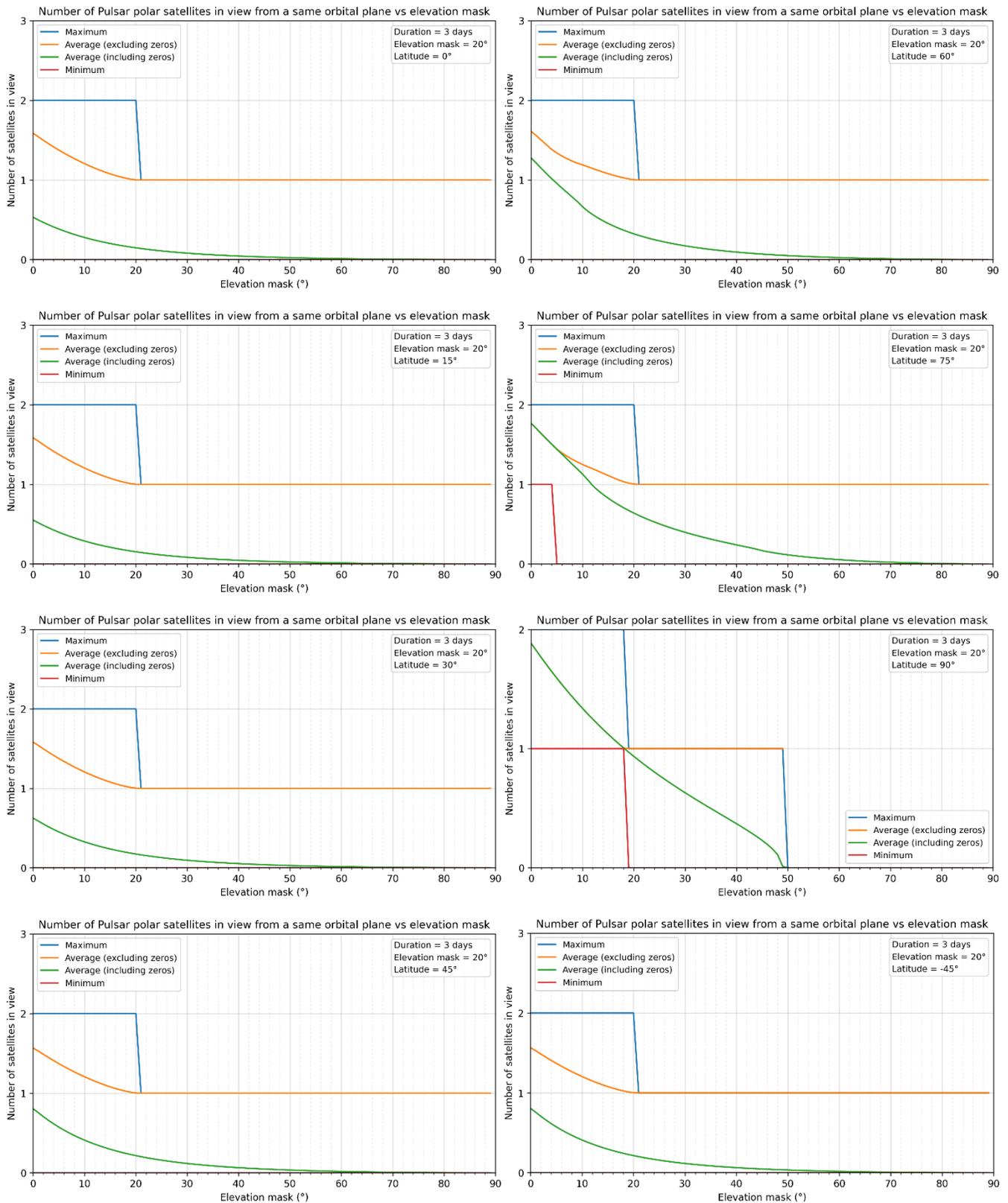

**Figure A.25** Number of Pulsar polar satellites in view from a same orbital plane vs elevation mask for different latitudes

## B. DETAILED PLOTS RELATED TO THE SIGNALS

### B.1. Carrier Doppler Over Time

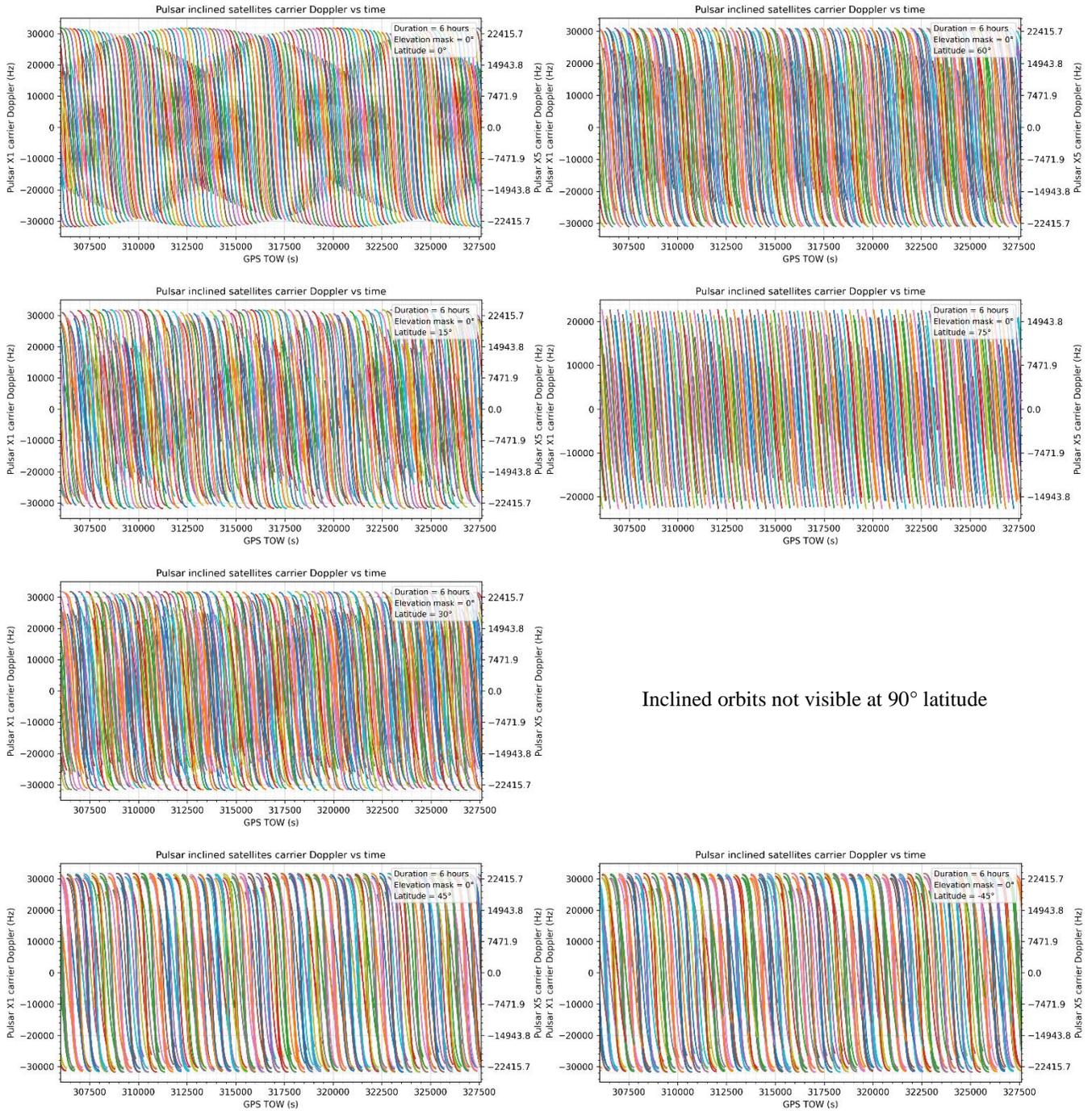

**Figure B.1** Carrier Doppler over time of Pulsar inclined satellites for different latitudes

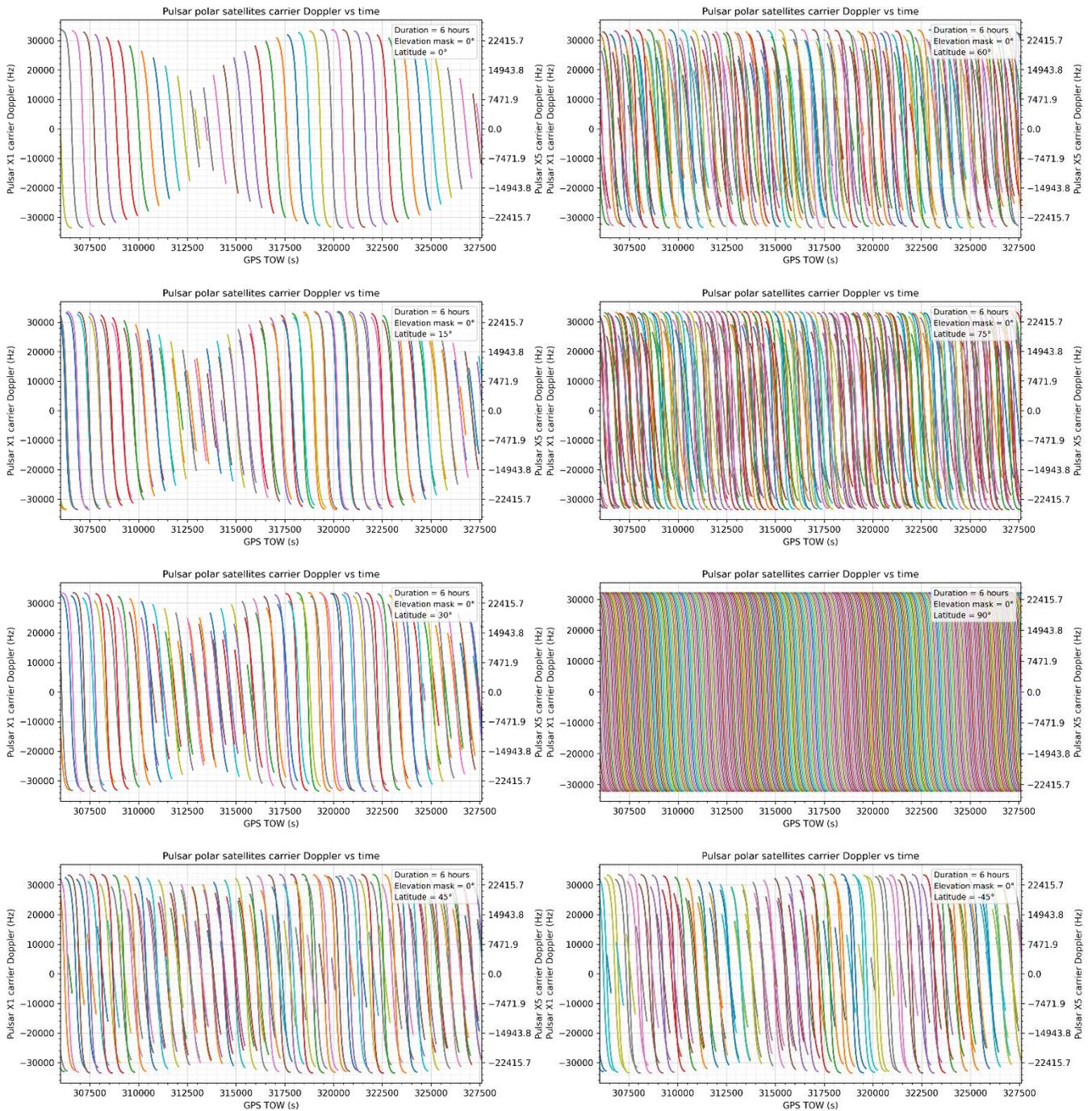

**Figure B.2** Carrier Doppler over time of Pulsar polar satellites for different latitudes

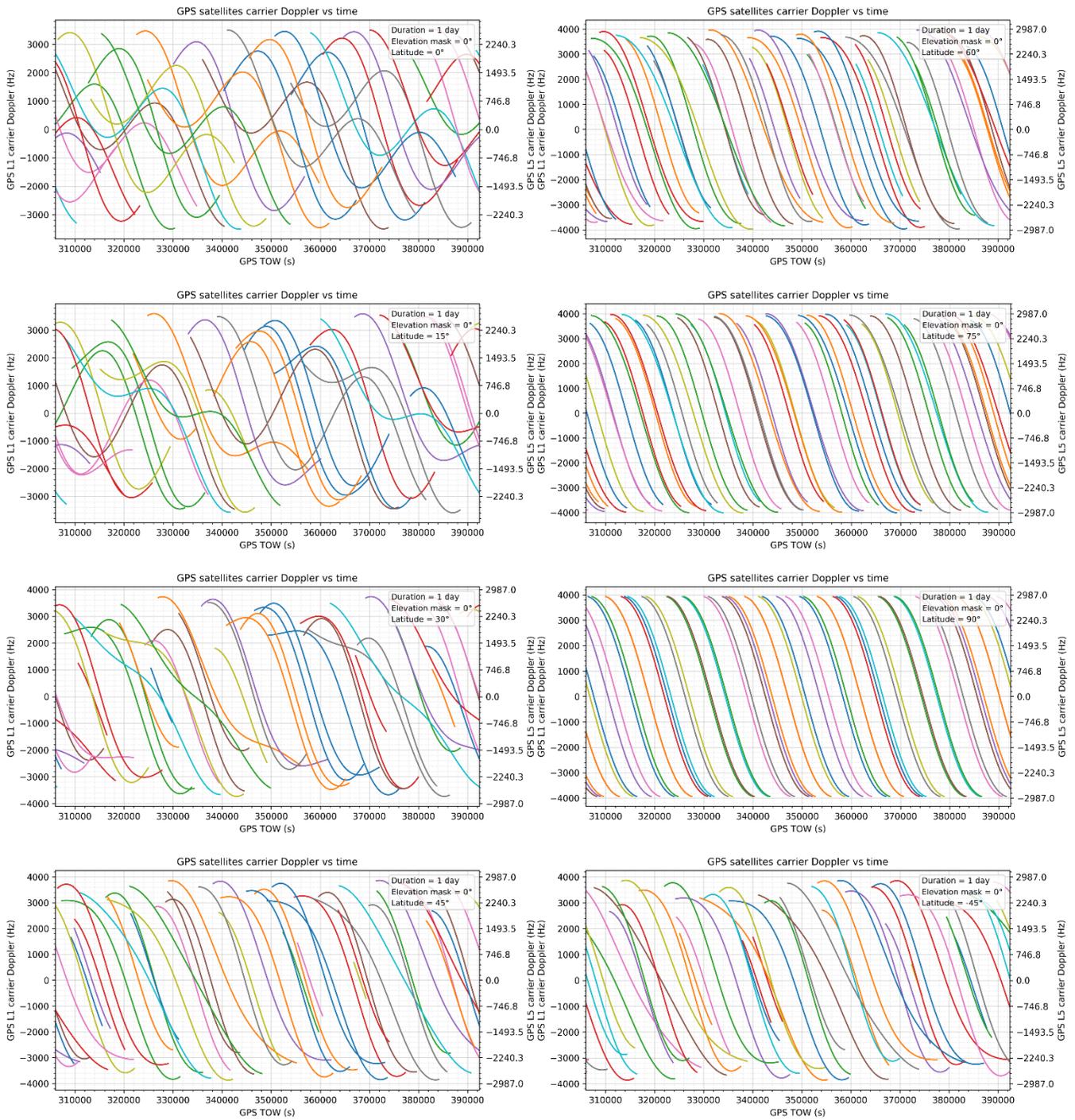

**Figure B.3** Carrier Doppler over time of GPS satellites for different latitudes

## B.2. Carrier Doppler Distribution

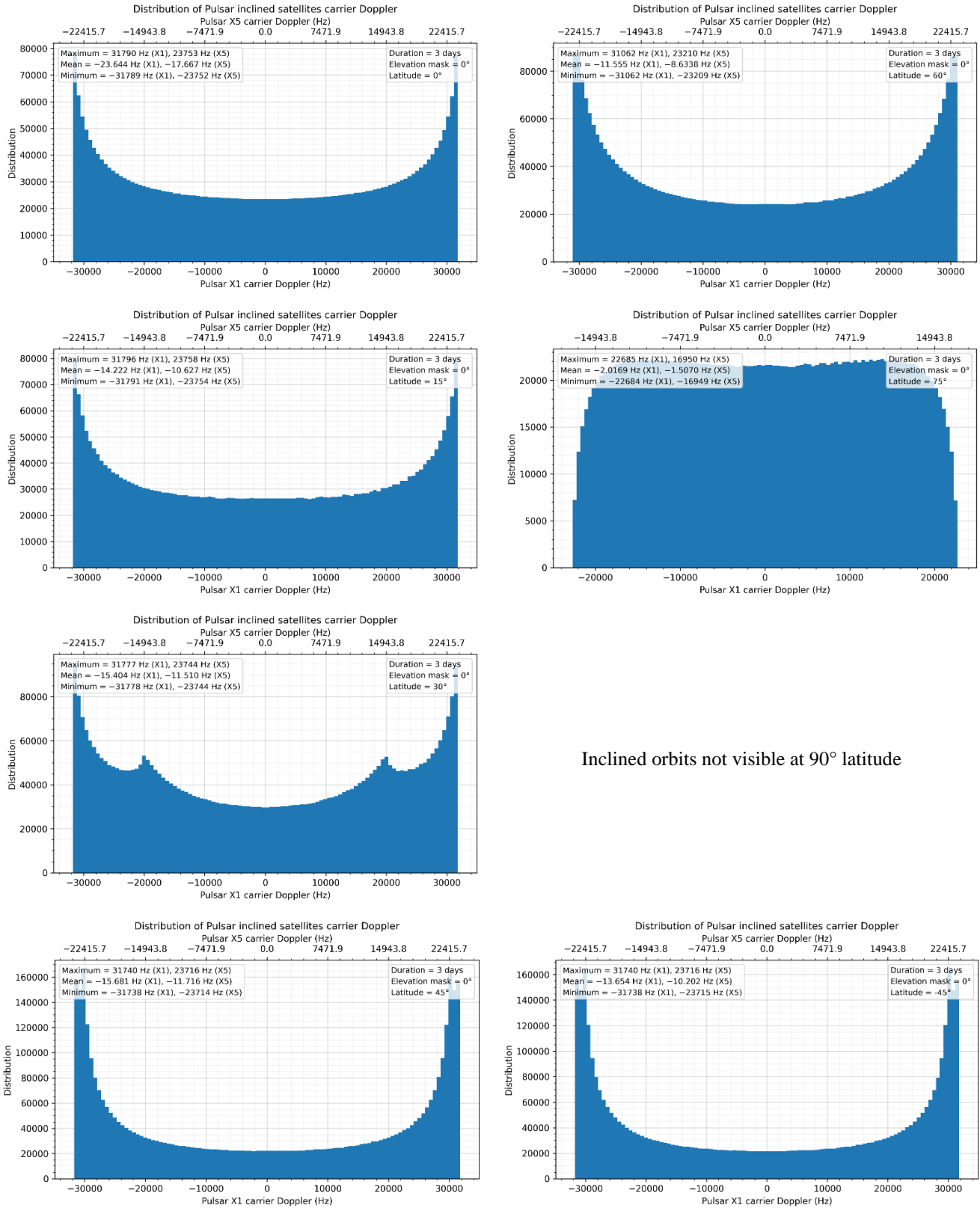

**Figure B.4** Carrier Doppler distribution of Pulsar inclined satellites for different latitudes

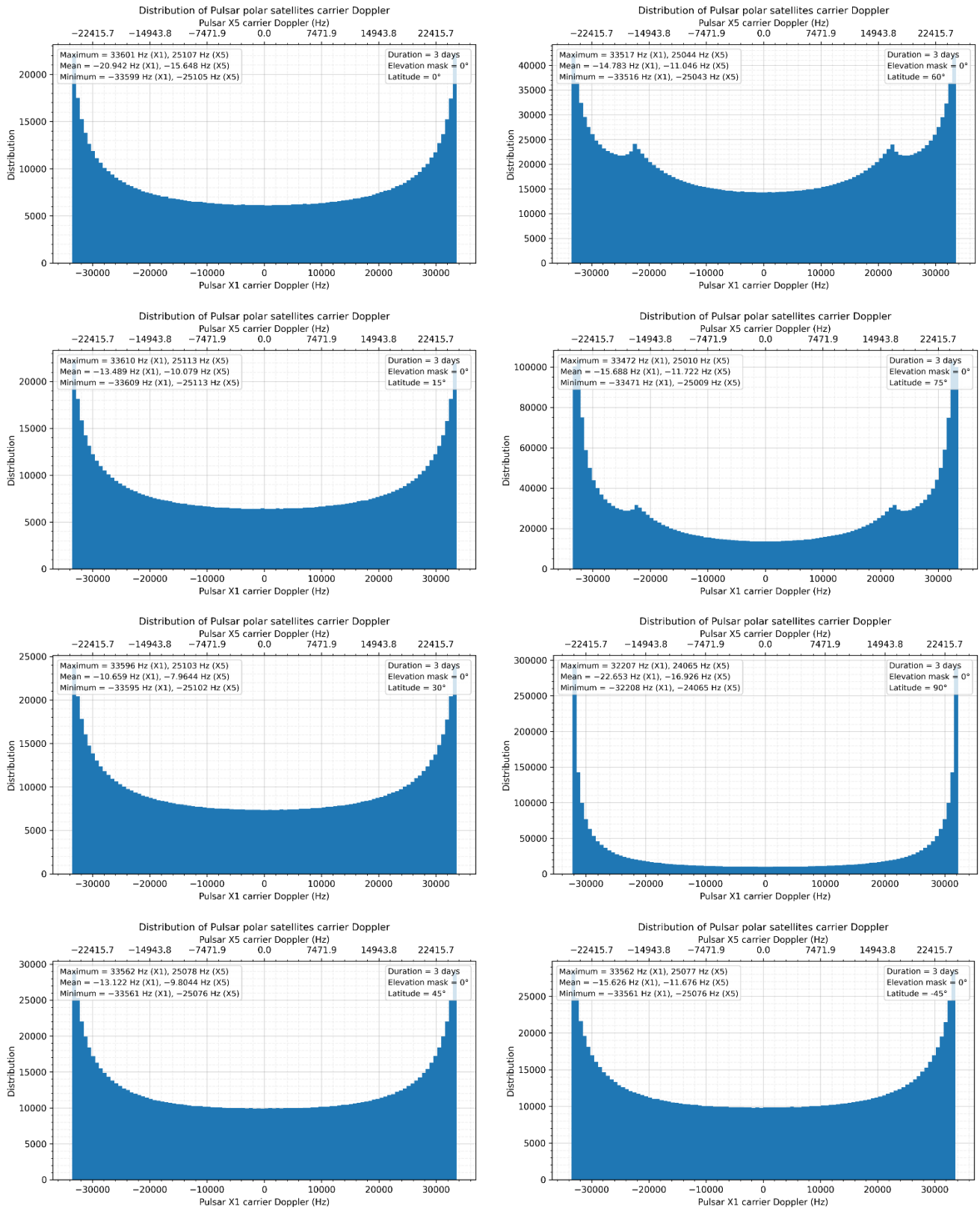

**Figure B.5** Carrier Doppler distribution of Pulsar polar satellites for different latitudes

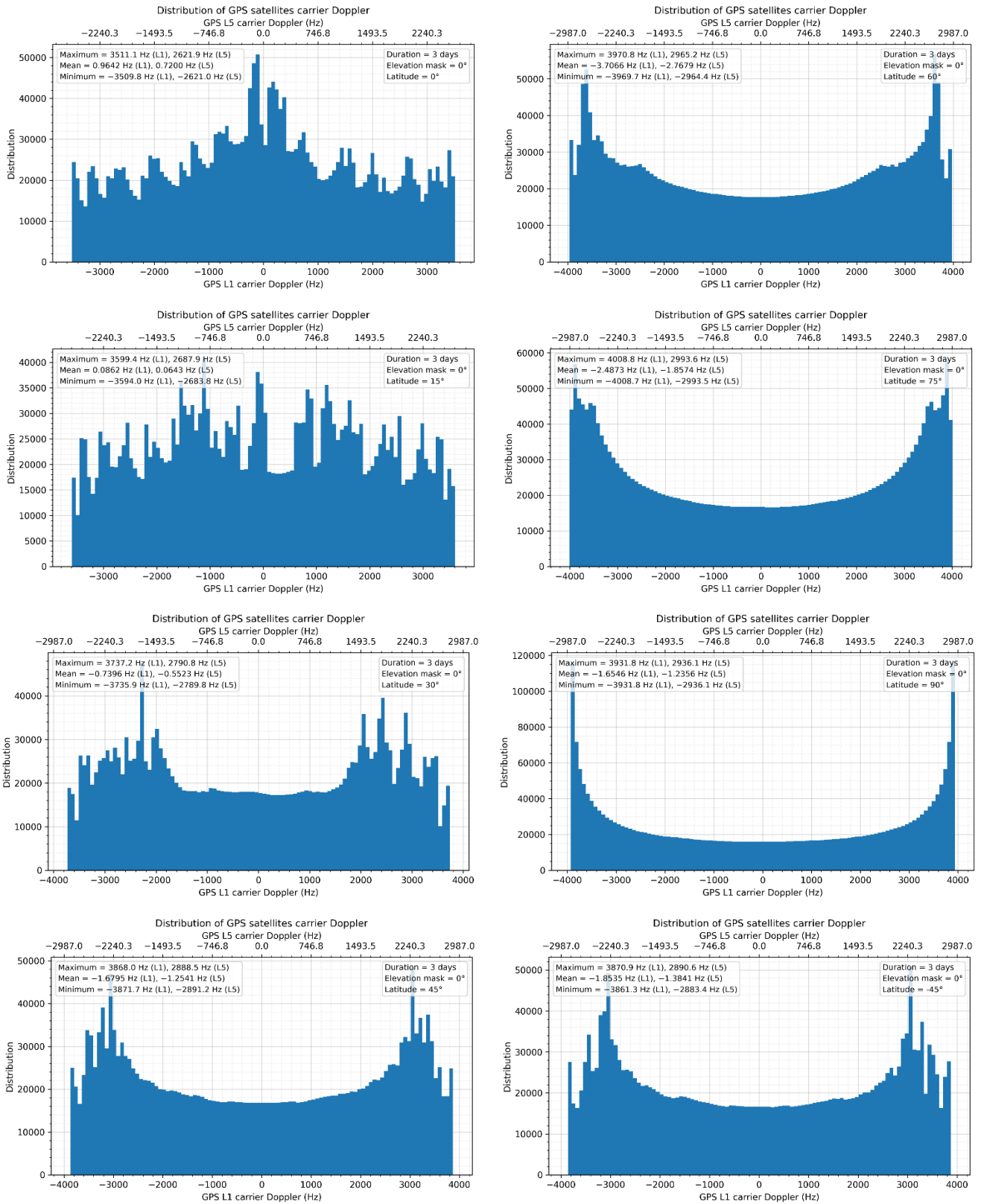

**Figure B.6** Carrier Doppler distribution of GPS satellites for different latitudes

## B.3. Carrier Doppler vs Elevation

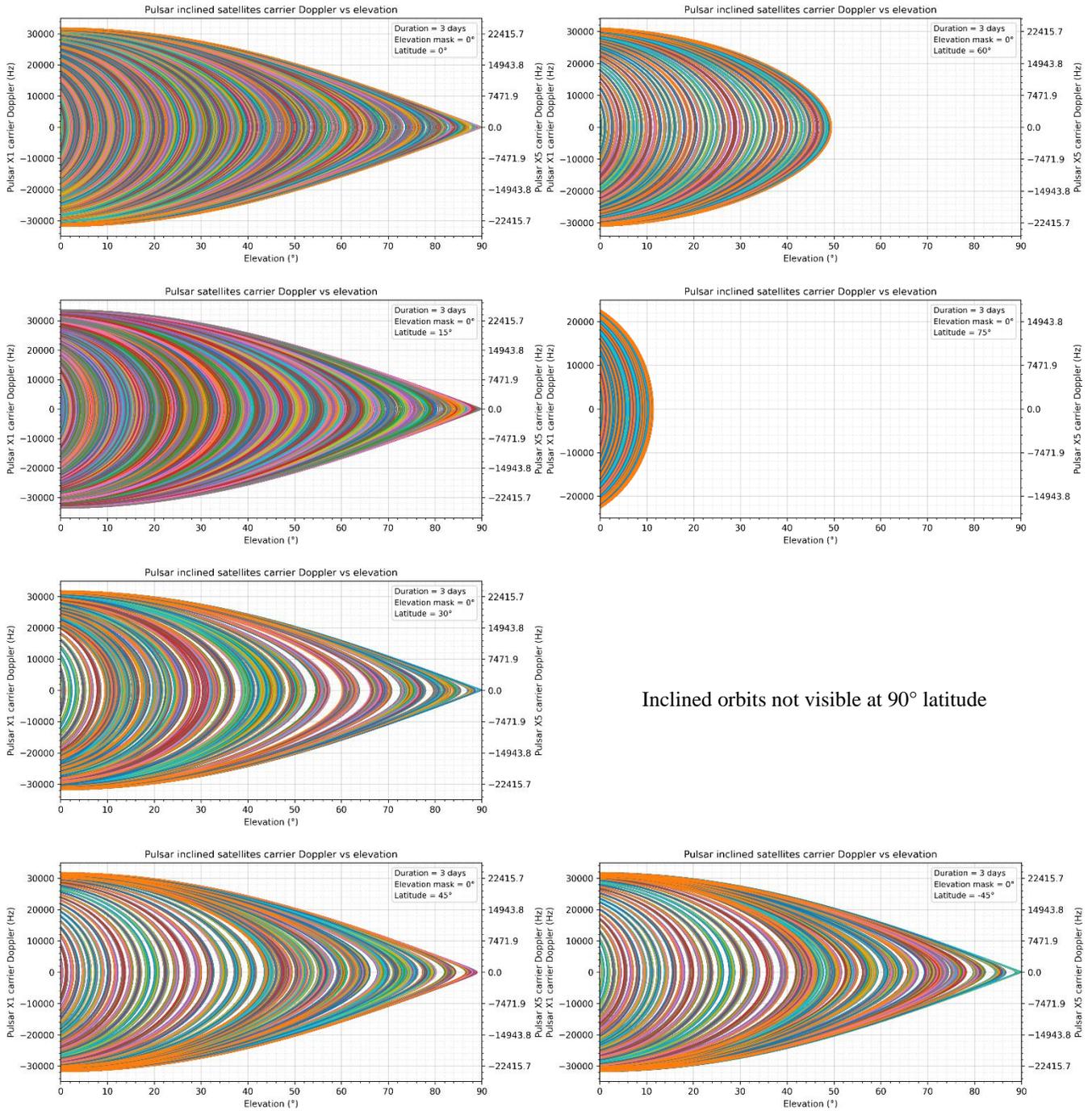

**Figure B.7** Carrier Doppler of Pulsar inclined satellites as a function of elevation for different latitudes

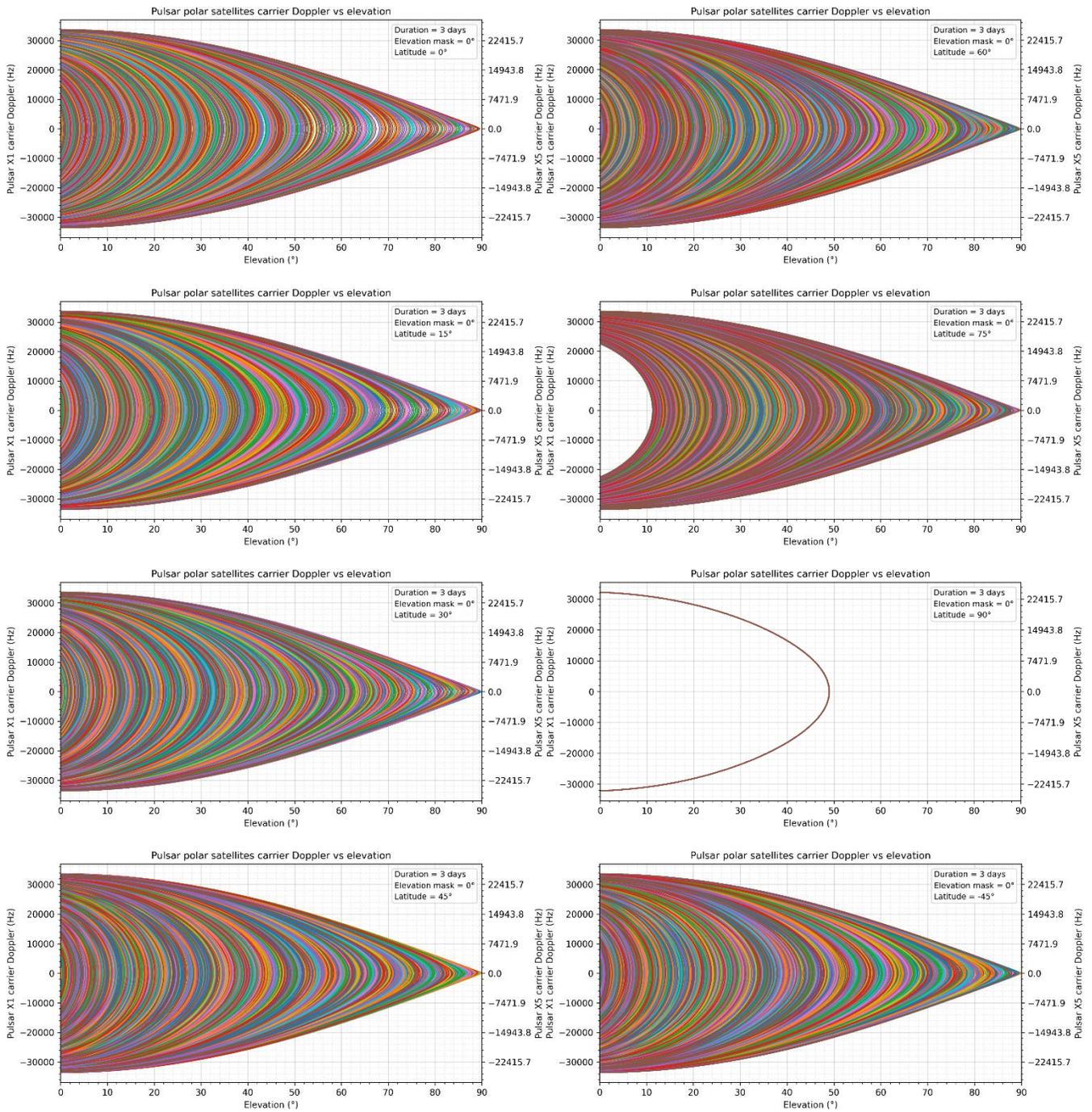

**Figure B.8** Carrier Doppler of Pulsar polar satellites as a function of elevation for different latitudes

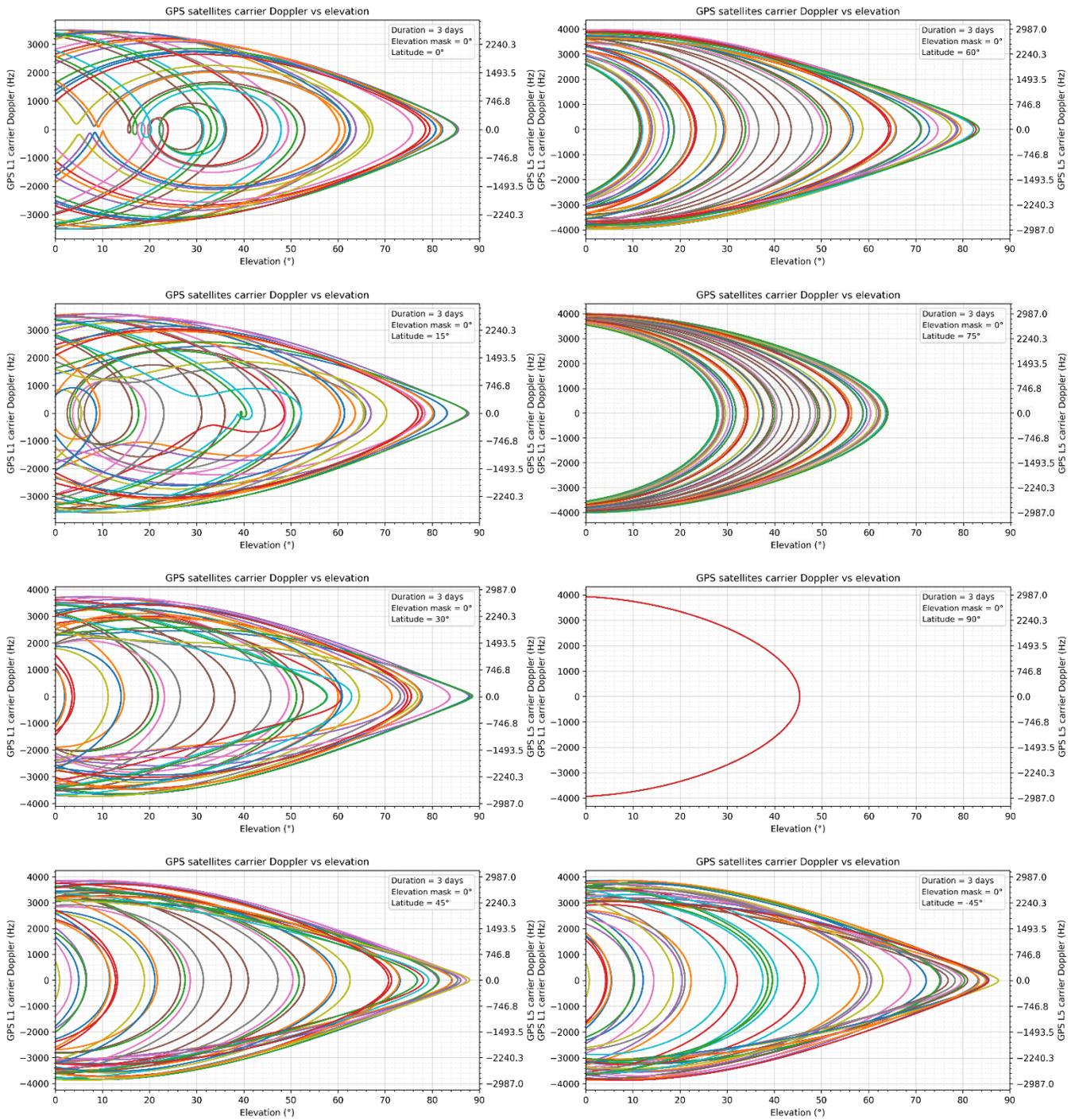

**Figure B.9** Carrier Doppler of GPS satellites as a function of elevation for different latitudes

**B.4. Carrier Doppler Rate Over Time**

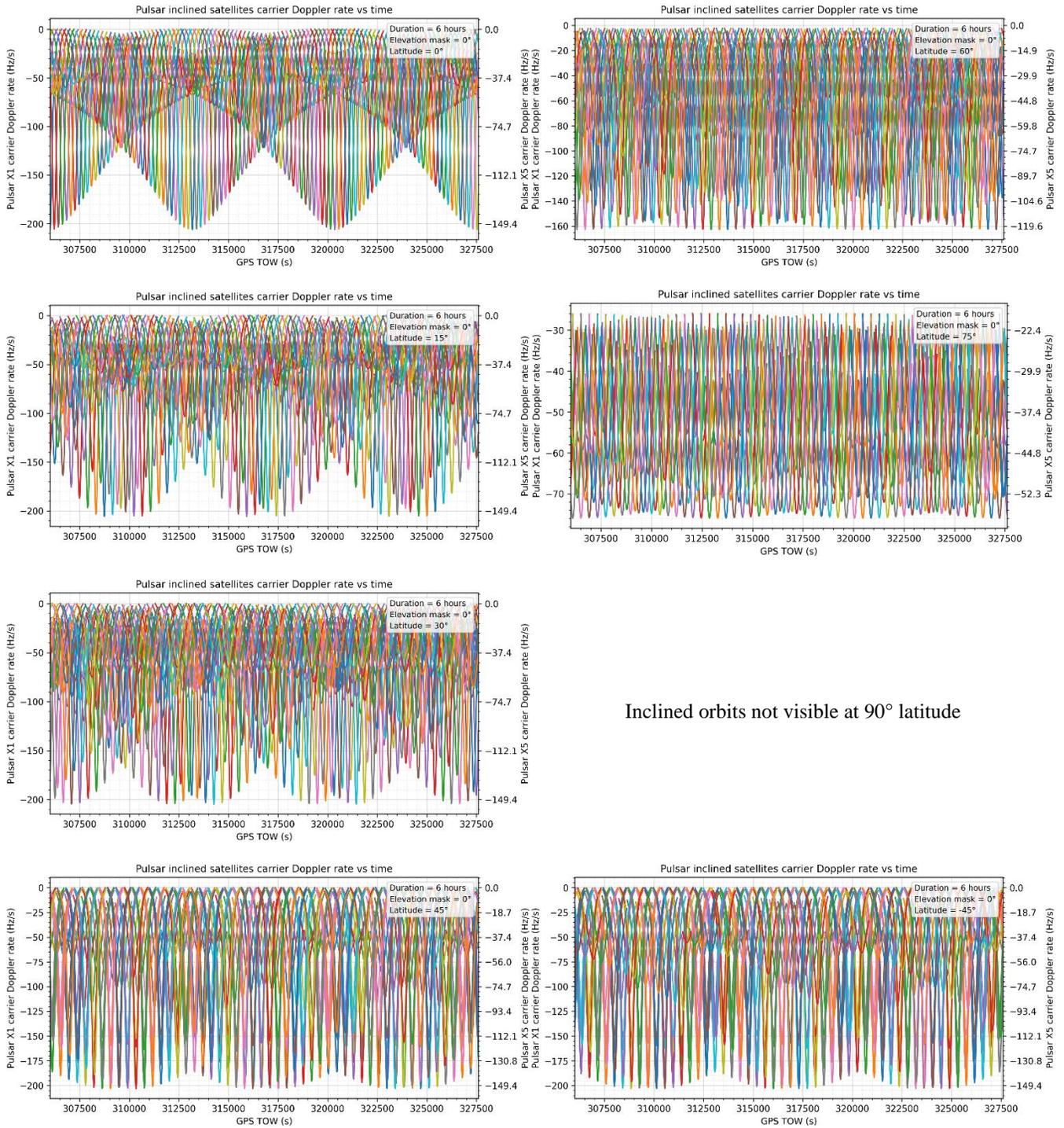

**Figure B.10** Carrier Doppler rate over time of Pulsar inclined satellites for different latitudes

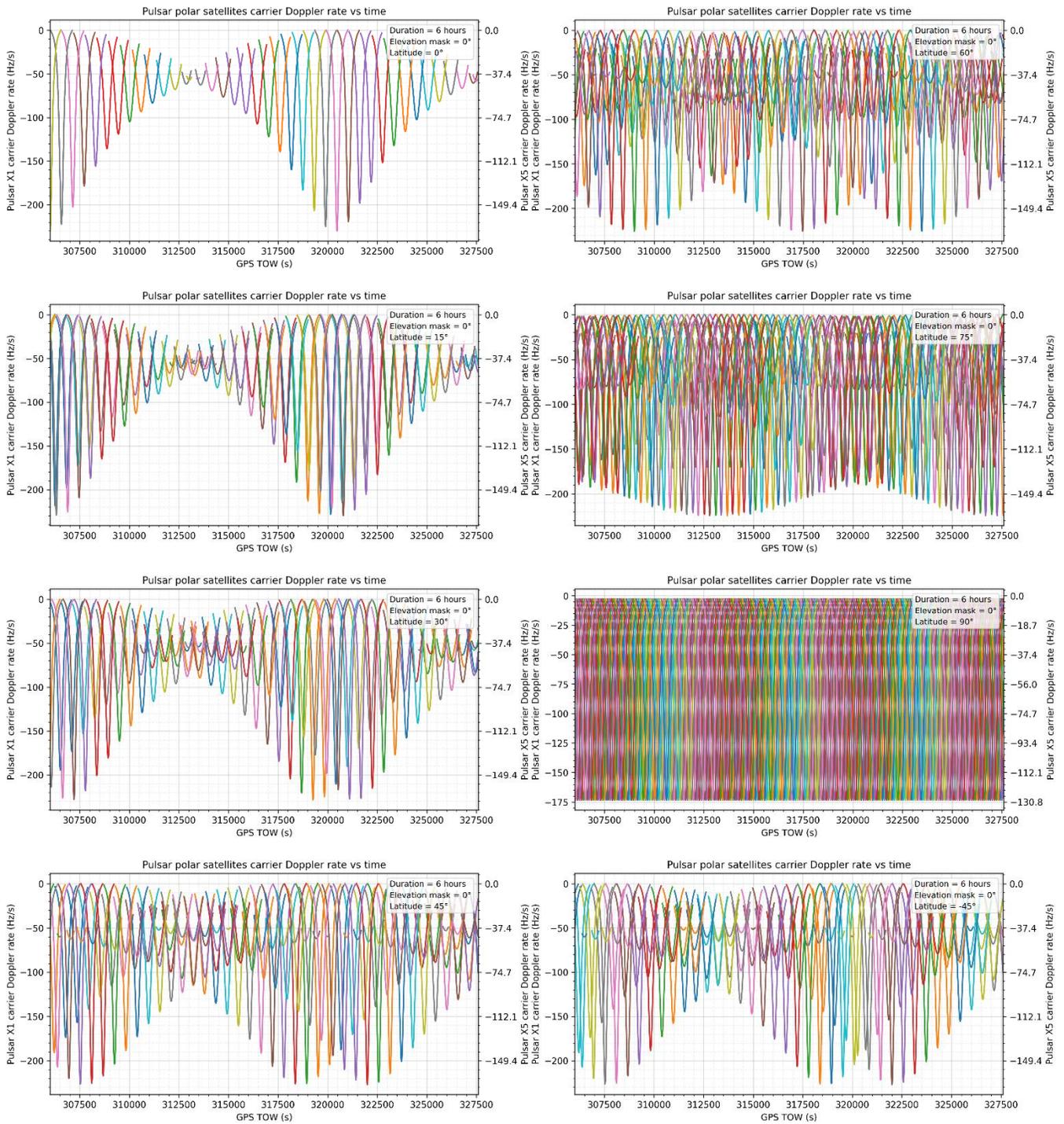

**Figure B.11** Carrier Doppler rate over time of Pulsar polar satellites for different latitudes

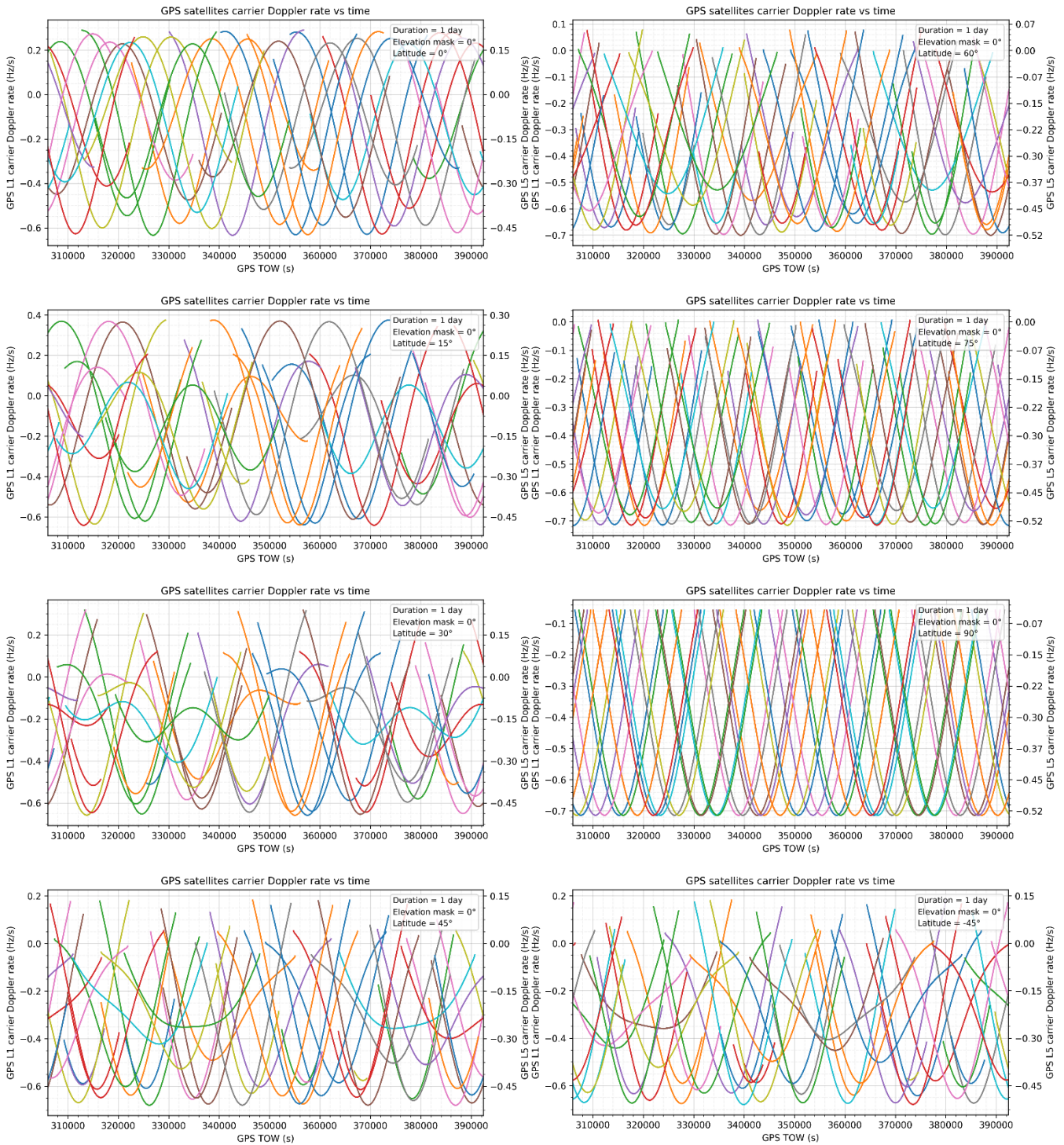

**Figure B.12** Carrier Doppler rate over time of GPS satellites for different latitudes

## B.5. Carrier Doppler Rate Distribution

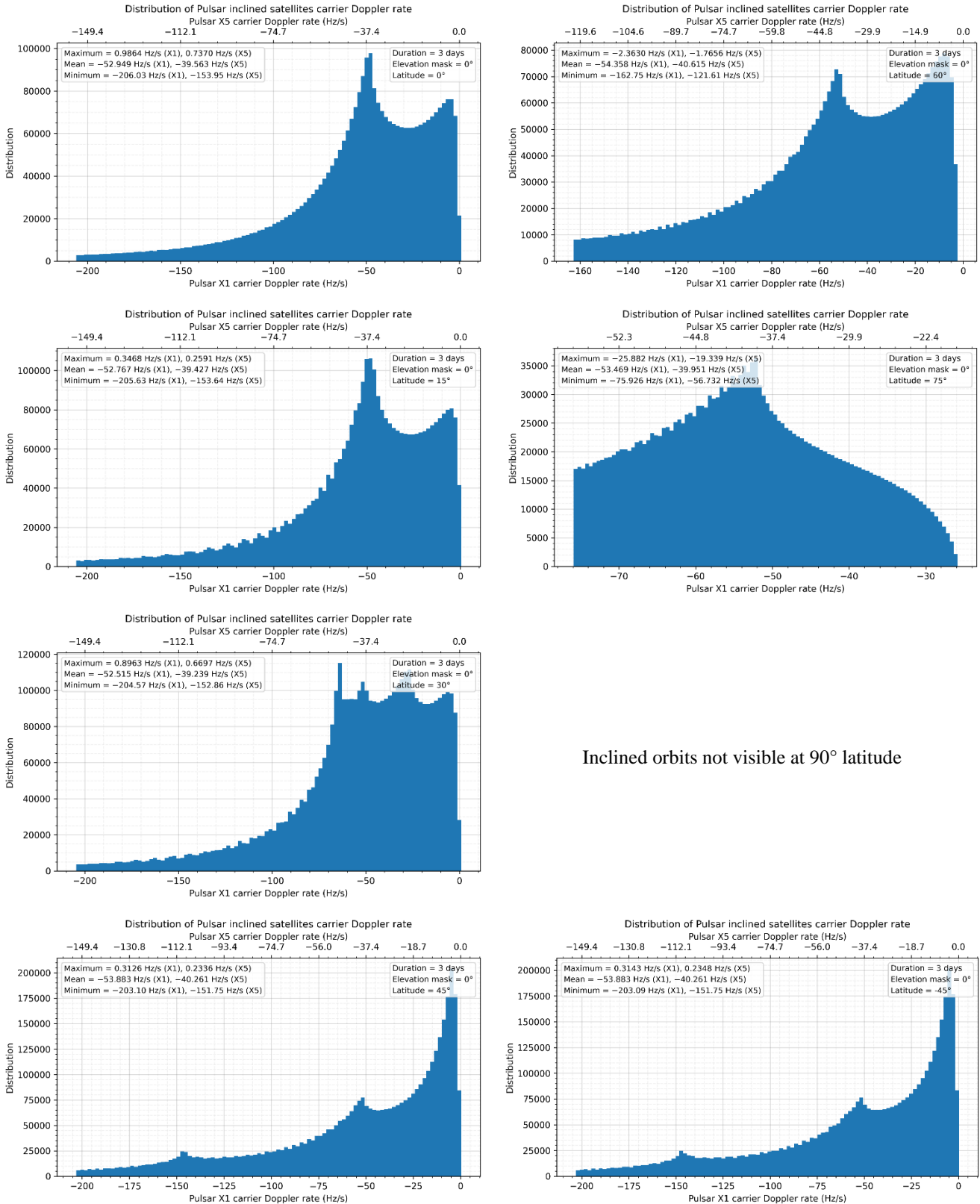

**Figure B.13** Carrier Doppler rate distribution of Pulsar inclined satellites for different latitudes

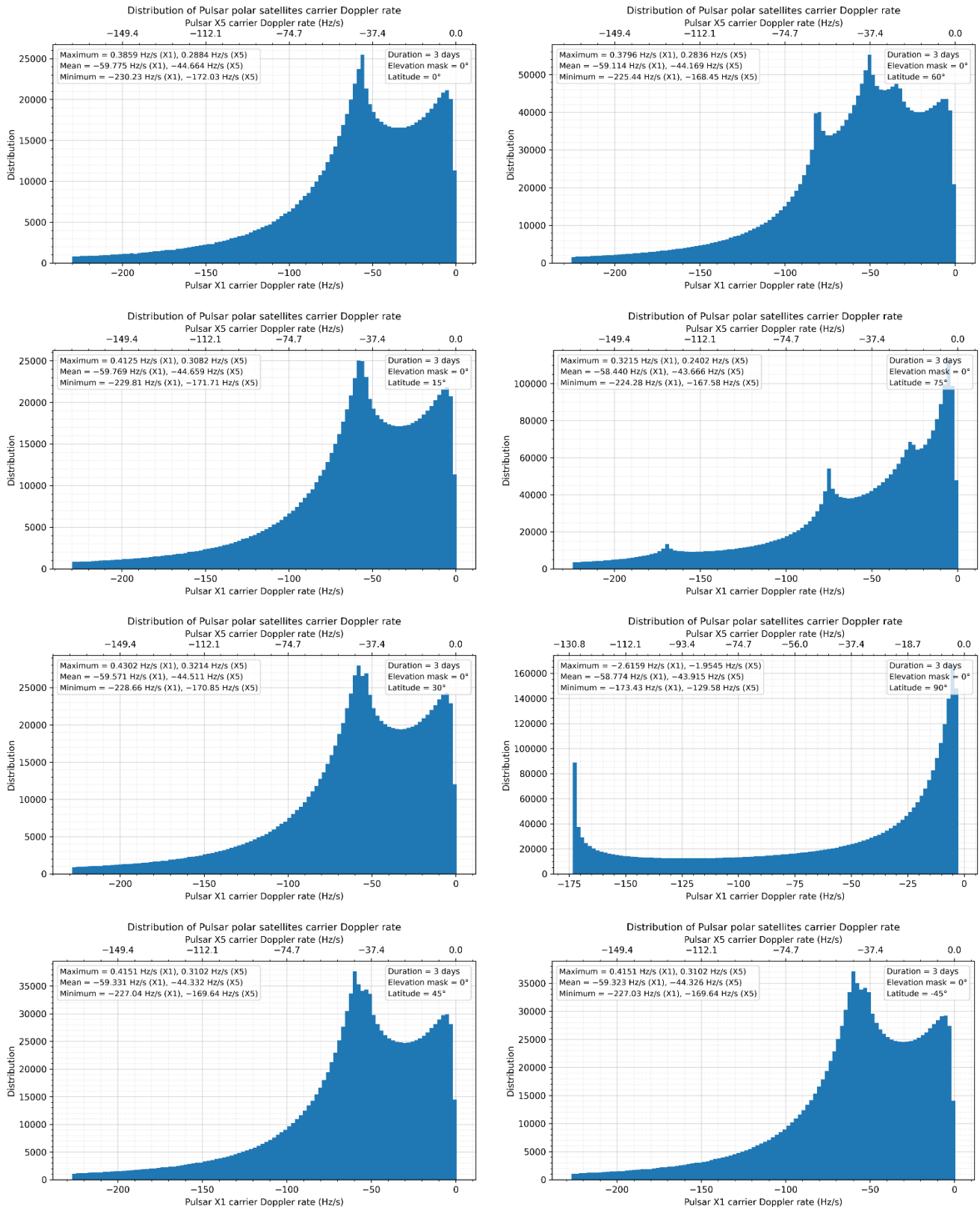

**Figure B.14** Carrier Doppler rate distribution of Pulsar polar satellites for different latitudes

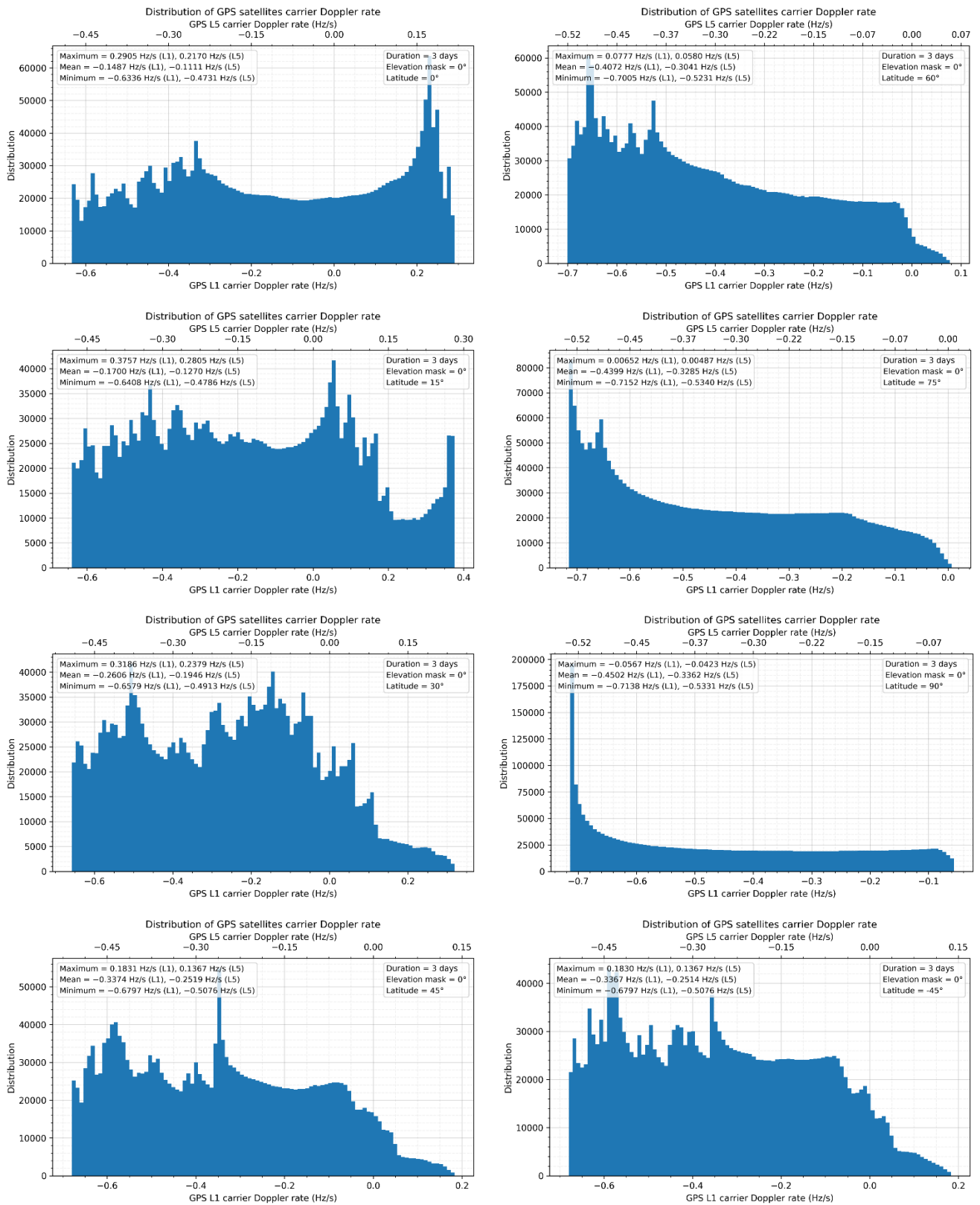

**Figure B.15** Carrier Doppler rate distribution of GPS satellites for different latitudes

## B.6. Carrier Doppler Rate vs Elevation

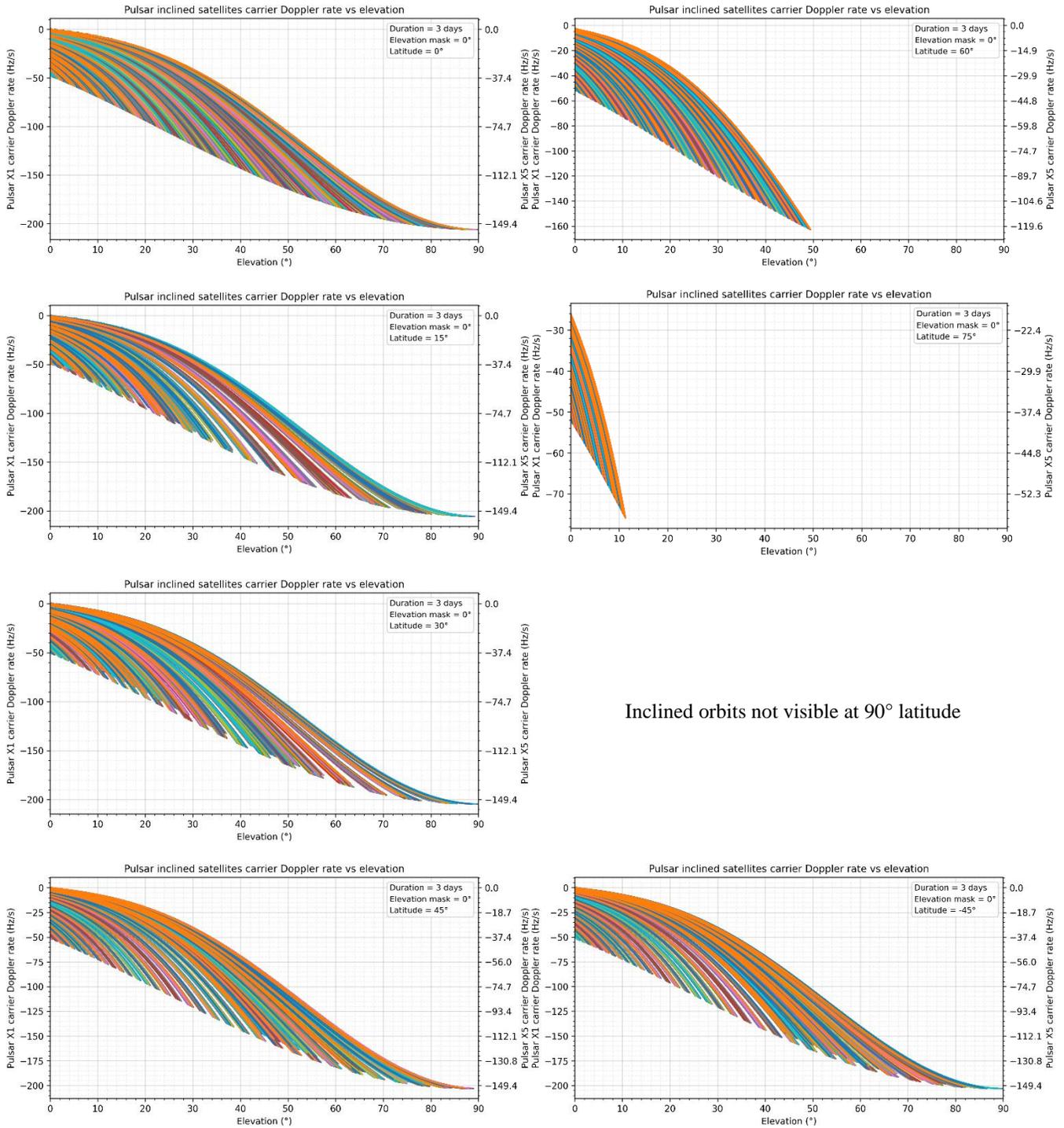

**Figure B.16** Carrier Doppler rate of Pulsar inclined satellites as a function of elevation for different latitudes

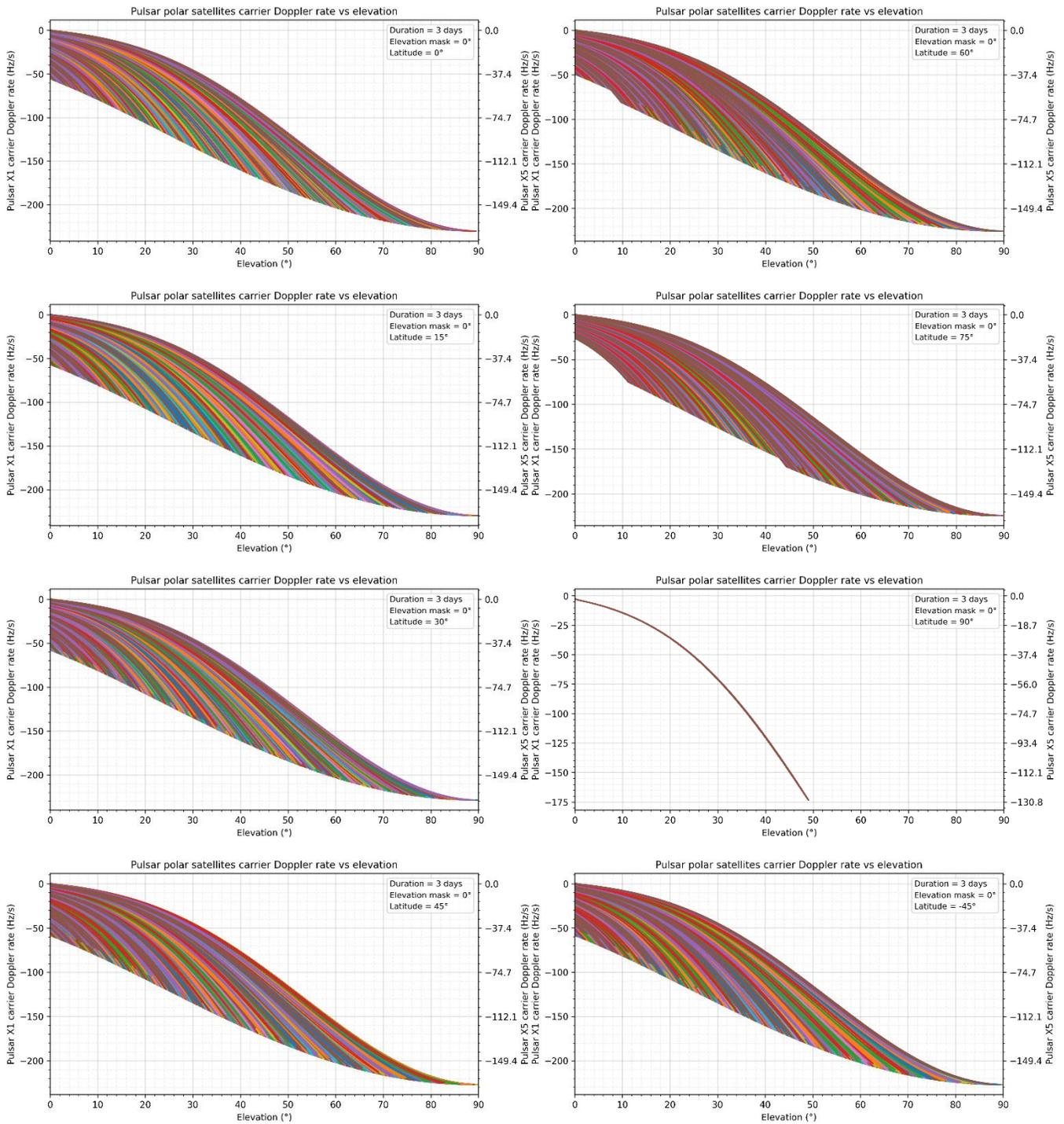

**Figure B.17** Carrier Doppler rate of Pulsar polar satellites as a function of elevation for different latitudes

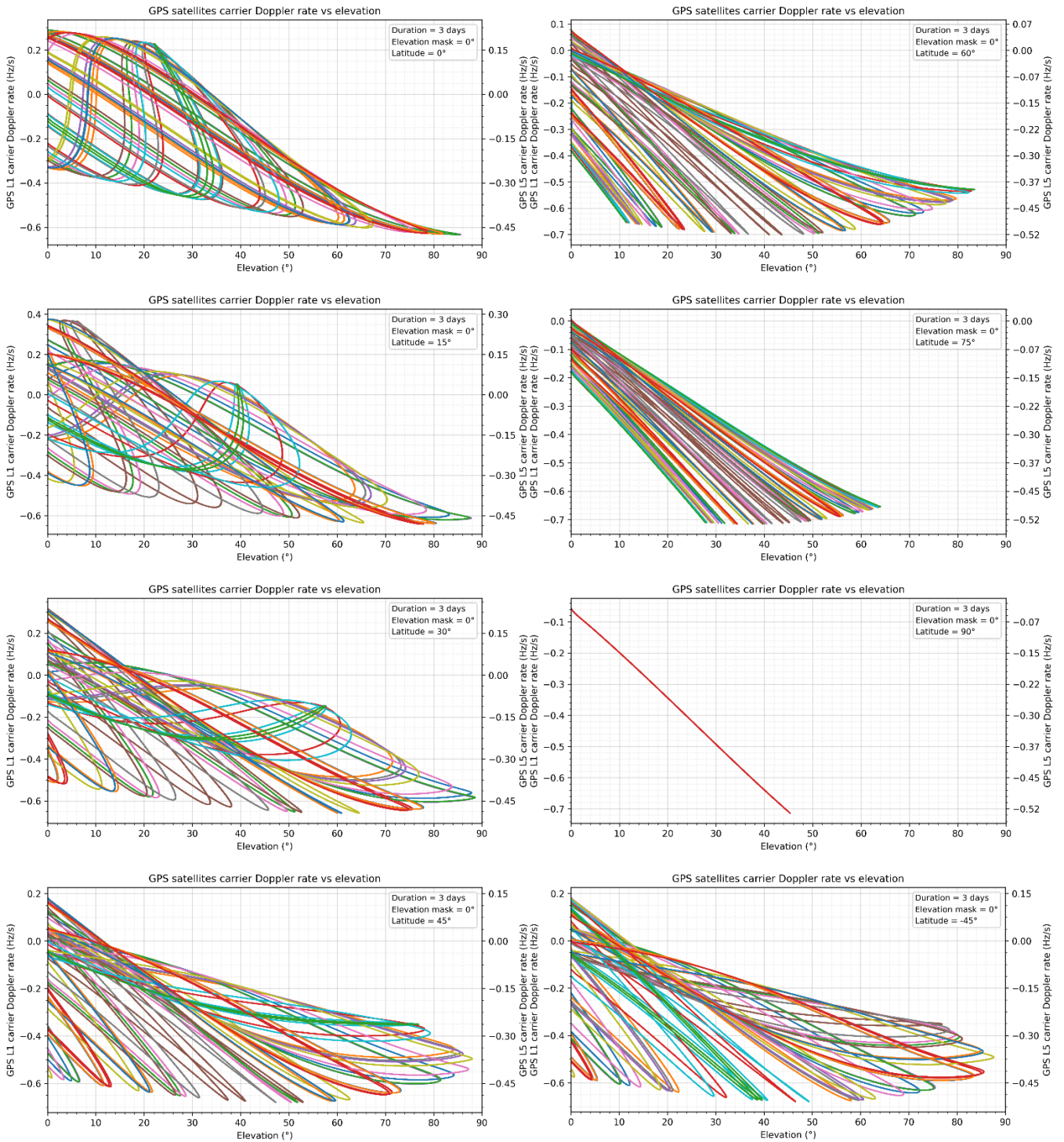

**Figure B.18** Carrier Doppler rate of GPS satellites as a function of elevation for different latitudes

## B.7. Carrier Doppler Rate vs Carrier Doppler

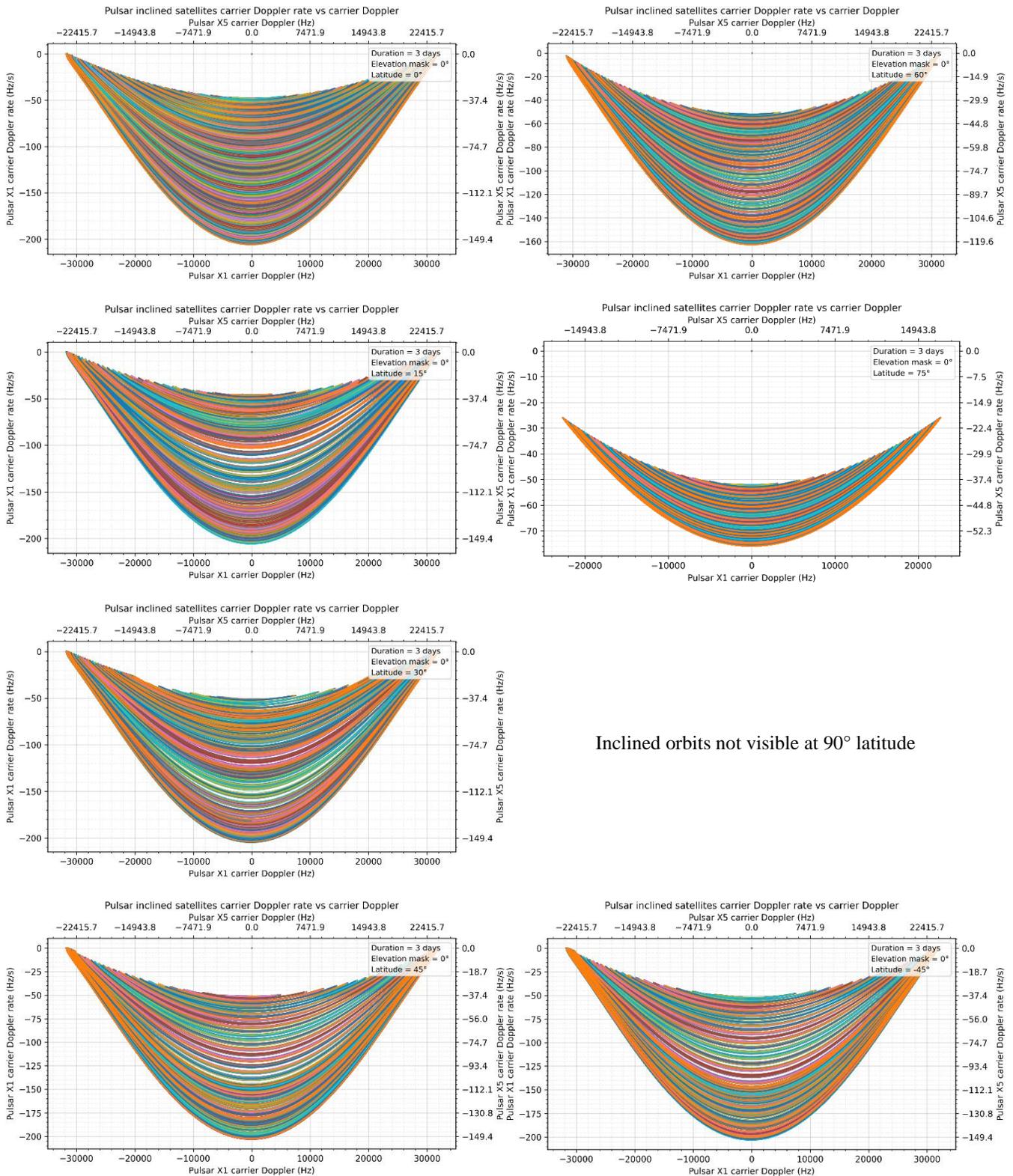

**Figure B.19** Carrier Doppler rate as a function of carrier Doppler for Pulsar inclined satellites for different latitudes

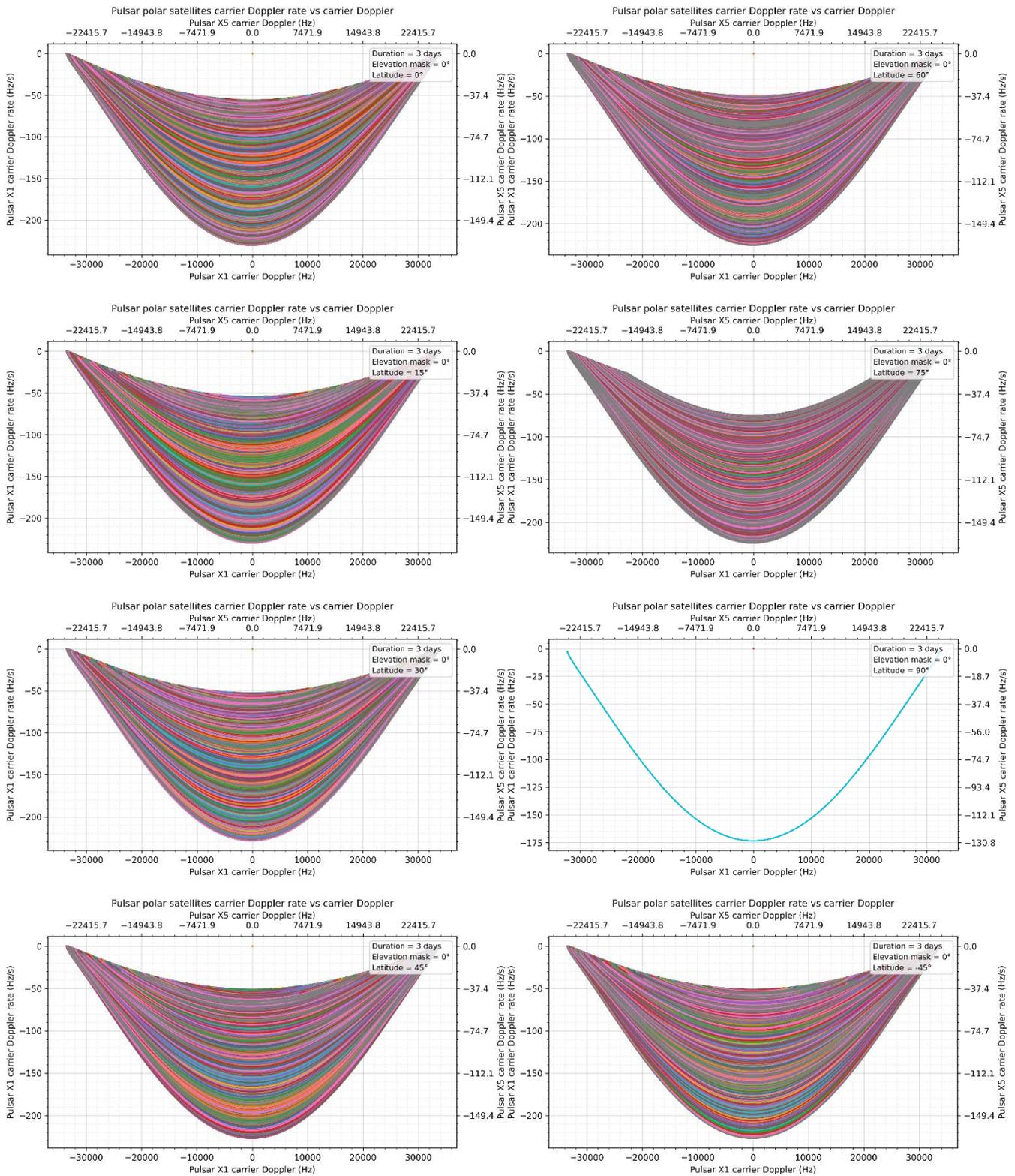

**Figure B.20** Carrier Doppler rate as a function of carrier Doppler for Pulsar polar satellites for different latitudes

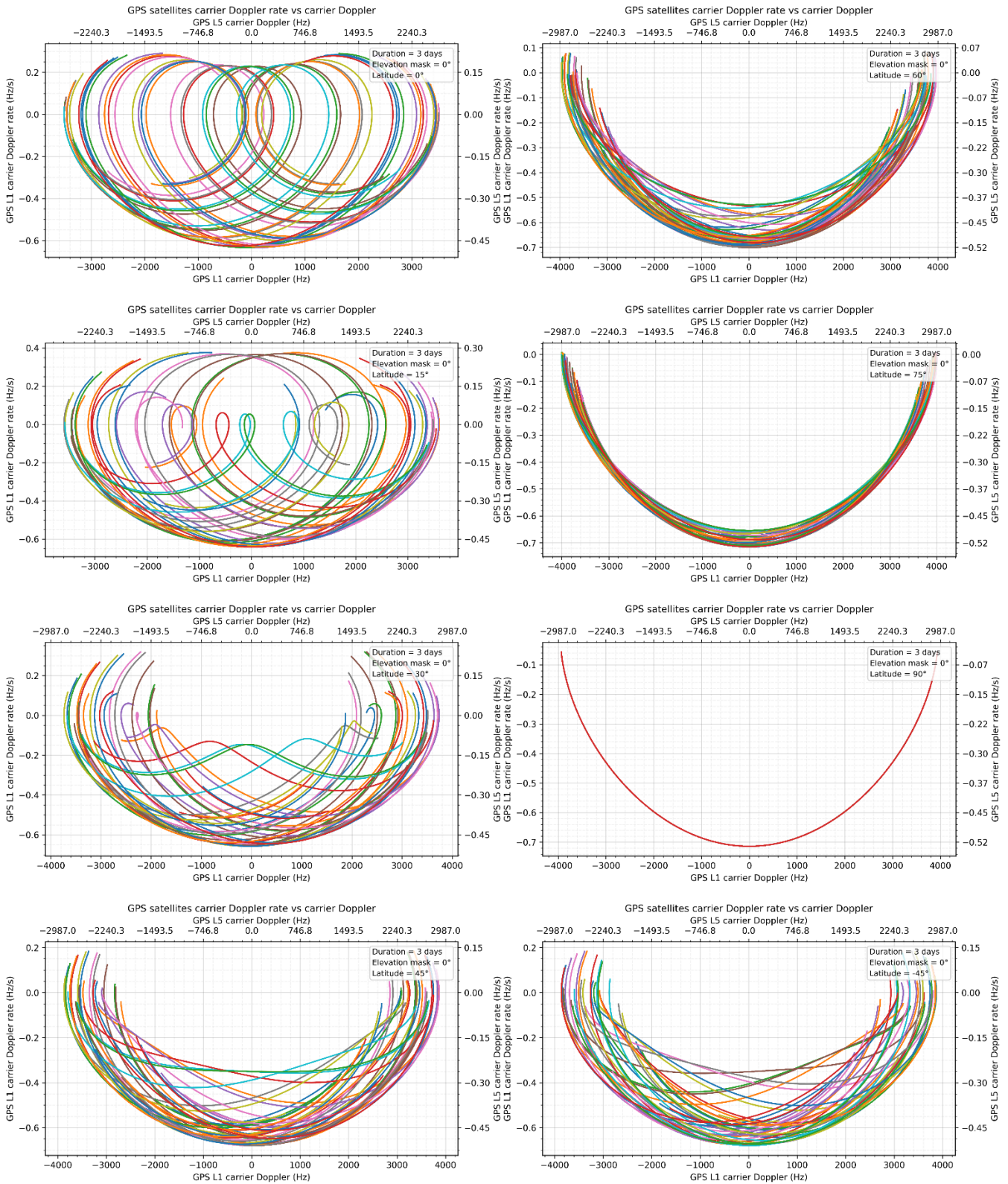

**Figure B.21** Carrier Doppler rate as a function of carrier Doppler for GPS satellites for different latitudes

## B.8. Range Difference Distribution

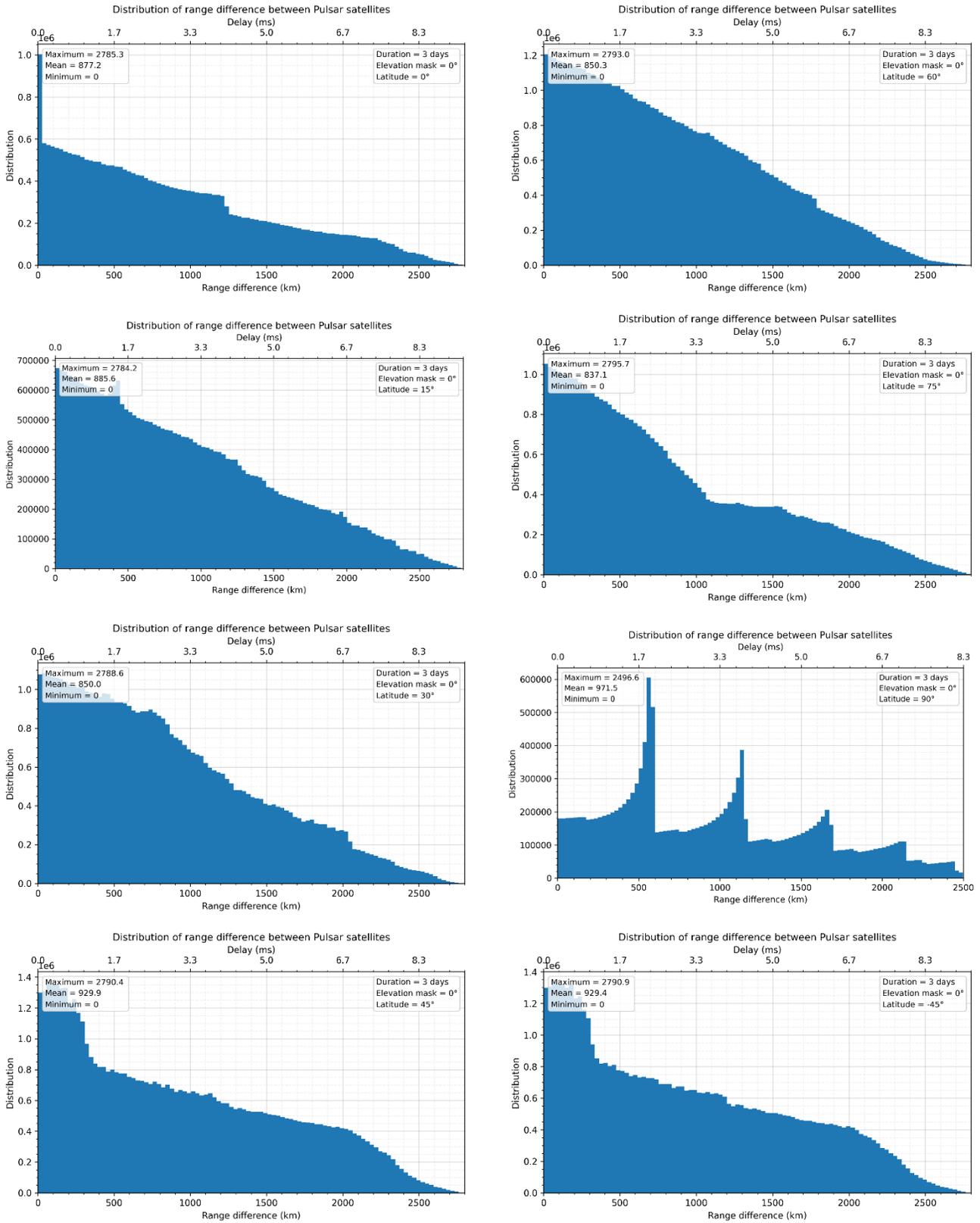

**Figure B.22** Distribution of range difference between Pulsar satellites for different latitudes

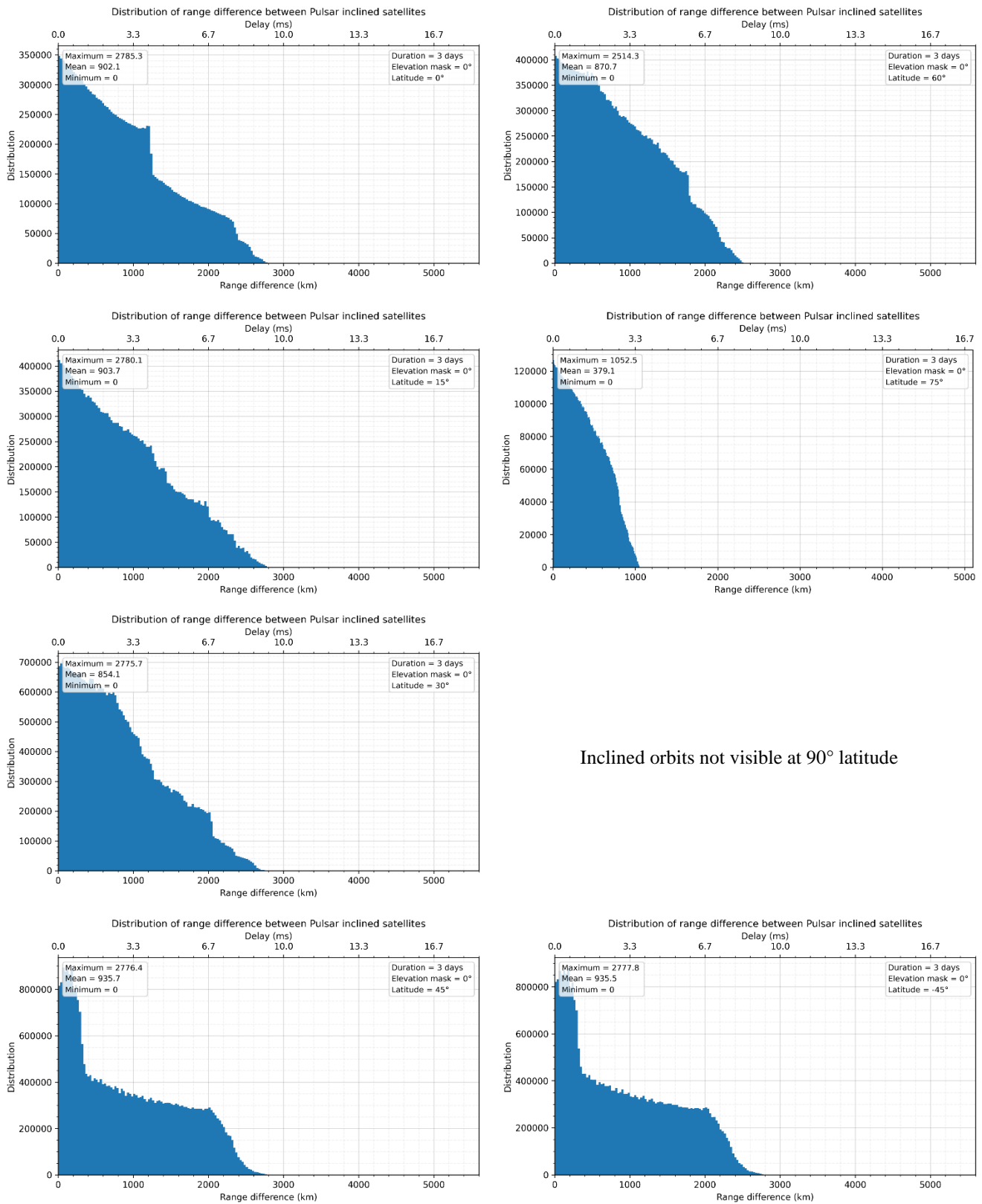

**Figure B.23** Distribution of range difference between Pulsar inclined satellites for different latitudes

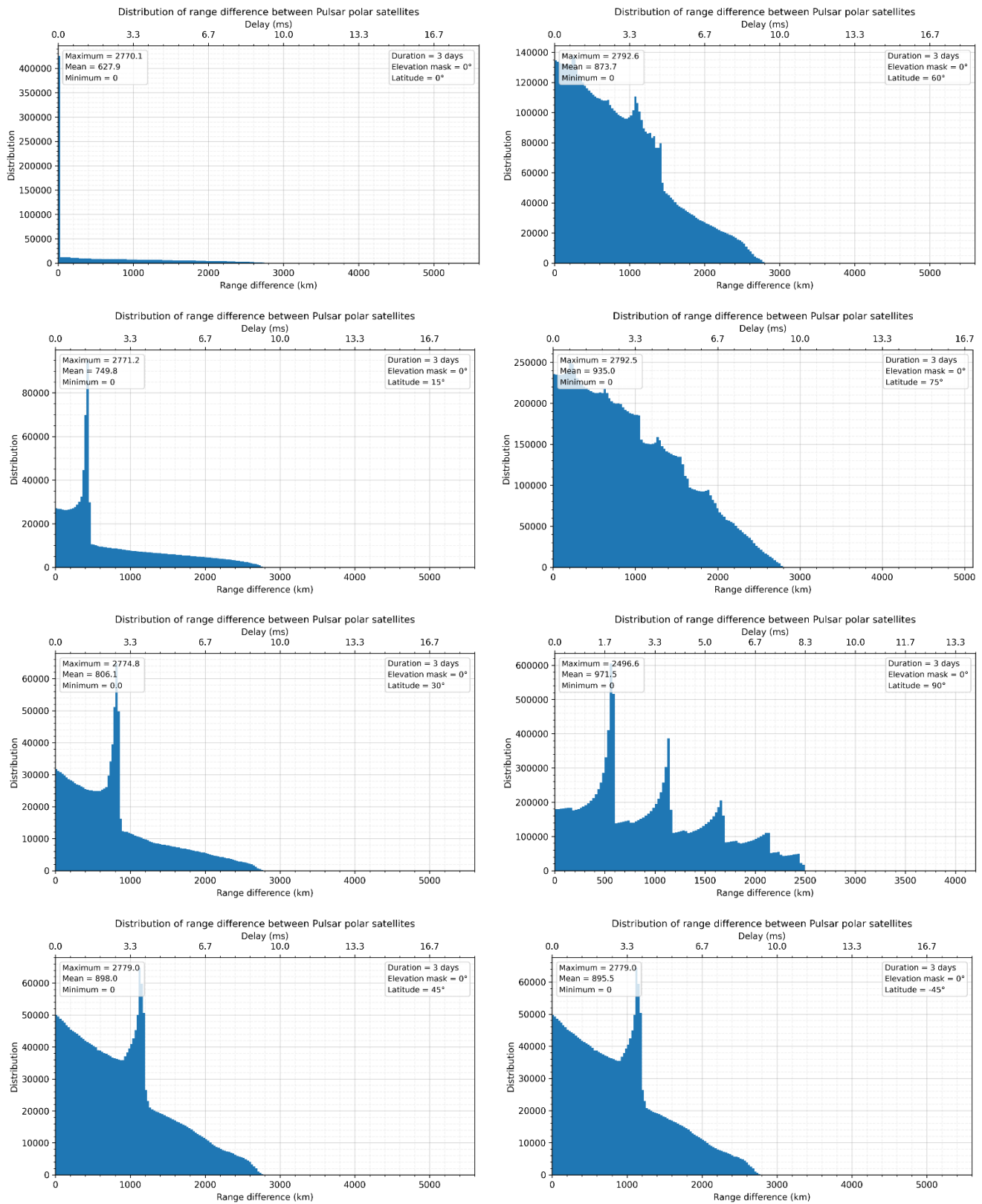

**Figure B.24** Distribution of range difference between Pulsar polar satellites for different latitudes

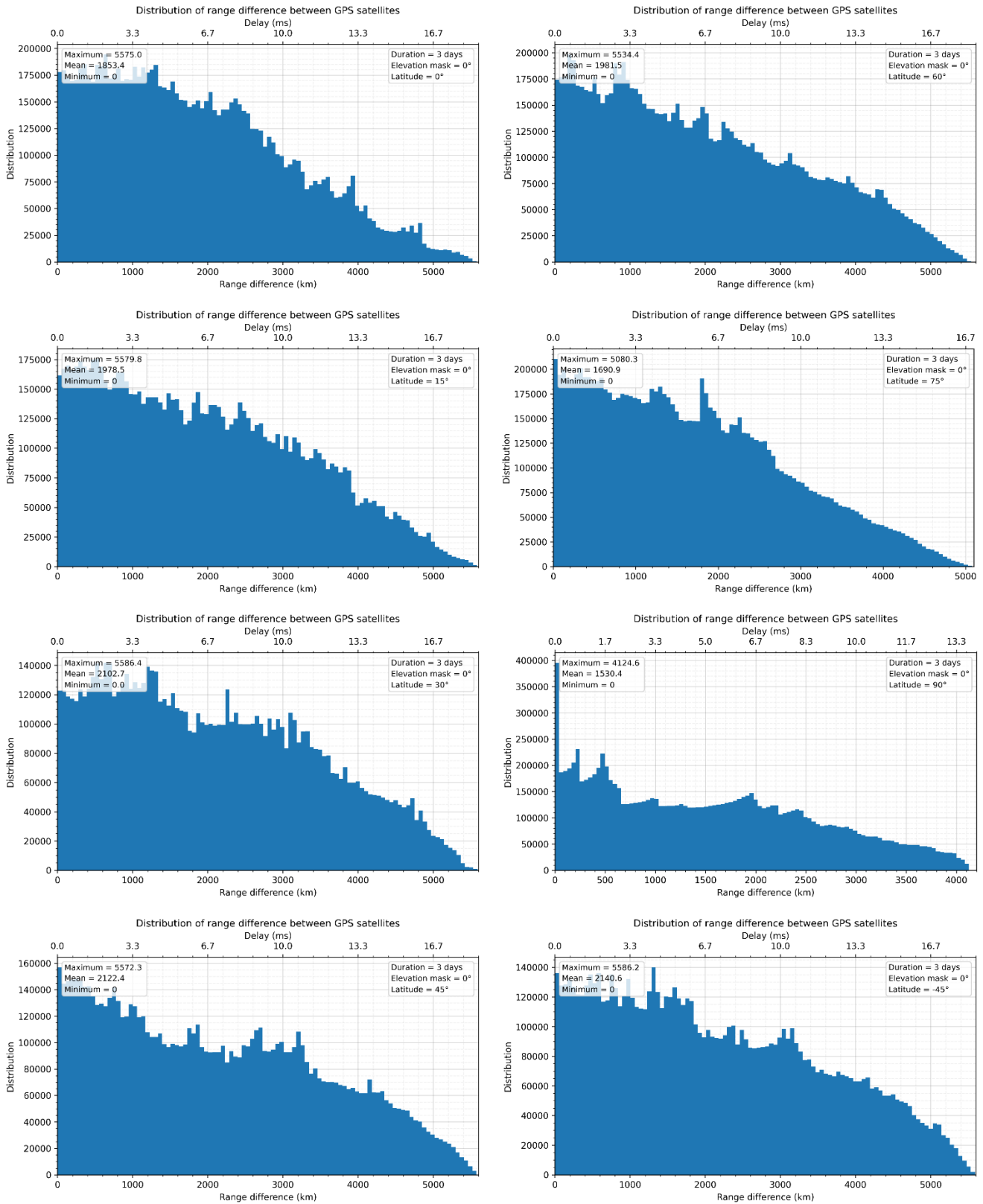

**Figure B.25** Distribution of range difference between GPS satellites for different latitudes

**B.9. Range Difference vs Elevation Mask**

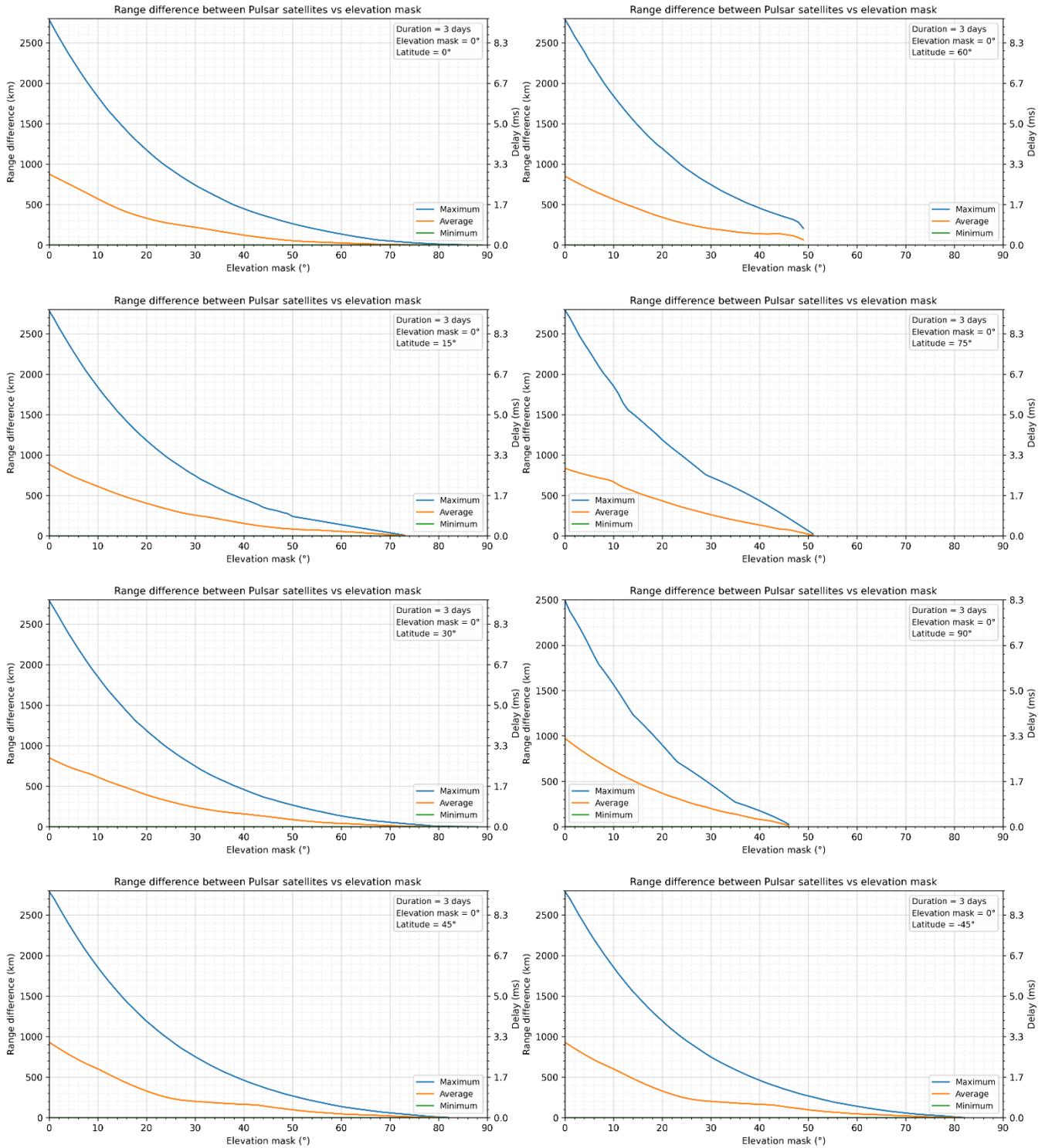

**Figure B.26** Range difference between Pulsar satellites vs elevation mask for different latitudes

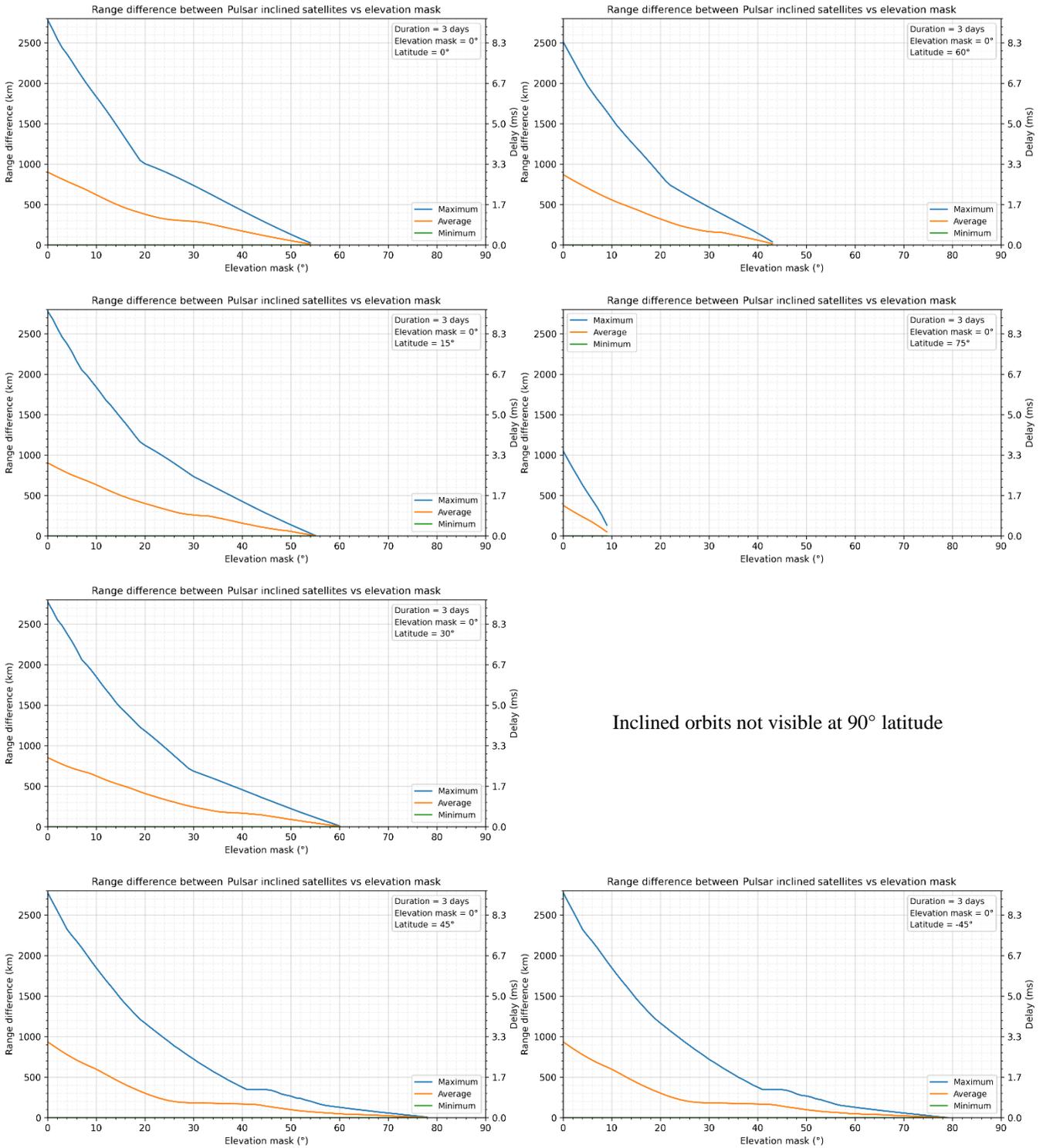

**Figure B.27** Range difference between Pulsar inclined satellites vs elevation mask for different latitudes

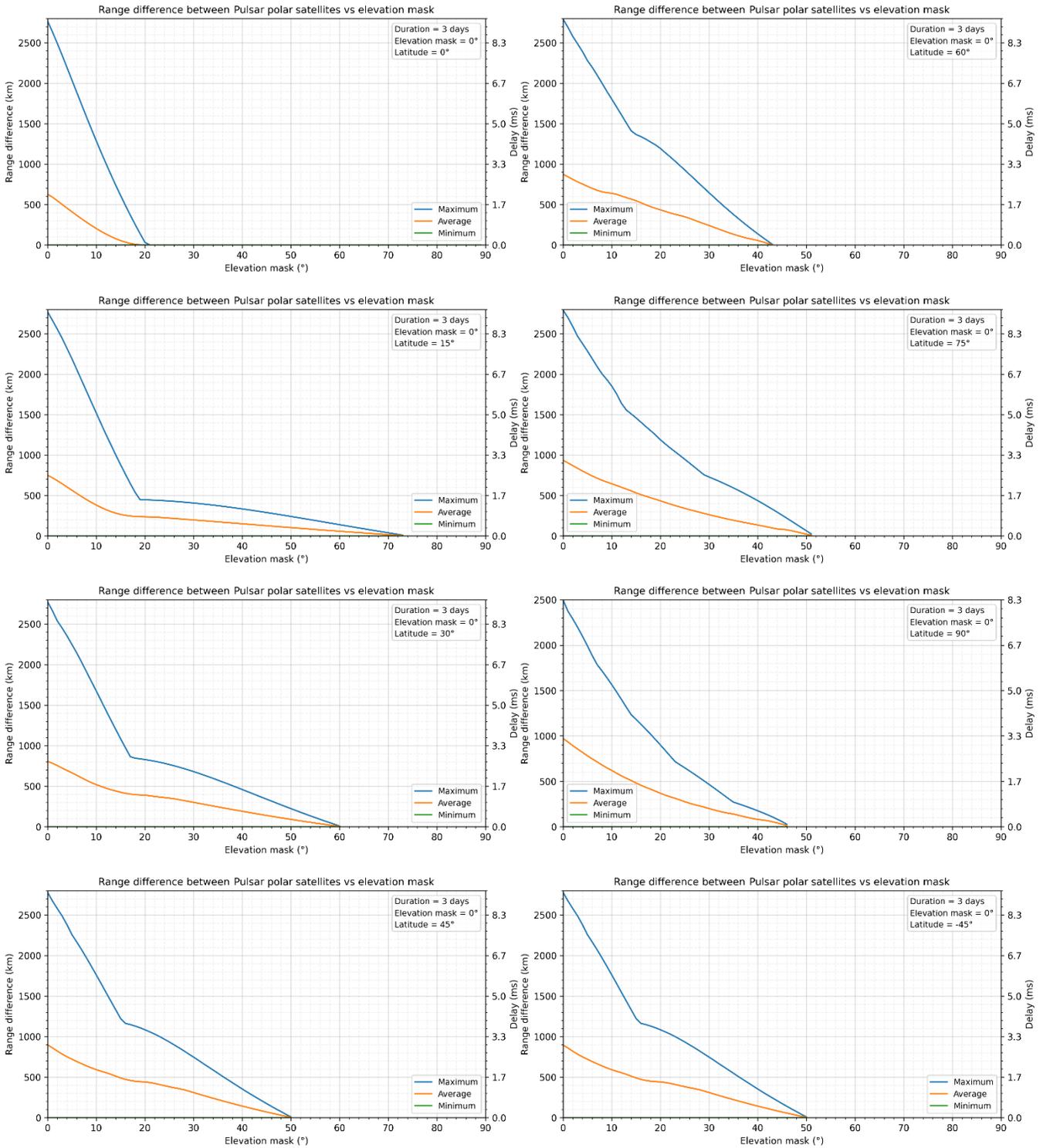

**Figure B.28** Range difference between Pulsar polar satellites vs elevation mask for different latitudes

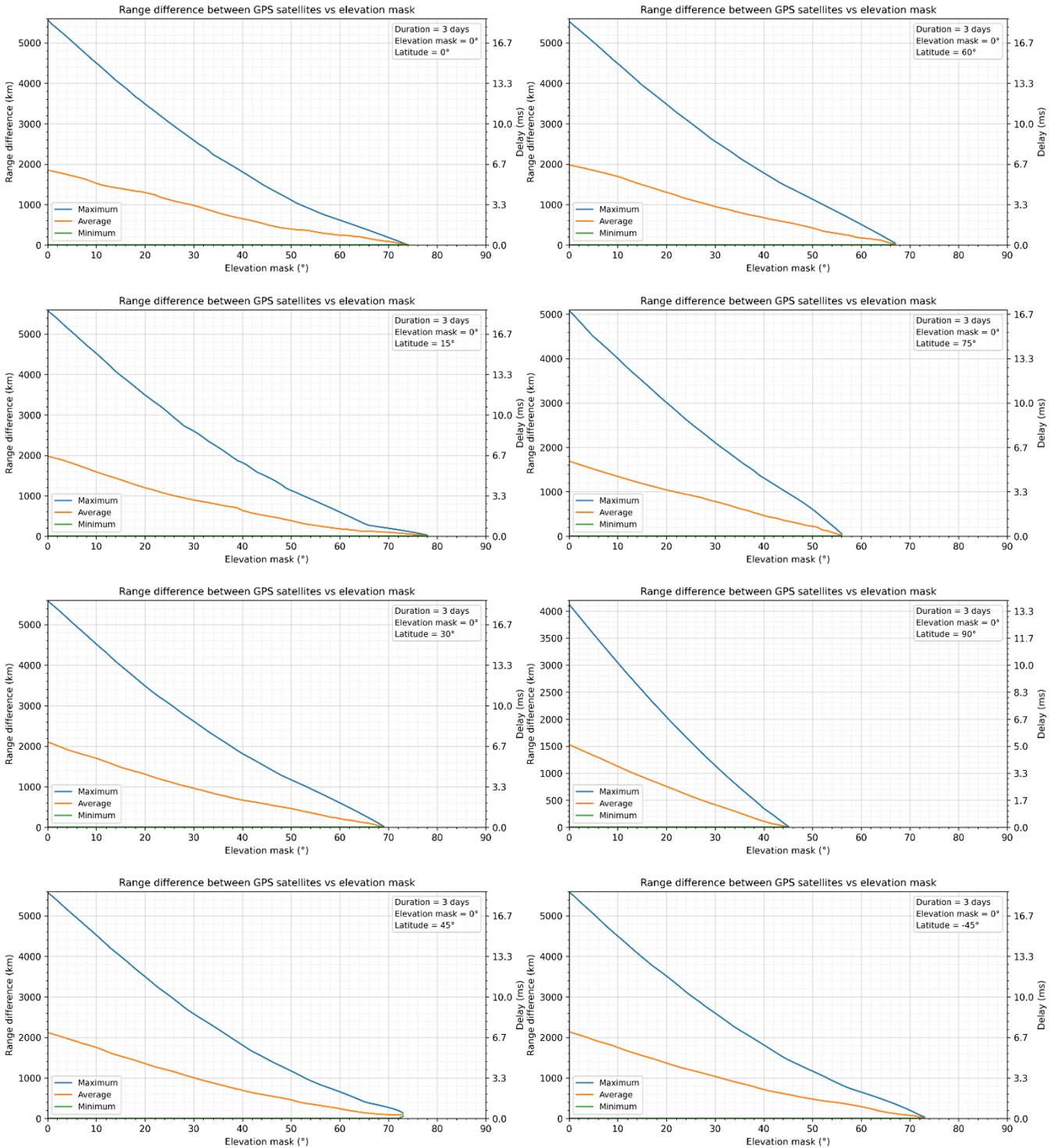

**Figure B.29** Range difference between GPS satellites vs elevation mask for different latitudes

## B.10. Range Difference (Same Orbital Plane) Distribution

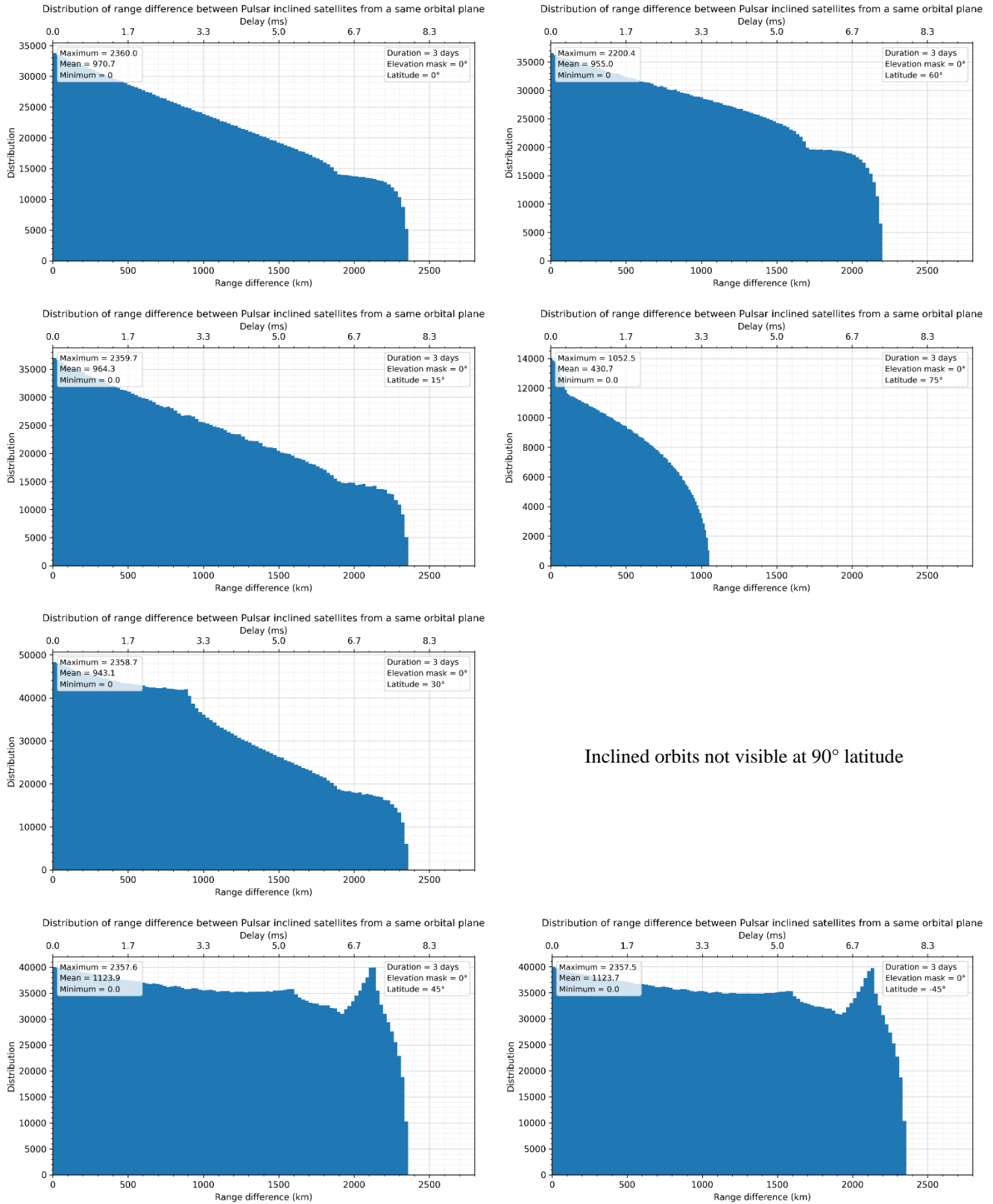

**Figure B.30** Distribution of range difference between Pulsar inclined satellites from a same orbital plane for different latitudes

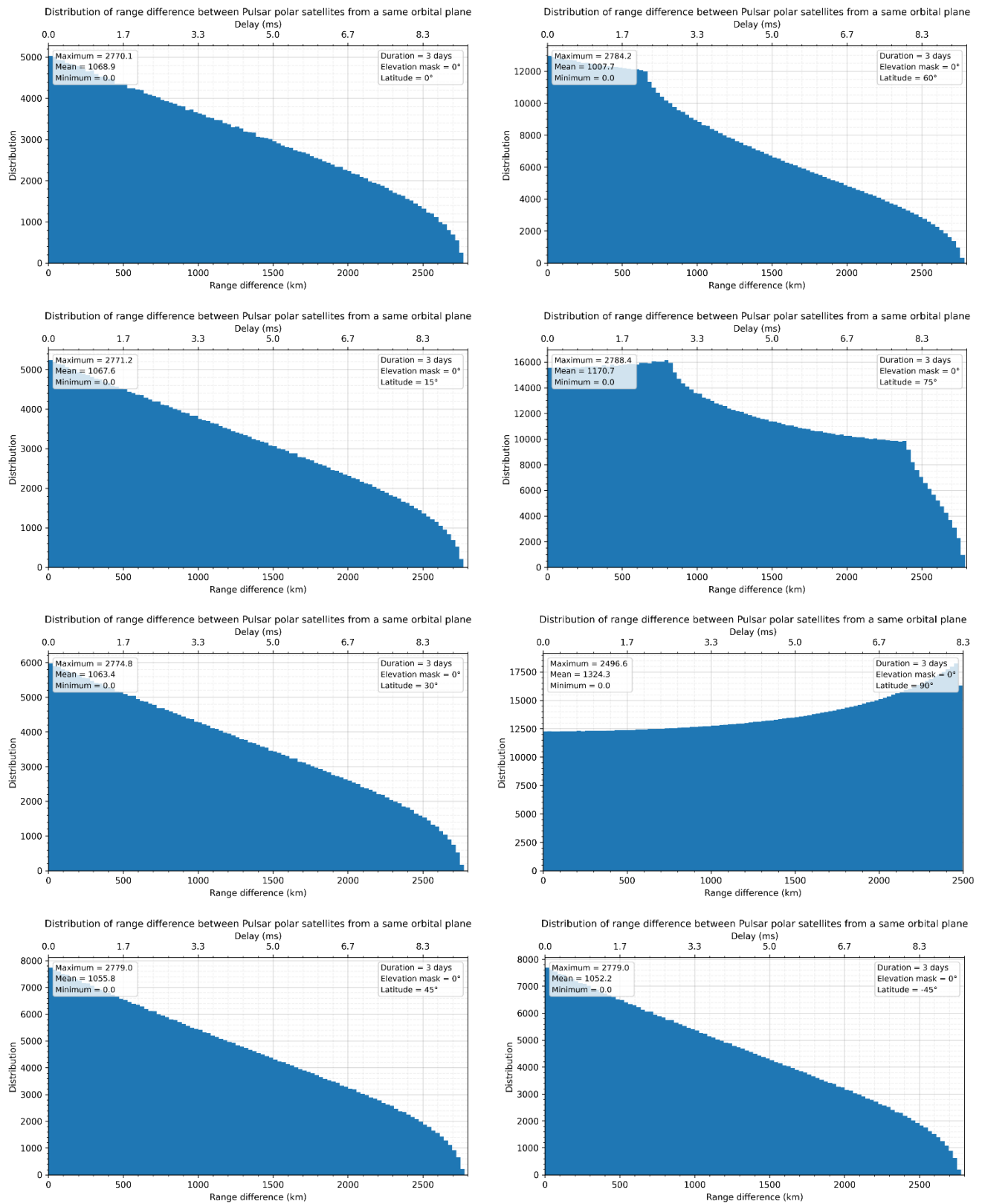

**Figure B.31** Distribution of range difference between Pulsar polar satellites from a same orbital plane for different latitudes

## B.11. Range Difference (Same Orbital Plane) vs Elevation Mask

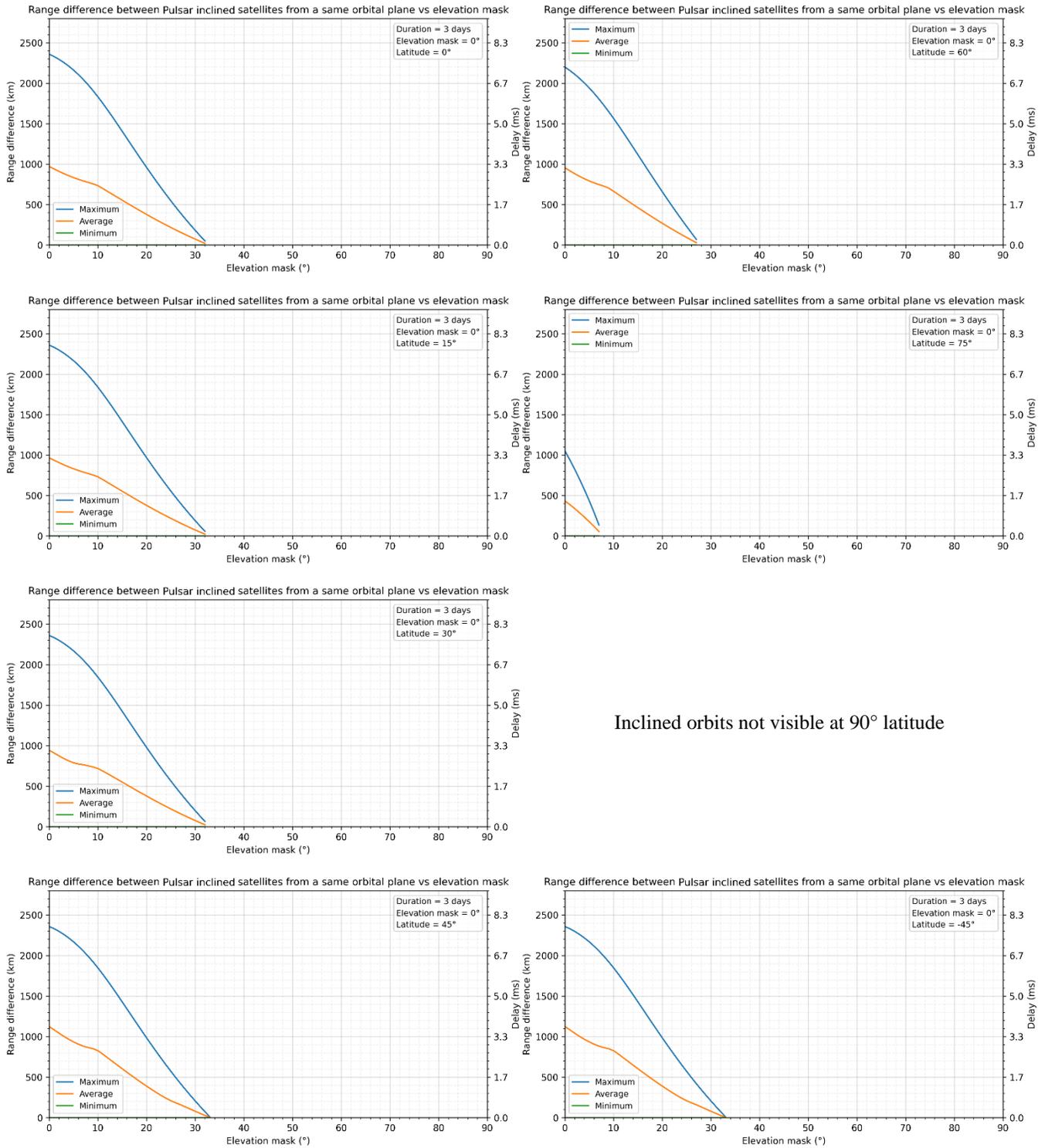

**Figure B.32** Range difference between Pulsar inclined satellites of a same orbital plane vs elevation mask for different latitudes

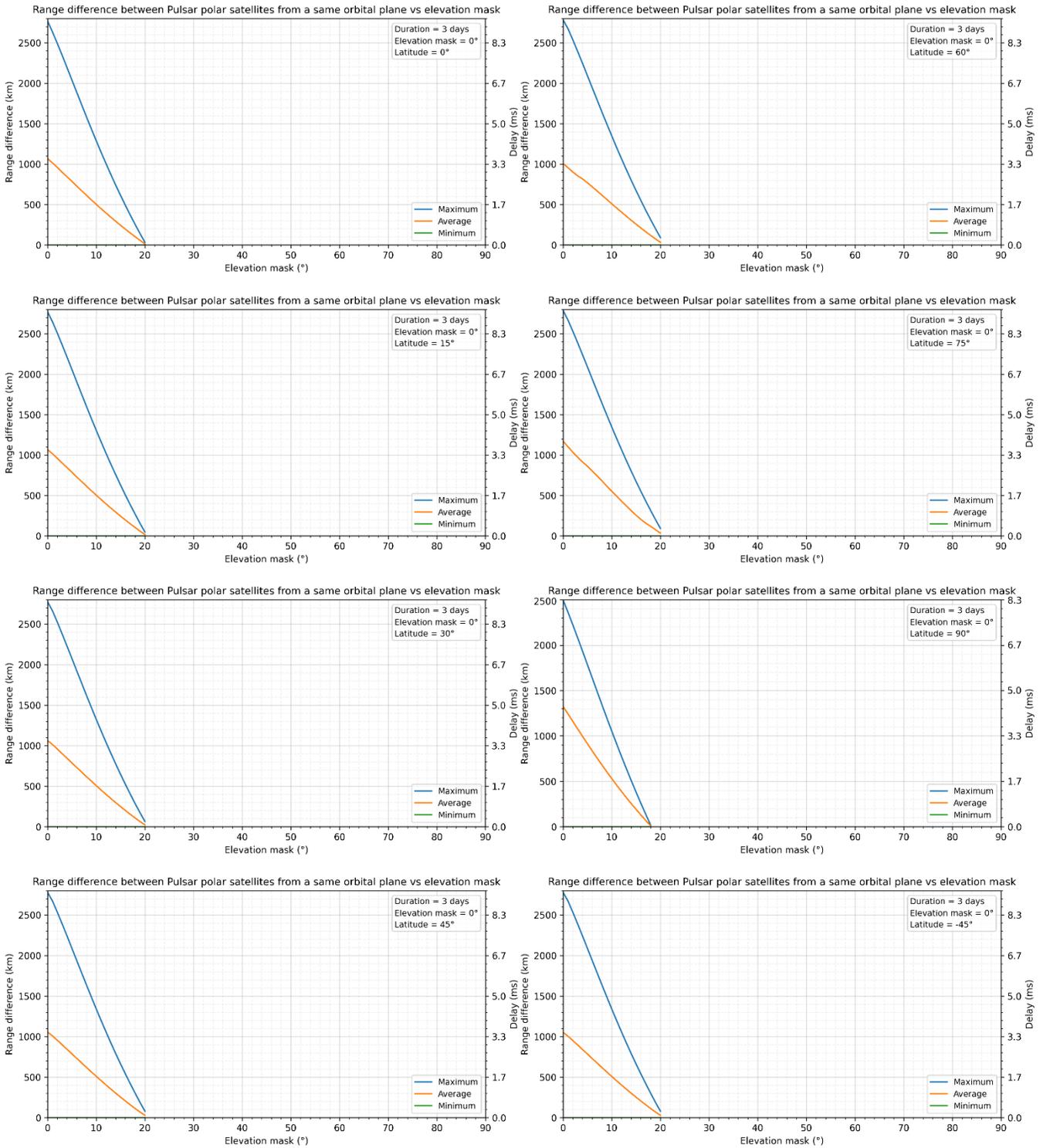

**Figure B.33** Range difference between Pulsar polar satellites of a same orbital plane vs elevation mask for different latitudes

## B.12. Doppler Difference (Same Orbital Plane) Over Time

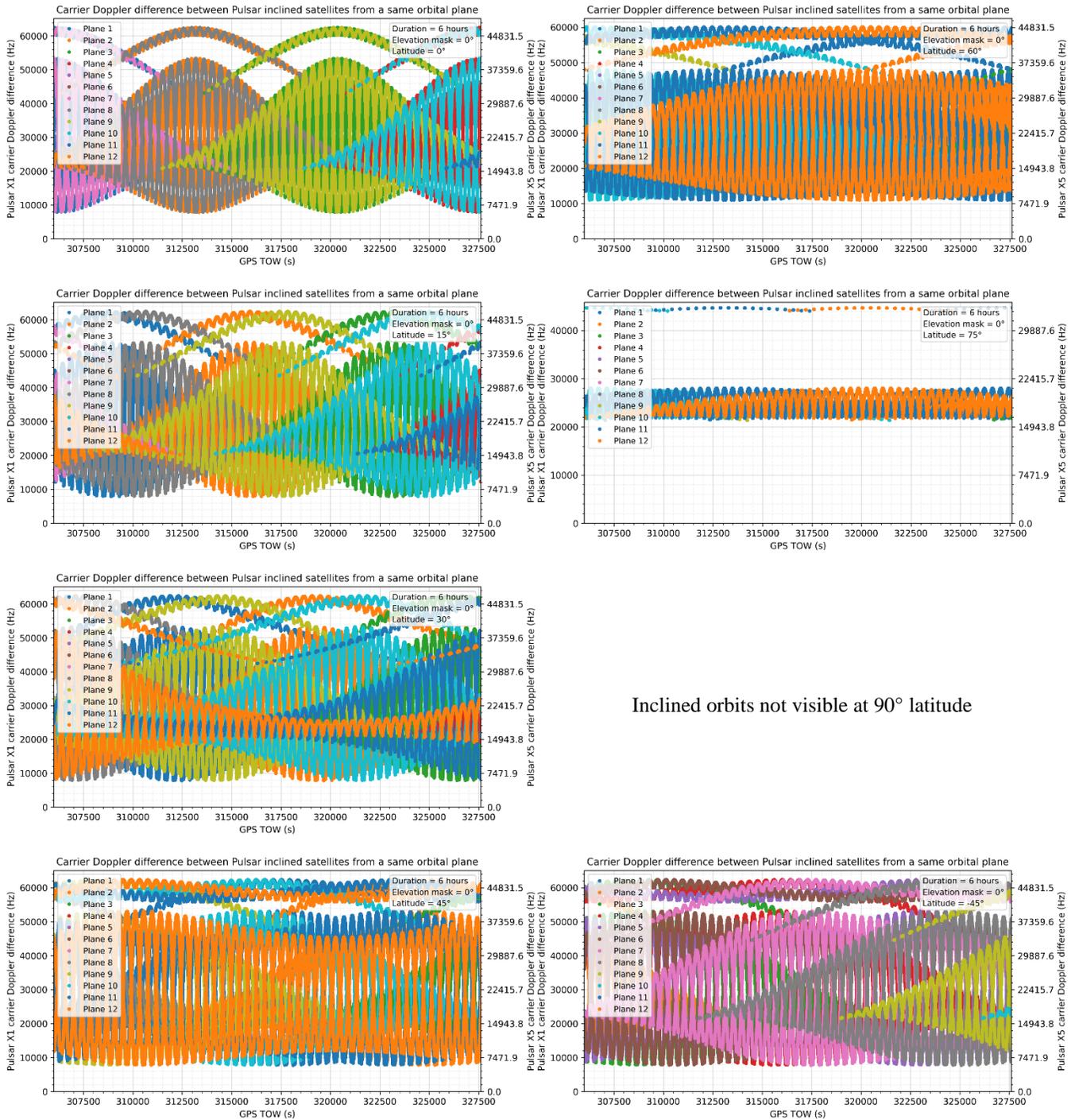

**Figure B.34** Doppler difference between Pulsar inclined satellites from a same orbital plane over time for different latitudes

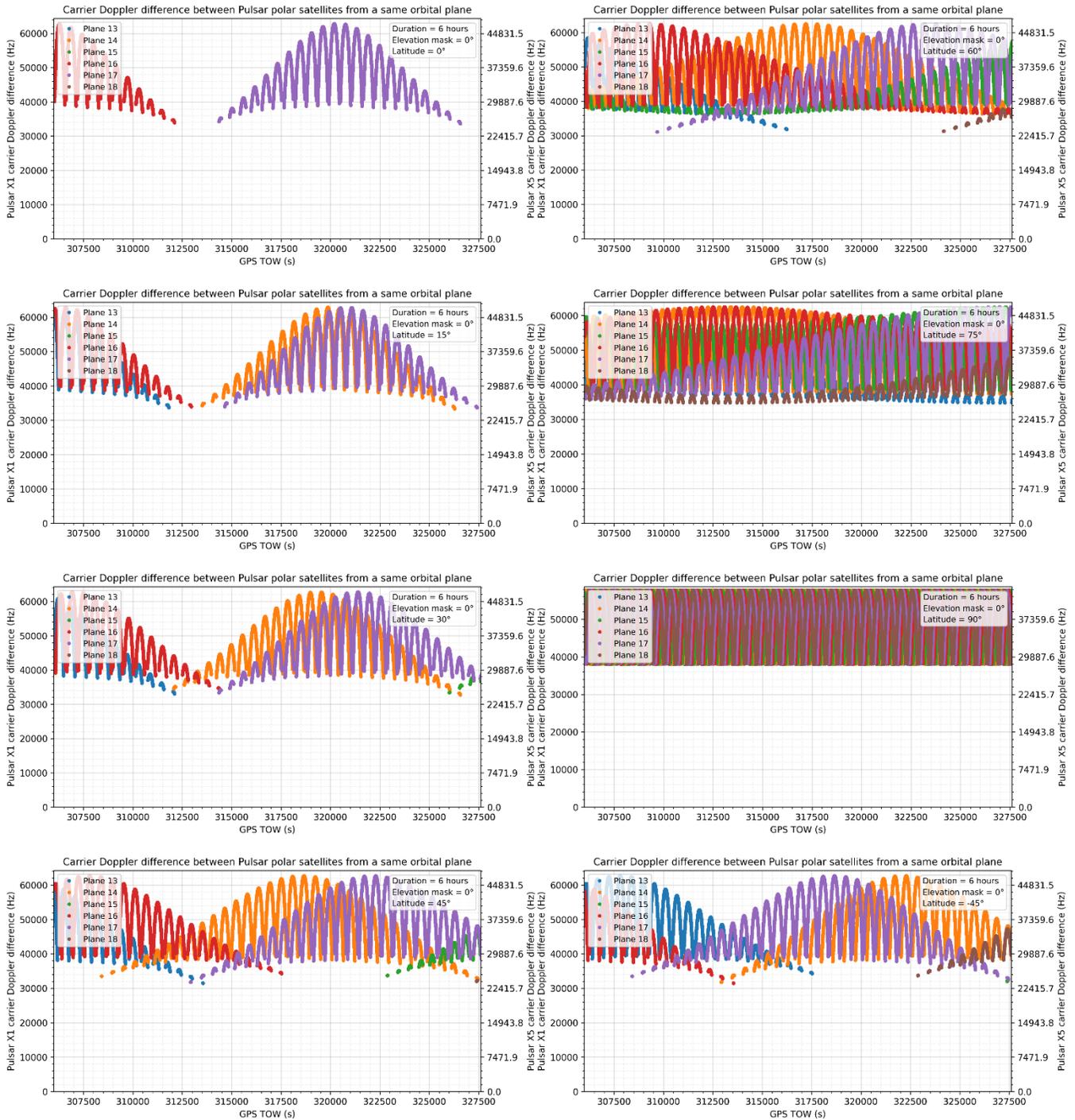

**Figure B.35** Doppler difference between Pulsar polar satellites from a same orbital plane over time for different latitudes

## B.13. Doppler Difference (Same Orbital Plane) vs Elevation Mask

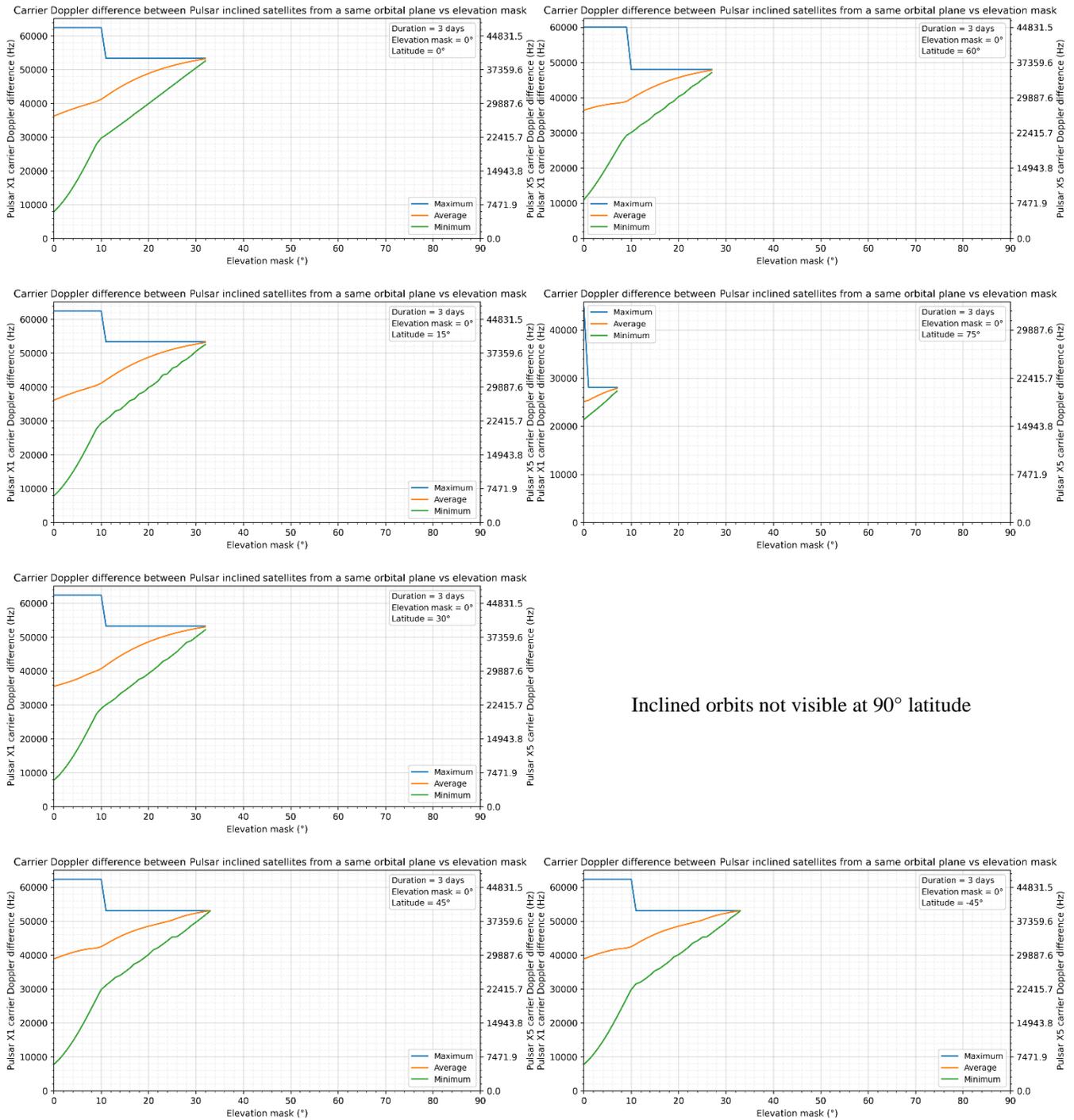

**Figure B.36** Doppler difference between Pulsar inclined satellites of a same orbital plane vs elevation mask for different latitudes

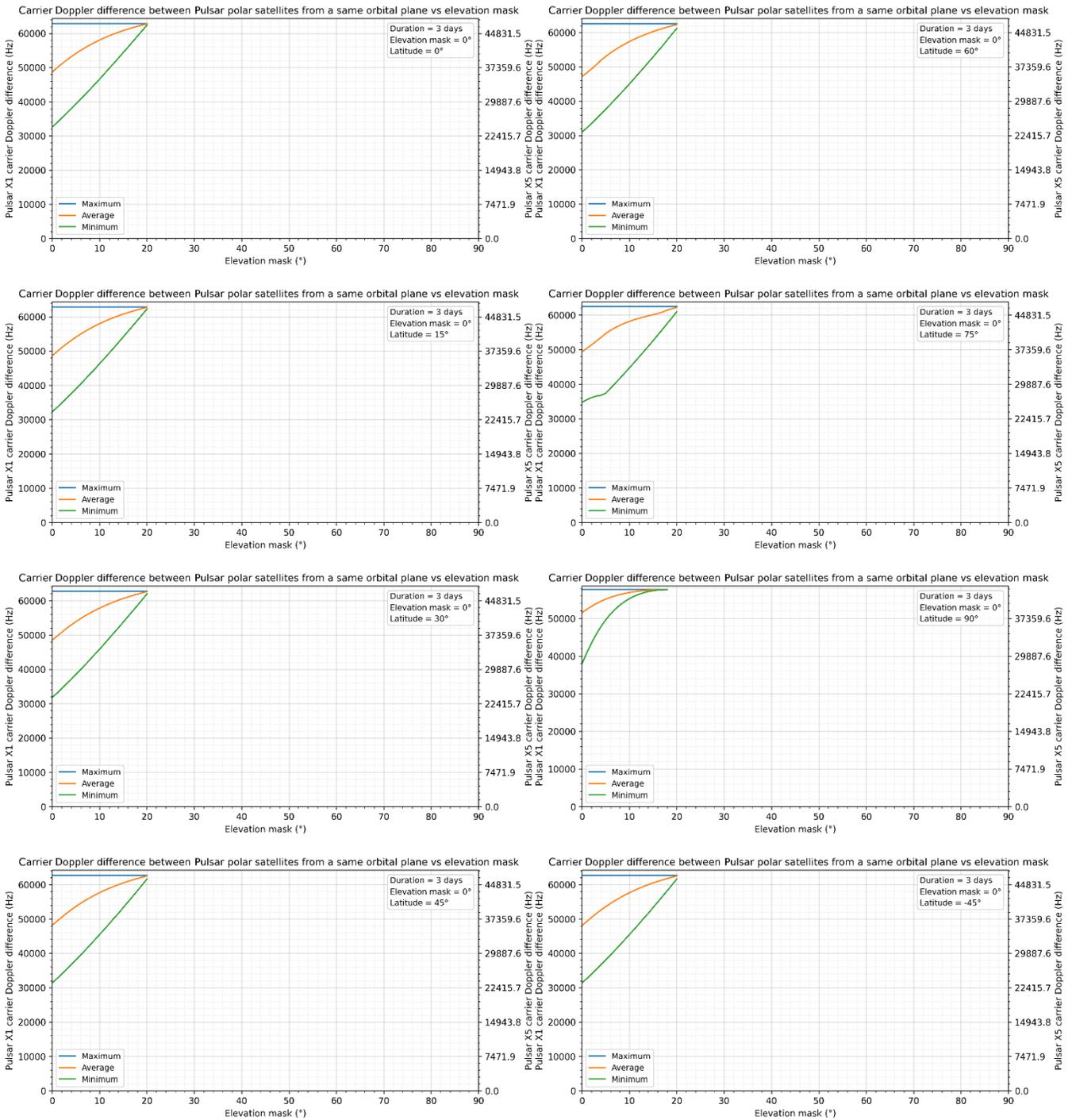

**Figure B.37** Doppler difference between Pulsar polar satellites of a same orbital plane vs elevation mask for different latitudes

## B.14. Doppler Rate Difference (Same Orbital Plane) Over Time

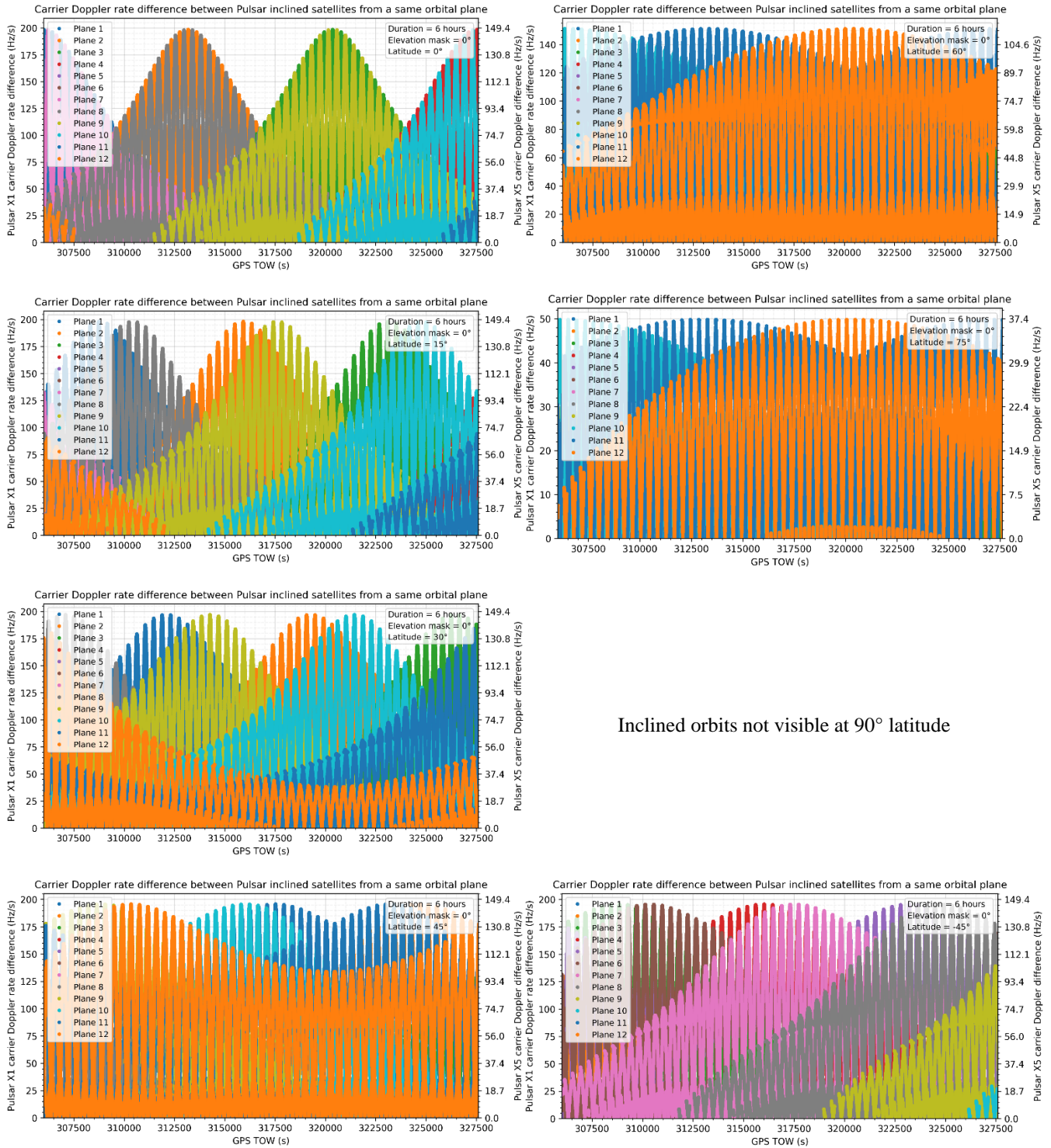

**Figure B.38** Doppler rate difference between Pulsar inclined satellites from a same orbital plane over time for different latitudes

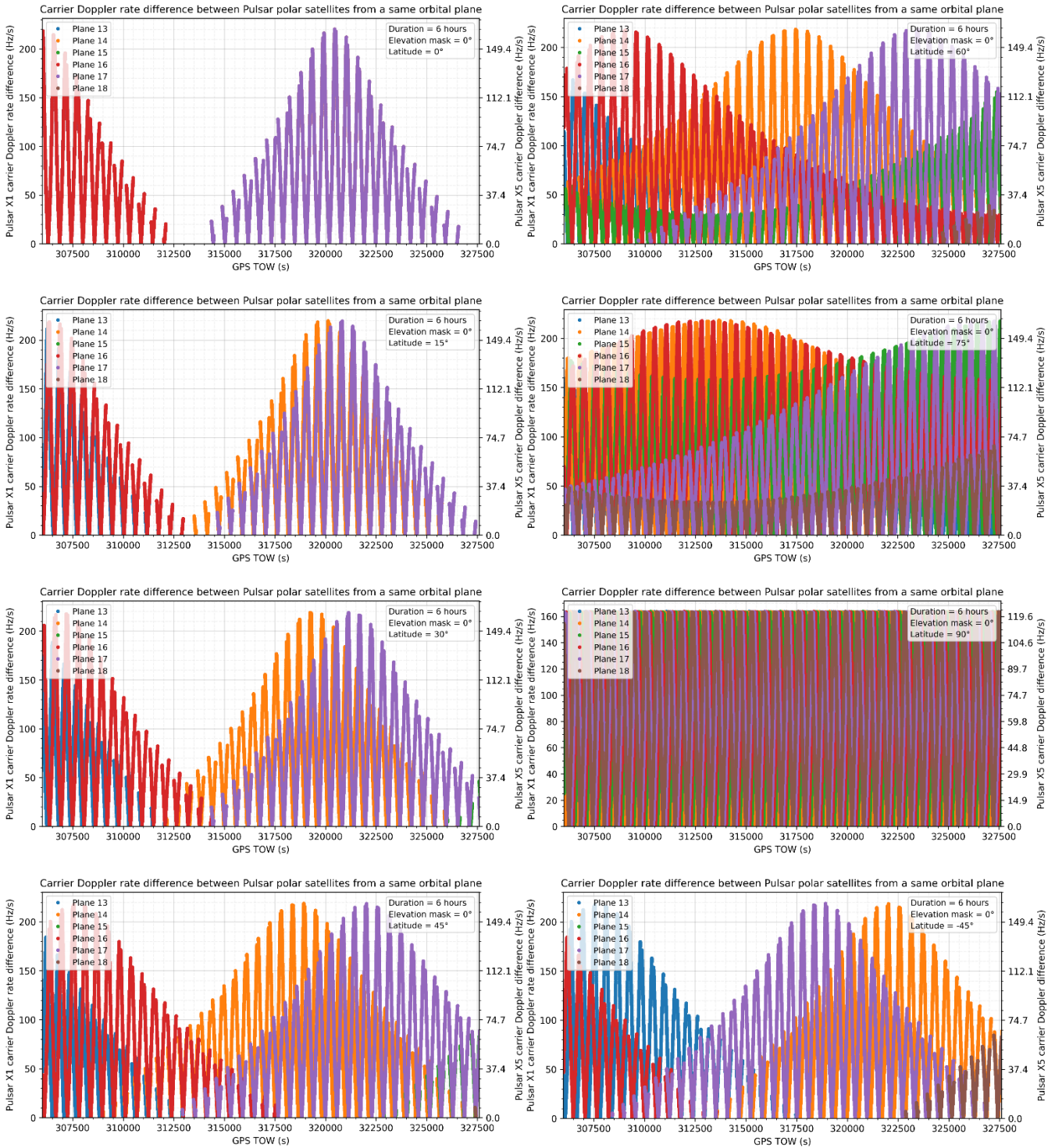

**Figure B.39** Doppler rate difference between Pulsar polar satellites from a same orbital plane over time for different latitudes

## B.15. Doppler Rate Difference (Same Orbital Plane) vs Elevation Mask

**Figure B.40** Doppler rate difference between Pulsar inclined satellites of a same orbital plane vs elevation mask for different latitudes

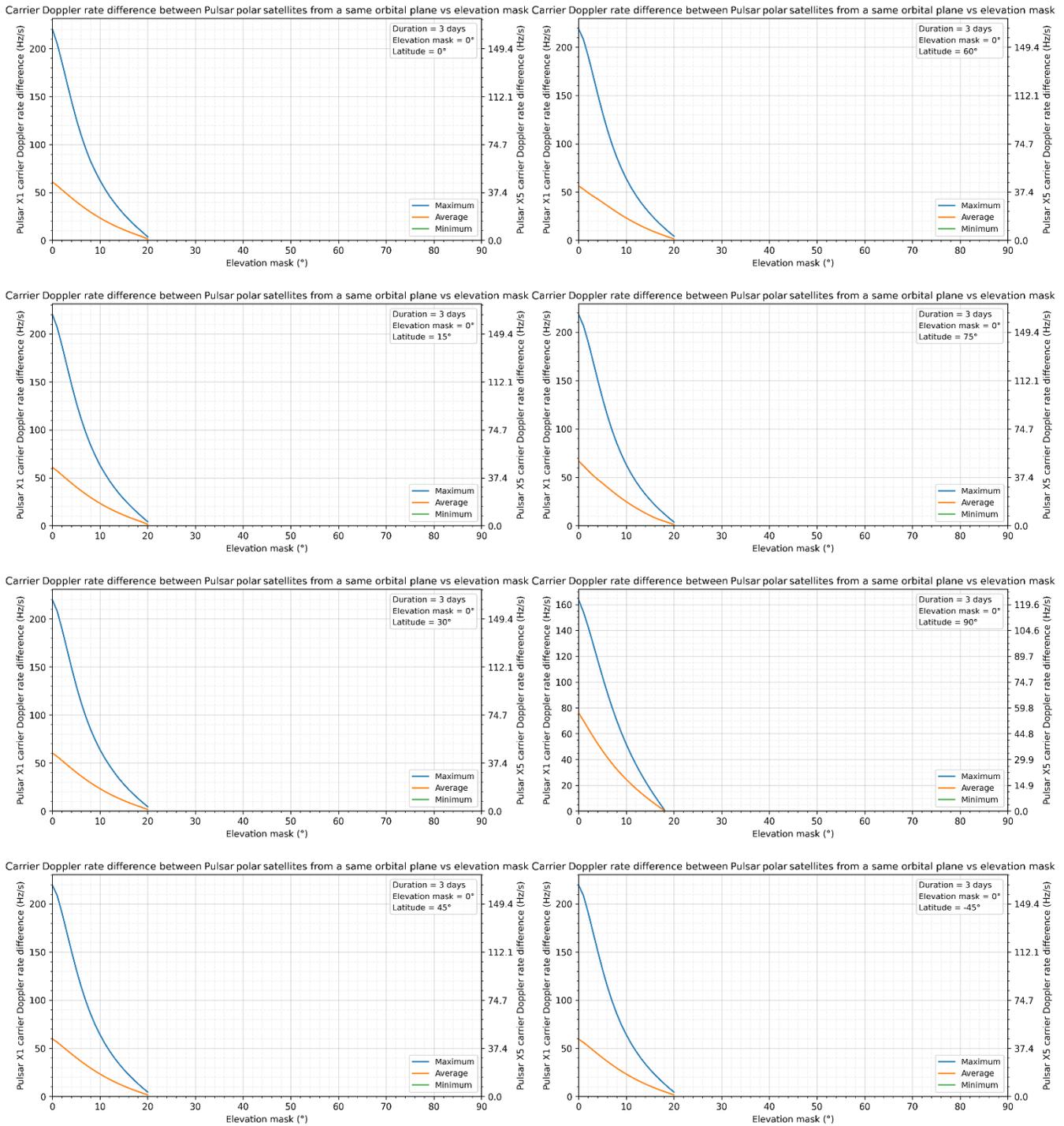

**Figure B.41** Doppler rate difference between Pulsar polar satellites of a same orbital plane vs elevation mask for different latitudes